\documentclass[%
 reprint,
groupedaddress,
amsmath,amssymb,
aps,
pra,
]{revtex4-1}
\usepackage{graphicx}% Include figure files
\usepackage{dcolumn}% Align table columns on decimal point
\usepackage{bm}% bold math
\usepackage{float}
\usepackage{multirow}

\usepackage[dvipdfm,colorlinks,linkcolor=blue, urlcolor=blue, anchorcolor=blue, citecolor=blue]{hyperref}
\begin{document}

\title{Cavity-Heisenberg spin-$j$ chain quantum battery and reinforcement learning optimization}

%\title{Cavity-Heisenberg spin-$j$ chain quantum battery and the charging performance optimization through reinforcement learning}

%\title{Reinforcement learning optimization of charging performance in cavity-Heisenberg large-spin chain quantum battery}

\author{Peng-Yu Sun}
\author{Hang Zhou}
\author{Fu-Quan Dou}
\email{doufq@nwnu.edu.cn}
\affiliation{College of Physics and Electronic Engineering, Northwest Normal University, Lanzhou, 730070, China}
%Institute of Quantum Physics, Hunan Key Laboratory of Nanophotonics and Devices, and Hunan Key Laboratory of Super-Microstructure and Ultrafast Process, School of Physics, Central South University, Changsha 410083, China

%%%%%%%%%%%%%%%%%%%%%%%%%%%%%%%%%%%%%%%%%%%%%%%%%%%%%%%%%%%%%%%%%%%%%%%%%%%%%%%%%%%%
\begin{abstract}
Machine learning offers a promising methodology to tackle complex challenges in quantum physics. In the realm of quantum batteries (QBs), model construction and performance optimization are central tasks. Here, we propose a cavity-Heisenberg spin chain quantum battery (QB) model with spin-$j~(j=1/2,1,3/2)$ and investigate the charging performance under both closed and open quantum cases, considering spin-spin interactions, ambient temperature, and cavity dissipation. It is shown that the charging energy and power of QB are significantly improved with the spin size. By employing a reinforcement learning algorithm to modulate the cavity-battery coupling, we further optimize the QB performance, enabling the stored energy to even exceed the upper bound in the absence of spin-spin interaction. We analyze the optimization mechanism and find an intrinsic relationship between cavity-spin entanglement and charging performance: increased entanglement enhances the charging energy in closed systems, whereas the opposite effect occurs in open systems. Our results provide a possible scheme for design and optimization of QBs.
\end{abstract}
%%%%%%%%%%%%%%%%%%%%%%%%%%%%%%%%%%%%%%%%%%%%%%%%%%%%%%%%%%%%%%%%%%%%%%%%%%%%%%%%%%%%

\maketitle

\section{INTRODUCTION}
Quantum mechanics has attracted considerable attention due to its importance in driving scientific and technological progress, ranging from quantum communication \cite{RevModPhys.80.1083,PhysRevLett.102.240501}, quantum sensing \cite{PhysRevLett.131.150802,RevModPhys.89.035002}, to quantum computing \cite{PRXQuantum.2.020343,RevModPhys.79.135}. Among these, quantum thermodynamics has emerged as a field that aims to reconstruct thermodynamics through the fundamental laws of quantum mechanics, and one of its important tasks is to focus on work, heat, and entropy within a quantum framework \cite{RevModPhys.93.035008,RevModPhys.92.041002,Skrzypczyk2014}. In the realm of energy storage, the concept of the quantum battery (QB) has been proposed by applying the principles of quantum thermodynamics to revolutionize conventional battery technology \cite{campaioli2018quantum,PhysRevE.87.042123,RevModPhys.96.031001}. Experiments have also shown advances towards the exploration of quantum batteries (QBs) \cite{Hu_2022,quach2020,Zheng_2022,batteries8050043,PhysRevLett.131.260401,PhysRevA.106.042601}.

Model construction of a QB is prerequisite for its realization. With various QB models proposed \cite{PhysRevA.109.032201,PhysRevResearch.2.023095,PhysRevA.110.032205,Beleo_2024,PhysRevB.98.205423,
PhysRevResearch.2.023095,PhysRevB.109.235432,PhysRevLett.125.236402,PhysRevE.100.032107,Dou2020,PhysRevLett.132.210402,PhysRevA.106.022618,lu2024topologicalquantumbatteries}, two theoretical models have gained traction: cavity QBs \cite{PhysRevA.100.043833,PhysRevE.104.044116,PhysRevE.94.052122,PhysRevLett.122.047702,PhysRevLett.120.117702,zhang2018enhanced,PhysRevB.102.245407,PhysRevB.105.115405,https://doi.org/10.1002/qute.202400115,PhysRevA.108.062402} and spin chain QBs \cite{PhysRevLett.133.197001,PhysRevE.106.014138,PhysRevA.103.033715,PhysRevA.110.052601,PhysRevResearch.5.013155,PhysRevA.103.052220,PhysRevLett.129.130602,PhysRevA.97.022106,DouLMG2022,PhysRevA.105.022628,PhysRevA.110.052404,PhysRevA.101.032115,PhysRevE.104.024129,PhysRevB.100.115142,deMoraes_2024,PhysRevA.109.042207,PhysRevE.102.052109,PhysRevA.109.042619,PhysRevA.109.042411}. Cavity QBs rely on the properties of quantum cavities or optical resonators to store and release energy by controlling the interaction between the cavity and the battery which provides advantages in rapid charging  \cite{PhysRevB.102.245407,zhang2018enhanced,PhysRevLett.120.117702}. Spin chain QBs utilize quantum entanglement to enhance the efficiency and speed of energy storage. Large spin QBs further employ collective spin states in ensembles of magnetic ions or molecules, which provide high-energy storage \cite{PhysRevE.106.054119,PhysRevResearch.4.043150}. A significant development is the cavity-Heisenberg spin chain QB, which combines the benefits of a spin chain and quantum cavities. The integration enhances stored energy, increases charging power, and demonstrates a quantum advantage \cite{PhysRevA.106.032212}. Besides, open QBs consider factors such as dissipation and decoherence, and can be used to address issues related to stable charging and energy loss \cite{PhysRevE.104.064143,PhysRevB.99.035421,PhysRevE.104.044116,PhysRevA.109.062432,PhysRevA.107.023725,PhysRevB.99.035421,PhysRevA.102.052223,Caravelli2021energystorage,PhysRevE.104.054117,PhysRevA.110.032211,PhysRevA.109.042411}.

Performance optimization is also a crucial topic in QB research. Current optimization methods include the utilization of quantum resources \cite{PhysRevA.109.062432,PhysRevE.102.052109,PhysRevA.108.062402,PhysRevE.94.052122,PhysRevB.109.235432,PhysRevLett.111.240401,PhysRevLett.129.130602,Gumberidze2019,10.1116/5.0184903}, control of charging modes \cite{PhysRevA.103.033715,PhysRevE.106.014138,Dou2020,PhysRevA.110.052601,PhysRevLett.131.240401,Dou2021,PhysRevLett.128.140501,PhysRevE.99.052106,PhysRevA.109.052206,PhysRevA.107.032218,Rodrguez_2024,PhysRevResearch.6.023091}, using of model characteristics \cite{PhysRevA.105.022628,PhysRevE.104.024129,PhysRevA.101.032115,PhysRevA.97.022106,DouLMG2022,PhysRevA.110.052404,PhysRevA.110.012227,PhysRevA.109.012204,PhysRevB.105.115405}, and consideration non-Markovian dynamics \cite{PhysRevE.104.054117,PhysRevA.102.052223,https://doi.org/10.1002/qute.202400115,PhysRevLett.132.090401,PhysRevE.109.054132,PhysRevA.109.012224}. However, precise control over complex systems often presents challenges for practical application. Fortunately, the rapid development of reinforcement learning (RL) has shown promising applications in the quantum domain \cite{RevModPhys.91.045002,PhysRevLett.120.066401,PhysRevResearch.3.043184,PhysRevResearch.2.032051,PRXQuantum.2.010328,PhysRevX.8.031084,PhysRevLett.127.190403,PhysRevLett.125.170501,PhysRevLett.127.110502}. Especially in QBs, RL has been applied to optimise the charging process in Dicke QBs, which leads to higher energy extraction and greater charging precision compared to conventional methods \cite{PhysRevLett.133.243602}. It has also been used to develop stable charging protocols for micromaser QBs, which significantly enhances their overall efficiency \cite{PhysRevA.108.042618}. In RL algorithms, the soft actor-critic (SAC) algorithm offers a more advanced solution that enables efficient and adaptive optimization of complex parameter spaces. This approach not only accelerates the exploration of possible configurations but yields more precise and reliable performance improvements  \cite{haarnoja2019softactorcriticalgorithmsapplications,haarnoja2018softactorcriticoffpolicymaximum}.

Inspired by the development of QBs, we focus on two main issues. One is how to construct a more efficient QB model by combining cavity QBs with large spin QBs. The other is whether the performance of this QB can be further optimized through RL. In this work, we propose a cavity-Heisenberg spin chain QB model with large spins, where the stored energy, charging power, and entanglement property of the QBs for the chain with spin-$1/2$, spin-$1$, and spin-$3/2$ configuration are explored. The charging performance of the QBs can be effectively modulated by means of spin size, cavity-spin coupling and spin-spin interactions. Furthermore, we study the cavity QBs in the case of open systems with ambient temperature and cavity dissipation. Based on the SAC algorithm, the charging process in both closed and open systems are optimized through tuning the cavity-spin coupling parameter, where an intrinsic relationship between cavity-spin entanglement and charging performance is revealed that the increased entanglement enhances charging energy in a closed system, whereas the opposite effect occurs in the open system.

The rest of paper is organized as follows. In Sec. \ref{section2} we introduce the cavity-Heisenberg large-spin chain QB, performance metrics, and the RL optimization algorithm. In Sec. \ref{section3} the charging process of the QB with three different spin configurations in a closed system are investigated, where the influence of entanglement and the RL optimization on the cavity QB are studied. Furthermore, we examine the charging dynamics of the QB in an open system and explore the corresponding entanglement properties and the RL optimization in Sec. \ref{section4}. Finally, a brief conclusion is given in Sec. \ref{section5}.

\section{MODEL AND APPROACH}\label{section2}

\begin{figure}[tbp]
\centering
\includegraphics[width=0.445\textwidth]{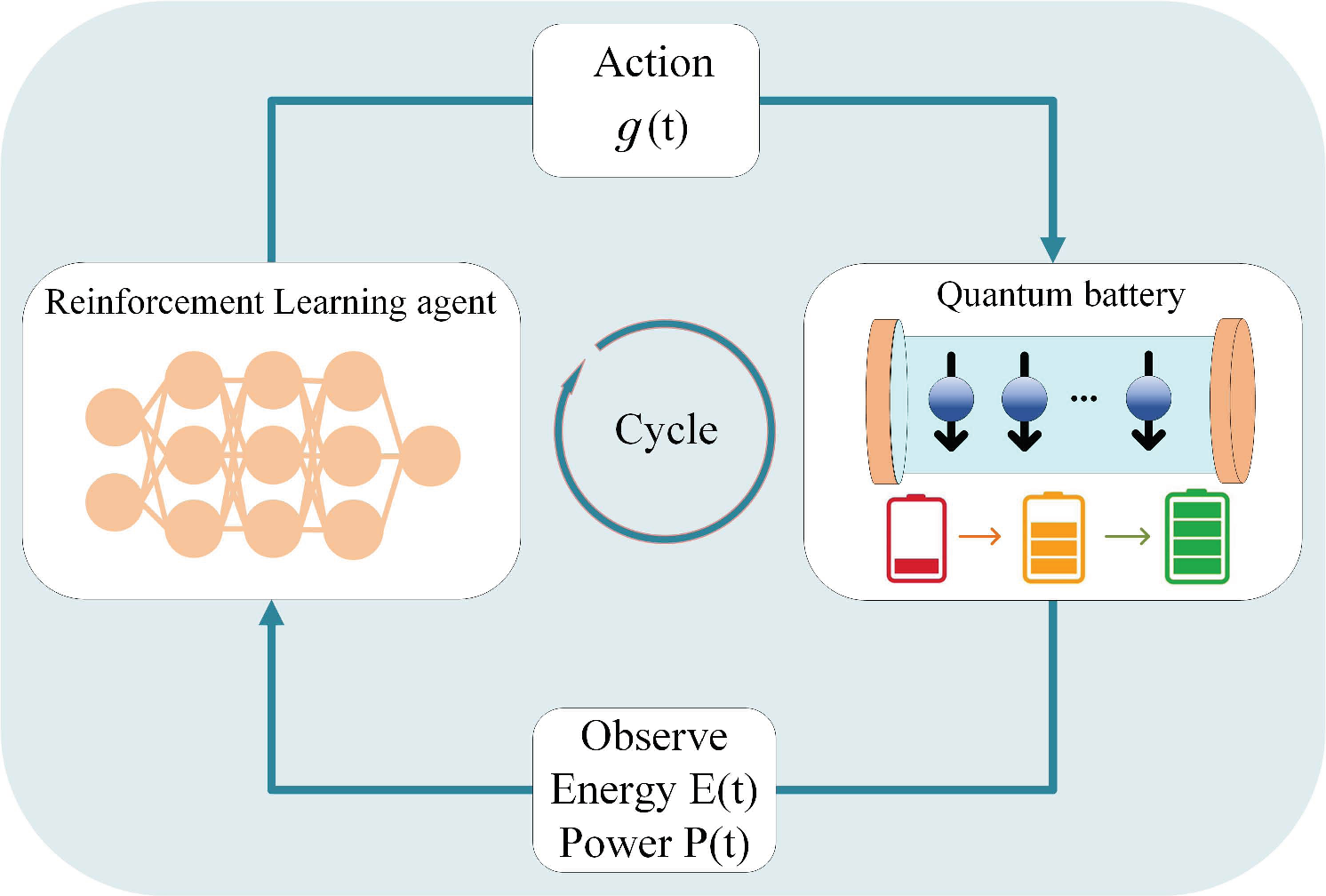}
\caption{Schematic diagram of an RL algorithm for optimising the charging performance of a cavity-Heisenberg spin chain QB. An RL agent determines the external control action of the cavity-spin coupling $g(t)$ by observing the current state of stored energy $E(t)$ and average charging power $P(t)$ of the QB, thereby maximizing the stored energy and achieving a relatively high power. The optimization process consists of numerous iterations between the RL algorithm and the QB. Through this cycle, QB charging efficiency is refined to its optimal level.}
\label{fig1}
\end{figure}

We consider a cavity-Heisenberg large-spin chain QB model, which consists of single-mode cavity as the charger and a Heisenberg spin chain with spin-spin interactions as the battery, as shown in the QB part of Fig.~\ref{fig1}. The whole system can be described by the Hamiltonian
\begin{eqnarray}\label{H}
H=H_{C}+H_{B}+\lambda(t)H_{I},
\end{eqnarray}
where $H_{C}$ and $H_B$ represent the charger and the battery, and $H_I$ is the interaction term with the charging time interval $\lambda(t)$ given by a step function equal
 to $1$ for $t\in[0,T]$ ($T$ is the total charging time) and zero elsewhere. The various terms (hereafter we set $\hbar= 1$) can be expressed as
\begin{eqnarray}\label{HC}
H_{C}=\omega_{c}\hat{a}^{\dag}\hat{a},
\end{eqnarray}
\begin{eqnarray}\label{HB}
\begin{split}
H_{B}=\omega_{a}\sum_{n=1}^{N}\hat{S}^{z}_{n}&+\omega_{a}J\sum_{n=1}^{N-1}[(1+\gamma)\hat{S}^{x}_{n}\hat{S}^{x}_{n+1}\\
&+(1-\gamma)\hat{S}^{y}_{n}\hat{S}^{y}_{n+1}+\Delta\hat{S}^{z}_{n}\hat{S}^{z}_{n+1}],
\end{split}
\end{eqnarray}
\begin{eqnarray}\label{HI}
H_{I}=g\sum_{n=1}^{N}(\hat{S}^{+}_{n}+\hat{S}^{-}_{n})(\hat{a}^{\dag}+\hat{a}),
\end{eqnarray}
where $\hat{a}~(\hat{a}^{\dag})$ is annihilation (creation) operator and the cavity field frequency is $\omega_{c}$. $\hat{S}^{i}_{n}$ with $i = x, y, z$ are the spin operators of the site $n$ and $J$ is the nearest-neighbor interaction between spins. $\omega_a$ is the frequency of spins and the strength of the spin-cavity coupling is given by the parameter $g$. $\gamma$ and $\Delta$ are the anisotropy coefficients and $N$ is the number of spins. $\hat{S}^{+}_{n}~(\hat{S}^{-}_{n})$ represents the raising (lowering) operator. In the case of spin-$1/2$ particles, $\hat{S}^{i}_{n}$ is the spin Pauli operators on site $n$. In order to ensure the maximum energy transfer, we will focus on the resonance regime (\emph{i.e.}, $\omega_{a}=\omega_{c}=1$), and the off-resonance case $\omega_{a} \neq \omega_{c}$ will not be considered since it characterizes a less efficient energy transfer between the cavity and spins. In all calculations, for simplicity, we take the parameters  $N=3, \gamma=0.4, \Delta=1$, and the maximum number of photons $N_{\text{Fock}}=4N+1$. Numerical work has been performed by using PyTorch \cite{Paszke2019} and QuTiP2 toolbox \cite{JOHANSSON20131234}.

At time $t\leq 0$, the QB is prepared in the ground state of $H_{B}$ and coupled to a single-mode cavity in the $N$ photons' Fock-state. Thus, the initial state of the total system is
\begin{eqnarray}
|\psi(0)\rangle = |G\rangle_{B}\otimes|N\rangle_{C}.
\end{eqnarray}

When environmental factors are taken into account, the system is treated as open, and the dynamic process of the QB charging can be described by solving the Lindblad master equation
\begin{eqnarray}\label{lind}
\frac{\mathrm{d} \rho(t)}{\mathrm{d} t}=-\frac{\mathrm{i}}{\hbar}\left[H,\rho(t)\right]+\mathfrak{D}[\rho(t)],
\end{eqnarray}
where $\rho(t)$ is the density matrix of the system at time $t$. $\mathfrak{D}[\cdot]$ represents the dissipative superoperator. In the open system, the QB has a practical significance only when the cavity dissipation $\kappa$ is much greater than spin dissipative $\kappa_{s}$, \emph{i.e.}, $\kappa\gg\kappa_{s}$. We only consider the effects of dissipation and ambient temperature on the cavity field, and ignore the interaction between spin and environment. Therefore, the dissipative superoperator $\mathfrak{D}[\cdot]$ can be expressed as
\begin{eqnarray}
\begin{split}
\mathfrak{D}[\rho(t)] &= \frac{1}{2}\kappa(n_{th}+1)\left[2a\rho(t) a^{\dagger}-a^{\dagger}\rho(t)a - \rho(t) a^{\dagger}a\right] \\
&+\frac{1}{2}\kappa n_{th}\left[2a^{\dagger}\rho(t)a - aa^{\dagger}\rho(t)-\rho(t)aa^{\dagger}\right],
\end{split}
\end{eqnarray}
where $n_{th}=1/\left\{\mbox{exp}[(\hbar\omega_{c})/(k_{B}T)]-1\right\}$ is the mean occupation number of the boson heat bath. $k_{B}$ is the Boltzmann constant and $T$ is the ambient temperature. When $\kappa=0$, the environment has no influence on the system and the system is a closed one.

The stored energy $E(t)$ and the average charging power $P(t)$ are two typical metrics for charging performance of QB, which can be defined as
\begin{eqnarray}\label{E}
E(t)=\mbox{Tr}[H_{B}\rho_{B}(t)]-\mbox{Tr}[H_{B}\rho_{B}(0)],
\end{eqnarray}
\begin{eqnarray}\label{P}
P(t)=E(t)/t,
\end{eqnarray}
where $\rho_{B}(t)$ is the reduced density matrix of the QB at the time $t$. The entanglement between the cavity and the spin can be given by the logarithmic negativity \cite{PhysRevLett.95.090503,PhysRevA.106.032212}
\begin{equation}
E_\mathcal{N}= \mbox{log}_2\Vert\rho^{T_B}\Vert_{1},
\end{equation}
where the $\rho^{T_B}$ denotes the partial matrix of $\rho$ with respect to the subsystem $B$.

The SAC algorithm is one of outstanding RL algorithms and has already been applied in the field of quantum physics such as seeking improved control policies in quantum thermal machines \cite{Erdman2022}. We will employ the SAC algorithm to optimize the charging performance of the QB. As shown in Fig.~\ref{fig1}, the optimization process is illustrated, where the RL agent is a neutral network to optimize the cavity QB by tuning the interaction between charger and battery. The RL agent manages an external control function $g(t)$, which influences the cavity-battery coupling and its action is based on the current status of charging performance by observing $E(t)$ and $P(t)$. The observed results of the two functions are further fed back to the RL agent who would adjust the control parameter $g(t)$ in order to maximize the stored energy $E(t)$ and achieving a relatively high average charging power $P(t)$. The optimization procedure can be realized by a continuous cyclic process exploring the state-action space and refining policy, so that the charging efficiency of QBs can be continuously improved to the optimal level. The details of the SAC algorithm are presented in Appendix~\ref{appendix1}.

\section{CLOSED SYSTEM: $\kappa=0$}\label{section3}

\begin{figure}[tbp]
\centering
\includegraphics[width=0.485\textwidth]{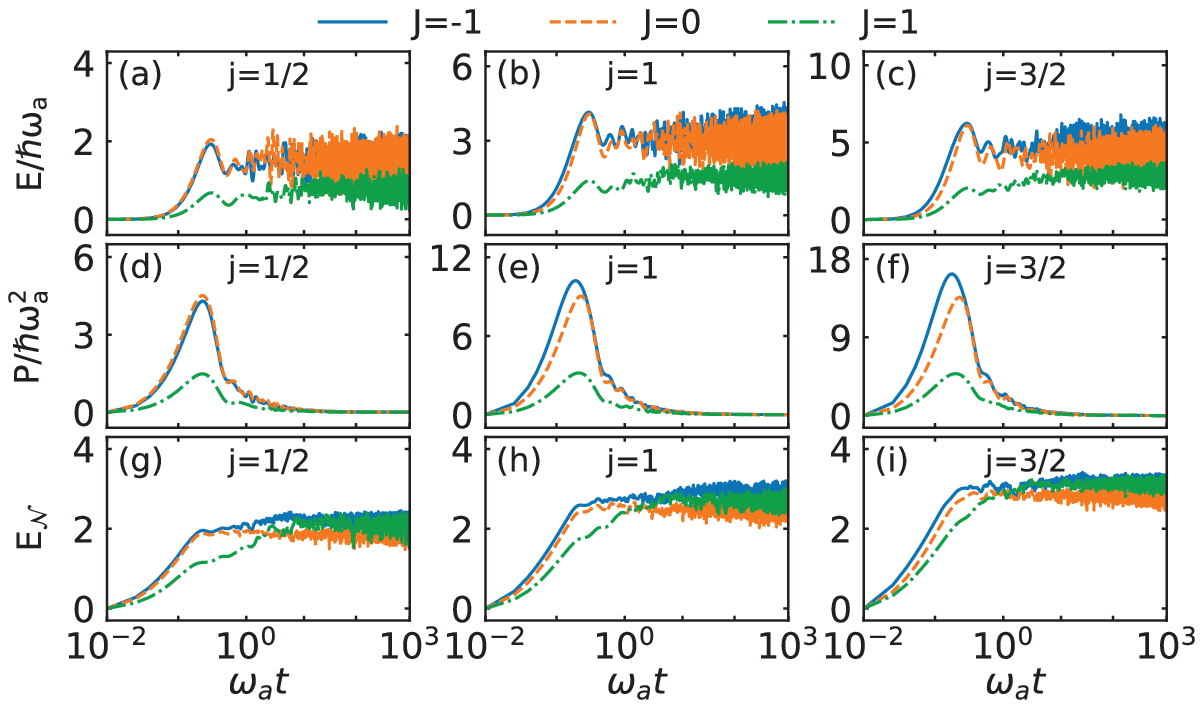}
\caption{The dependence of (a)-(c) the stored energy $E(t)$ (in units of $\hbar\omega_{a}$), (d)-(f) average charging power $P(t)$ (in units of $\hbar\omega^{2}_{a}$), and (g)-(i) logarithmic negativity $E_\mathcal{N}(t)$ of closed system QB as a function of $\omega_a t$ for different values of spin $j$. The different curves in these plots stand for various spin-spin interaction $J$, as indicated in the legends. The cavity-spin coupling is chosen as $g = 1$.}
\label{fig2}
\end{figure}

\begin{figure}[t]
\centering
\includegraphics[width=0.465\textwidth]{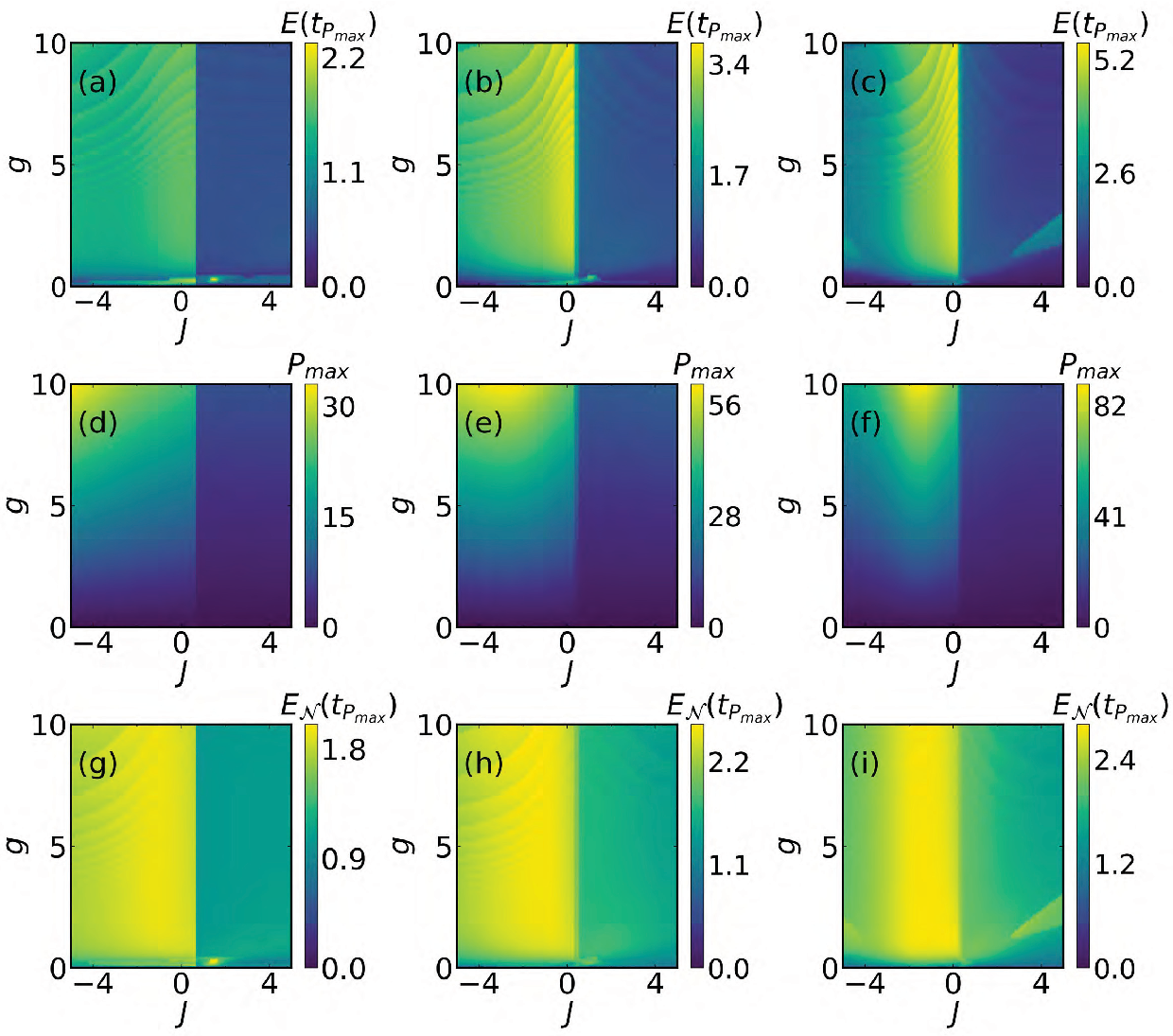}
\caption{The contour plots of closed system QB's (a)-(c) stored energy $E(t_{P_{max}})$ (in units of $\hbar\omega_{a}$), (d)-(f) maximum charging power $P_{max}$ (in units of $\hbar\omega_{a}^{2}$), and the logarithmic negativity $E_\mathcal{N}(t_{P_{max}})$ as functions of the cavity-spin coupling strength $g$ and spin-spin interaction strength $J$ for different spin $j$: (a), (d) and (g) spin-$1/2$, (b), (e) and (h) spin-$1$, (c), (f) and (i) spin-$3/2$.}
\label{fig3}
\end{figure}

We first study the charging properties of the QB in the case of closed system which corresponds to the dissipative parameter $\kappa=0$. To investigate the behavior of the QB during the charging process, we calculate the time-dependent of stored energy $E(t)$, average charging power $P(t)$, and the entanglement $E_{\mathcal{N}}(t)$ between the cavity and the spin chain with different spin-$j$ configurations, and the results are illustrated in Fig.~\ref{fig2} for the cavity-spin coupling $g = 1$. It shows that the larger the spin-$j$ of QB, the greater the energy $E(t)$, the power $P(t)$ as well as the entanglement $E_{\mathcal{N}}(t)$ between charger and battery. Here we focus on the qualitative relationship between stored energy and cavity-spin entanglement under the specific dynamics with the same spin-spin interaction. For a given spin-spin interaction, the stored energy and the cavity-spin entanglement evolve with similar behavior over time. For example, for the case of $J=1$, two green curves in Fig.~\ref{fig2}(a) and ~\ref{fig2}(g) exhibit the consistent behaviors (see also Fig.~\ref{fig3}). This property also holds in cases with other $j$. This is because that the energy occupancy of QB changes from the lowest energy state to some higher energy states (see Appendix~\ref{appendix2} for details), which results in the entanglement increasing correspondingly in the evolution of closed system. Moreover, the performance of QBs are also influenced by the spin-spin interaction, where the antiferromagnetic interaction ($J>0$) may diminish charging efficiency.

It is noted that the stored energy $E(t)$ of the QBs first begins to rise rapidly and then exhibits an oscillation phenomenon due to continuous exchange of energy between the cavity and the spin in the closed system, which presents a challenge to achieving the maximal stored energy. A potential solution is to cease the charging process when the average charging power $P(t)$ of the QB reaches its peak, and the energy, its corresponding power and the entanglement at this specific moment can be labeled as $E(t_{P_{max}})$, $P_{max}$ and $E_\mathcal{N}(t_{P_{max}})$, respectively. We analyze the influence of the cavity-spin coupling $g$ and the spin-spin interaction strength $J$ on the $E(t_{P_{max}})$, $P_{max}$ and $E_\mathcal{N}(t_{P_{max}})$ for different spin-$j$ configurations. The energy, the corresponding charging power and logarithmic negativity as the functions of the parameters $g$ and $J$ are shown in Fig.~\ref{fig3}. It is shown that the cavity-spin coupling and the spin-spin interaction can modulate effectively the energy $E(t_{P_{max}})$ and the power $P_{max}$, and the QB with higher spin configuration can achieve better charging performance. In addition, the strong antiferromagnetic spin-spin interaction results in the lower $E(t_{P_{max}})$ and $P_{max}$, and the enhanced cavity-spin coupling $g$ can boost the charging power $P_{max}$. The high stored energy range occurs in regions where the interaction is weak. When antiferromagnetic interaction approaches a critical value, the charging efficiency suddenly becomes low, i.e., both the stored energy and the charging power become smaller. Along with the increase of the spin size, on the one hand the energy $E(t_{P_{max}})$ and the power $P_{max}$ will increase, and on the other hand the zone of the maximum of average charging power $P_{max}$ will move in the parameter space of $J$ and $g$ (yellow regions in Fig.~\ref{fig3}(d-f)). This means that for a small spin system, a large ferromagnetic spin-spin interaction is necessary to achieve a high charging power, and for a large spin QB, only a weak ferromagnetic spin-spin interaction is required to obtain a high charging power. By comparing Fig.~\ref{fig3}(g-i) and Fig.~\ref{fig3}(a-c), we further verify that the cavity-spin entanglement $E_\mathcal{N}(t_{P_{max}})$ and the maximal stored energy $E(t_{P_{max}})$ have the consistent behaviors. %, where the maximal entanglement can be obtained without resorting to the strong spin-spin interaction along with the increasing of spin size.
 This indicates that the stored energy is positively related to the cavity-spin entanglement in the closed QB system.

%\begin{figure}[b]
%\centering
%\includegraphics[width=0.485\textwidth]{fig4.eps}
%\caption{The logarithmic negativity $E_\mathcal{N}(t_{P_{max}})$ as functions of the cavity-spin coupling strength $g$ and spin-spin interaction strength $J$ for different spin $j$: (a) spin-$1/2$, (b) spin-$1$, and (c) spin-$3/2$.}
%\label{fig4}
%\end{figure}

%In order to analyze the above property of the maximal stored energy, it is necessary to investigate the entanglement between the charger and the battery since the stored energy and the cavity-spin entanglement exhibits the similar dynamical behaviours in Fig.~\ref{fig2}. Figure~\ref{fig4} illustrates the cavity-spin entanglement $E_\mathcal{N}(t_{P_{max}})$ at maximal average power for different spin-$j$ configurations, where the entanglements are functions of cavity-spin coupling $g$ and spin-spin interaction $J$. It is obvious that the entanglement $E_\mathcal{N}(t_{P_{max}})$ in Fig.~\ref{fig4} have the consistent behaviors in comparison with those of maximal stored energy $E(t_{P_{max}})$ in Fig.~\ref{fig3}, where the maximal cavity-spin entanglement can be obtained without resorting to the strong spin-spin interaction along with the increasing of spin size. This indicates that the stored energy is positively related to the cavity-spin entanglement in the closed QB system.
%This indicates a positive correlation between the cavity-spin entanglement and the stored energy in closed QB system.

\begin{figure}[bp]
\centering
\includegraphics[width=0.485\textwidth]{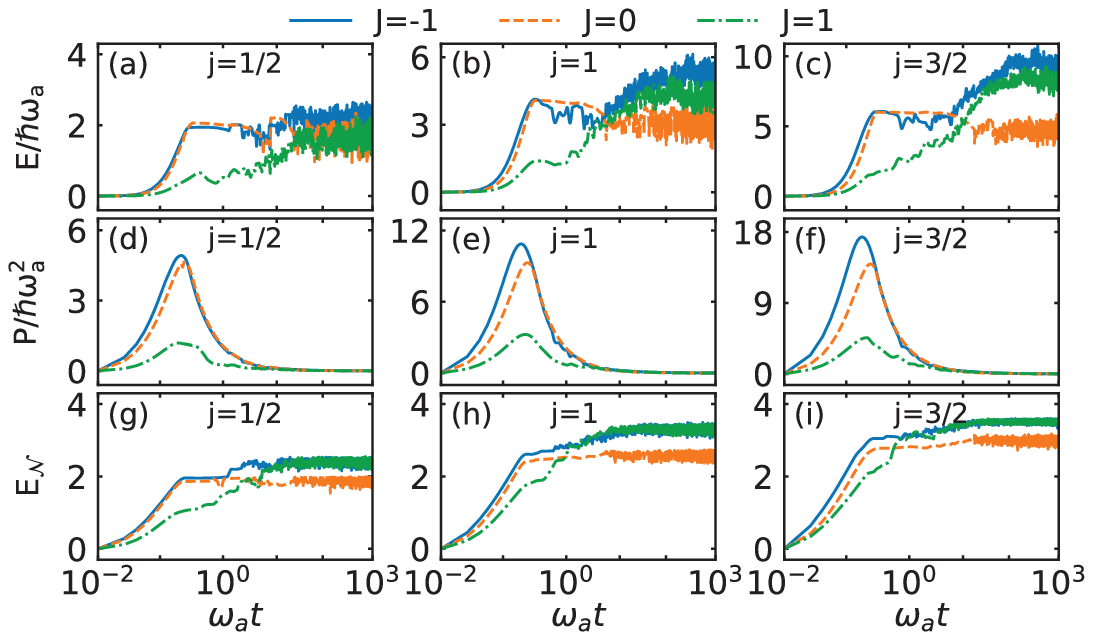}
\caption{Optimized results: (a)-(c) The dependence of the stored energy $E(t)$ (in units of $\hbar\omega_{a}$), (d)-(f) average charging power $P(t)$ (in units of $\hbar\omega^{2}_{a}$), and (g)-(i) logarithmic negativity $E_\mathcal{N}(t)$ of closed system QB as a function of $\omega_a t$ for different spin $j$.} %The different curves in these plots stand for various spin-spin interaction $J$, as indicated in the legends.}
\label{fig4}
\end{figure}
%The optimization pathways of the cavity-spin coupling $g$ for (a) $j=1/2$, (b) $j=1$, (c) $j=3/2$.

The SAC algorithm represents an exemplary approach to machine learning, exhibiting remarkable capabilities in addressing complex tasks. %As an outstanding method of machine learning, the SAC algorithm is capable of far more complex task.
 Here, we employ this algorithm to optimize the performance of the QB by adjusting the cavity-spin coupling $g(t)$.
 %Considering positive correlation between the coupling strength and charging power within a reasonable range
 Considering that the coupling strength shows a positive relationship with the charging power within a reasonable range, and in order to facilitate a comparison with the results obtained prior to optimisation, the coupling range is selected to be within the interval $[0,1]$. In fact, the range can be even wider in our SAC algorithm. The optimization process is part of a multi-objective optimization framework. The RL agent learns to maximize the $E(t)$ and then ensure a relatively high $P(t)$ (for details see Appendix~\ref{appendix1}). The $E(t)$ and $P(t)$ of the QB serve as the observed state input of the agent, which enables the agent to modulate the coupling strength between the cavity and the battery as the action output of the QB. Through continuous iteration, the pathway of cavity spin coupling is continuously updated and optimized. As a result, the performance of QB is enhanced.

\begin{figure}[tp]
\centering
\includegraphics[width=0.425\textwidth]{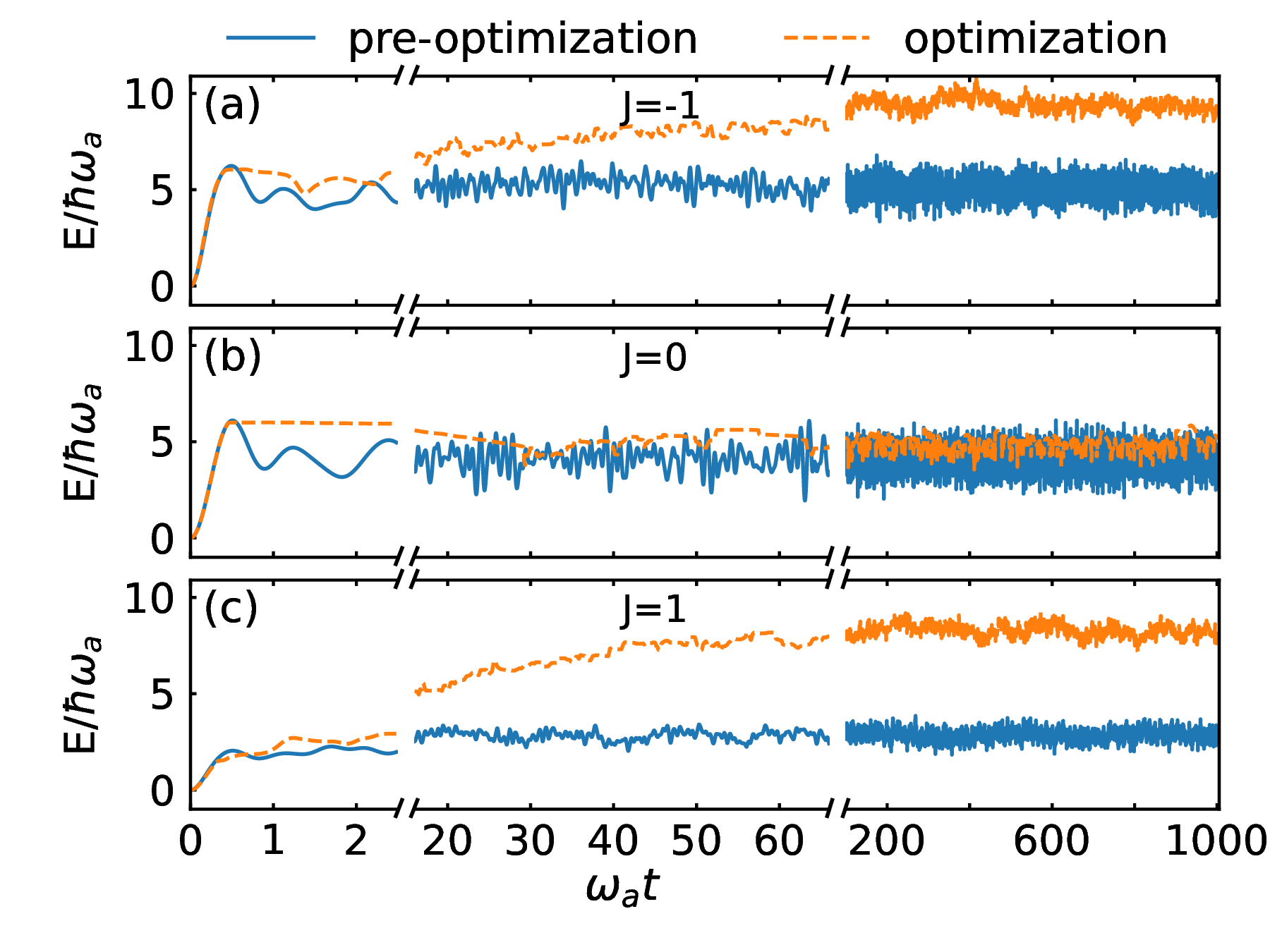}
\caption{The time evolution of stored energy in pre-optimization and optimization for spin-$3/2$ under different interaction (a) $J=-1$, (b) $J=0$, and (c) $J=1$.}
\label{fig5}
\end{figure}

\begin{figure}[tp]
\centering
\includegraphics[width=0.425\textwidth]{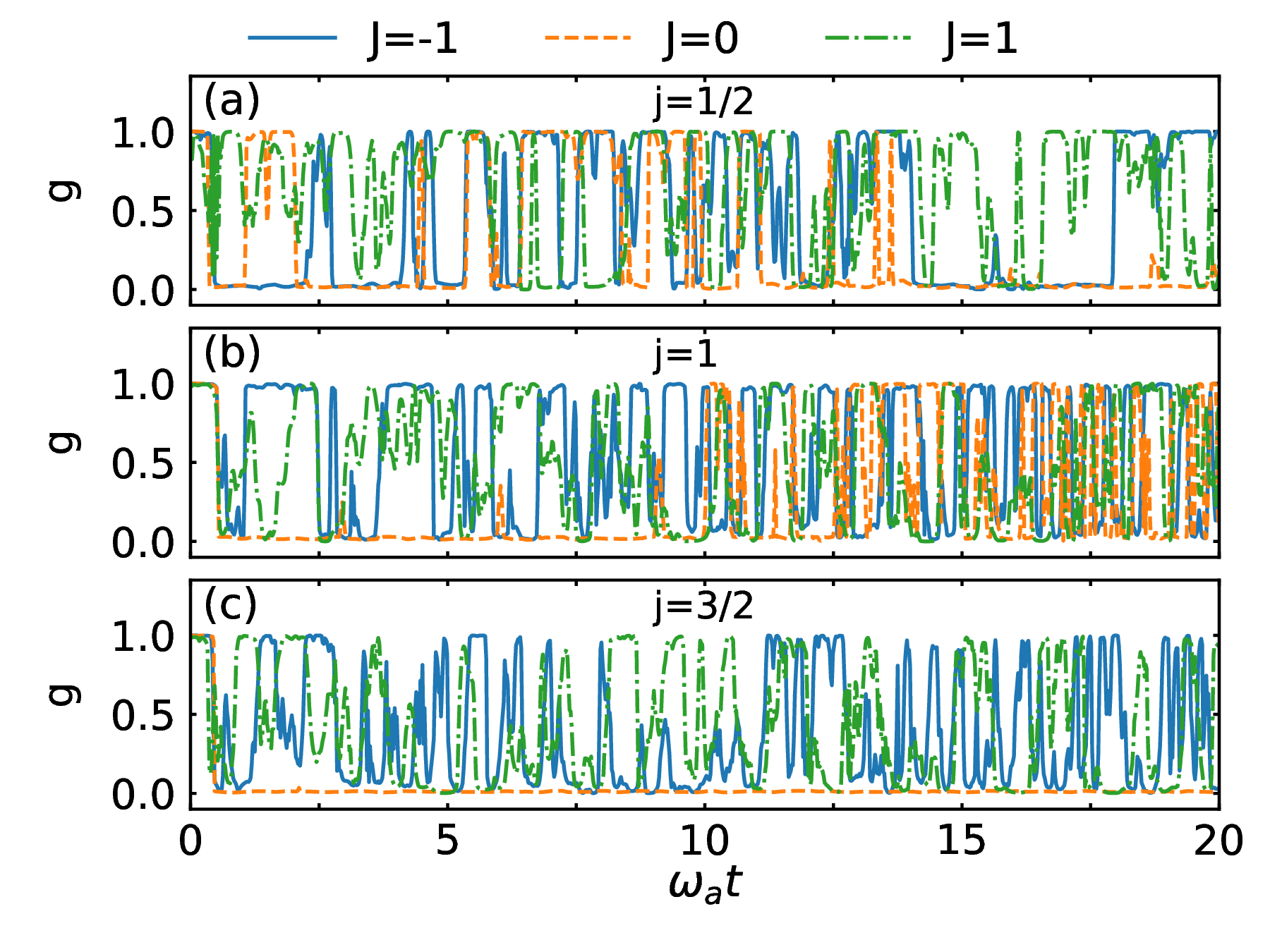}
\caption{The optimization pathways of the cavity-spin coupling $g$ for (a) $j=1/2$, (b) $j=1$, and (c) $j=3/2$.}
\label{fig6}
\end{figure}

The optimized results of the stored energy $E(t)$, average charging power $P(t)$, and the corresponding entanglement $E_\mathcal{N}(t)$  between the cavity and spin for different spin-$j$ configurations are presented in Fig.~\ref{fig4}. In comparison with the performance of the pre-optimization QB in Fig.~\ref{fig2}, we find that no matter which spin configurations, the QB's stored energy can be improved following the optimized process, and can even exceed the upper bound of stored energy for the case without spin-spin interaction.
 For spin-$j$, the upper bound values on the stored energy are given by $2jN\hbar\omega_a$, respectively. Since the upper bound of nonzero $J$ is currently unknown, to investigate the effect of spin-spin interaction on the performance of the QB, we choose the analytical upper bound for the $J=0$ case as a natural reference. Similarly, large spin corresponds to higher energy. Over time, the stored energy is divided into two stages: the rapid rise stage (basically consistent with the case without optimization), and the hold or lift stage. In the hold or lift stage, the stored energy is maintained or increased after reaching the pre-optimisation maximum. The system without spin-spin interactions belong to the former, while the systems with spin-spin interactions correspond to the latter. It is noted that for the system without spin-spin interaction, the optimal result is in the early stage, and the stored energy value will be slightly lower than its maximum value in the later stage of optimization, but it will always be better than the result of without optimization. More interestingly, different from the previous results without optimization, the final stored energy of both ferromagnetic and antiferromagnetic interactions is significantly increased, which is several times higher than that before optimization. We further demonstrate that the SAC algorithm is actually regulating the entanglement between the charger and the battery by adjusting the coupling between the cavity and the battery. In the closed QB system, the coupling between the cavity and the battery is constantly adjusted to increase the entanglement between the charger and the battery. During this process, the energy population distribution of the QB gradually transitions from the initial occupation of the lowest energy state to higher energy states. As a result, the QB's stored energy further increases. In order to more clarity, in the Fig. ~\ref{fig5}, we take spin-$3/2$ as an example to re-show the stored energy of pre-optimization and optimization by appropriate truncation in linear coordinates. Obviously, in the late stage of the pre-optimization process, the stored energy reaches a relatively stable value. Here, the relatively stable, we mean the stored energy oscillates with a relatively small amplitude. For spin-$3/2$, the time is around higher than $\omega_a t=20$. %, when the spin-spin interaction $J=-1, 1$, the time is arond higher than $20$  %(around $\omega_a t=35$), while the time is shorter for without interaction, $\omega t$ within $10$.
  In contrast, for the spin-$1/2$ and spin-$1$ cases, the system reaches the relatively stable interval at times shorter than $20$ (see also Fig.~\ref{fig2}). After optimization, the final stored energy of both ferromagnetic and antiferromagnetic interactions can reach a relatively stable and high value and this process does not take too long time. For the system without spin-spin interaction, the optimal time is shorter, i.e., $\omega_a t$ within $2$. In our all calculation, for consistency and comparison, the total time is taken as the time required for the pre-optimized charging energy to reach a stable maximum of large spin in the open system, i.e., $\omega_a t=1000$. The optimization pathways of the cavity-spin coupling $g$ for different QB configurations are illustrated in Fig.~\ref{fig6}, which presents only the time period till the stored energy reaches its relatively stable and high value for small spin. To enable a consistent comparison across all three scenarios, we set the total time on the $x$-axis to $20$ for all cases in Fig.~\ref{fig6}. In real applications, it is necessary to ensure a high average charging power as well as a high stored energy. It may leads to a relatively short time for optimization, and thus the pathways would be much more clear.
 
\section{OPEN SYSTEM: $\kappa \neq 0$}\label{section4}
In this section, we investigate the charging properties of the QB in the case of open system which introduces ambient temperature and cavity field dissipation.

\begin{figure}[bp]
\centering
\includegraphics[width=0.485\textwidth]{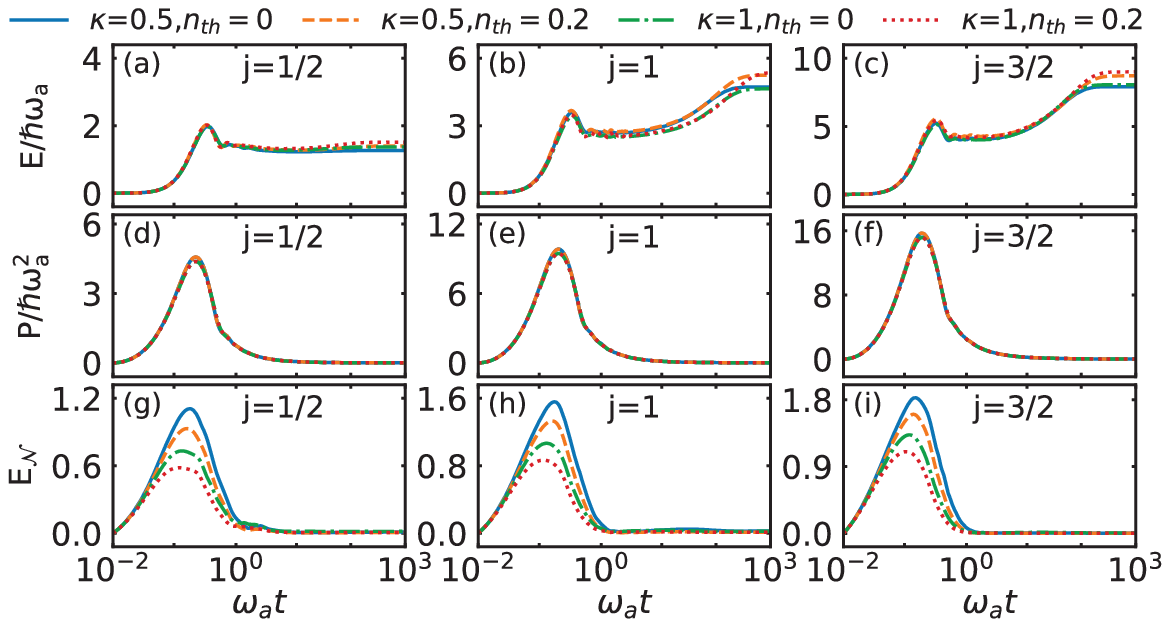}
\caption{The dependence of (a)-(c) the stored energy $E(t)$ (in units of $\hbar\omega_{a}$), and (d)-(f) average charging power $P(t)$ (in units of $\hbar\omega^{2}_{a}$), and (g)-(i) logarithmic negativity $E_\mathcal{N}(t)$ of open system QB as a function of $\omega_a t$ for different spin $j$. The different curves stand for various ambient temperature and cavity dissipation. The parameters are chosen as $g=1, J=-1$.}
\label{fig7}
\end{figure}

Figure~\ref{fig7} shows the time-dependent behaviour of the stored energy $E(t)$, average charging power $P(t)$, and the cavity-spin entanglement $E_\mathcal{N}(t)$ for QBs with different ambient temperatures and cavity field dissipations. In the open QB system, all QBs can achieve stable charging due to the environmental factors cancelling out the oscillation effect. Here, this energy transfer is purely quantum-mechanical effects, which are generated by the collective behaviour of the battery, charger, and surrounding environment \cite{PhysRevApplied.14.024092}. However, the final stable energy behaves differently for different spin configurations. For a spin-$1/2$ QB, the final stable energy is less than the maximum value, whereas for a spin-$1$ and spin-$3/2$ QB, the final stable energy is higher than the maximum value. Similarly, larger spins correspond to larger maximum stored energy and maximum average charging power, while the cavity-battery entanglement decreases over time in the open systems. During the earlier stage of the charging process, the cavity-spin interaction induces an increase in the entanglement between the charger and the battery, accompanied by a transition in the energy population of the QB from the lowest to higher energy states. In the middle stage, the charging energy $E(t)$ and the cavity-spin entanglement $E_{\mathcal{N}}(t)$ exhibit different behaviors where the charging energy maintains stability but the entanglement drops rapidly. This is due to the fact that the process is determined by the joint competition between the battery, the charger, and the environment. Although the energy population of QB remains essentially unchanged, the entanglement $E_{\mathcal{N}}(t)$ between the charger and the battery is changed as a result of the open system evolution. In the final stage of the charging process, the cavity-spin entanglement will gradually decrease due to the continuous interaction of environmental factors, until the entanglement eventually drops to zero. The energy population of QB tends to be at much higher energy states (see also Appendix~\ref{appendix2}), resulting in a further rise of the charging energy $E(t)$. %(\textcolor{red}{For the energy to increase, the population of higher energy levels needs to be further enhanced, which inevitably leads to a reduction of the population in lower energy levels. This redistribution of energy results in a decrease in the system¡¯s entropy and entanglement. Therefore, in the open system, a decrease in entanglement corresponds to an increase in the charging energy.}).%Meanwhile, we focus on the cavity-spin entanglement $E_\mathcal{N}(t)$ the QB and analyze a dynamic pattern of entanglement over time. Initially, as the QB evolved, the entanglement increased, signifying a growing interaction between the cavity and the spin. However, as time progressed, the entanglement subsequently decreased, ultimately stabilizing or disappearing altogether, which indicates the cavity-spin entanglement evolves and eventually reaches equilibrium or decouples.

\begin{figure}[tp]
\centering
\includegraphics[width=0.485\textwidth]{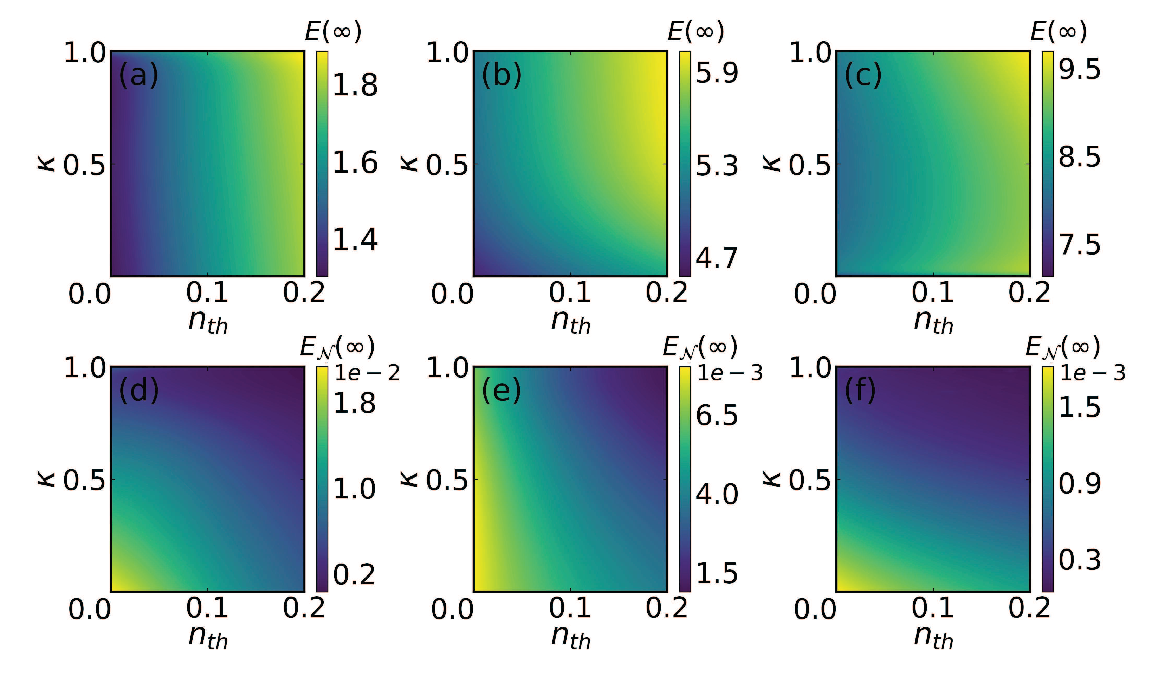}
\caption{ The contour plots of open system QB's (a)-(c) stable stored energy $E(\infty)$) (in units of $\hbar\omega_{a}$), and (d)-(f) logarithmic negativity $E_\mathcal{N}(\infty)$ as functions of $n_{th}$ and $\kappa$ with the spin-spin interaction $J=-1$.}
\label{fig8}
\end{figure}

% It shows that the energy $E(t)$ of the QB in the open system exhibited stable charging, which indicated that the QB could effectively store and maintain energy. The result indicates that QBs with lager spin exhibited greater $E(t)$ and $P(t)$. It is found that the increased ambient temperature enhances the QB's capacity to maintain higher $E(t)$, which suggests environmental factors can be harnessed to enhance the overall performance of QB's system. In addition, the greater the stable stored energy, the lower the cavity-spin entanglement.

\begin{figure}[htbp]
\centering
\includegraphics[width=0.485\textwidth]{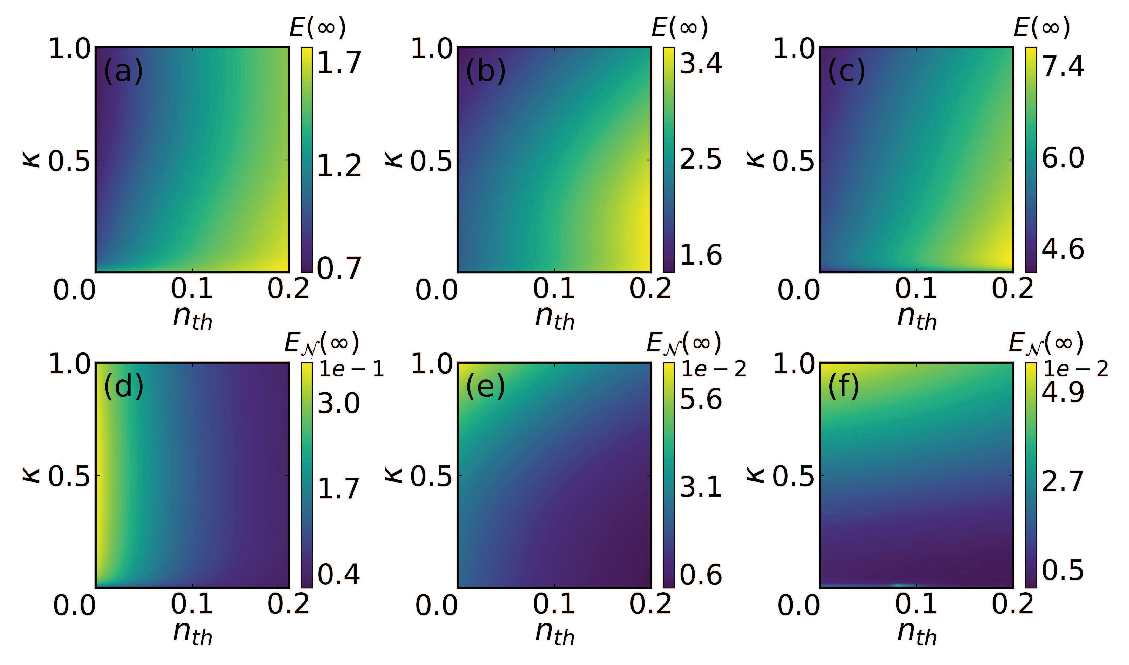}
\caption{The contour plots of open system QB's (a)-(c) stable stored energy $E(\infty)$) (in units of $\hbar\omega_{a}$), and (d)-(f) logarithmic negativity $E_\mathcal{N}(\infty)$ as functions of $n_{th}$ and $\kappa$ with the spin-spin interaction $J=1$.}
\label{fig9}
\end{figure}

To analyze the effect of ambient temperature and cavity dissipation on the charging energy and the physical mechanism of charging process, we calculate the stable stored energy (defined as $E(\infty)$) and the cavity-battery entanglement (defined as $E_\mathcal{N}(\infty)$) as a function of them, and these results are shown in Fig.~\ref{fig8} and Fig.~\ref{fig9} for different spins and spin-spin interactions. Here, we take a sufficiently long duration, ensuring that the stored energy stabilizes in our calculations. For all spin configurations QB, the final stable stored energy increases as the ambient temperature increases. Interestingly, the effect of cavity dissipation differs for different spin-spin interactions. In the ferromagnetic interaction, the strong dissipation shows a positive effect, and the large dissipation leads to the higher stored energy, while in the antiferromagnetic interaction, the dissipation suppresses the stable stored energy. Regardless of the spin-spin interactions and spin configurations, the final cavity-charger entanglement shows the opposite behavior to the stable stored energy, with small entanglement leading to large stable stored energy.

\begin{figure}[htbp]
\centering
\includegraphics[width=0.485\textwidth]{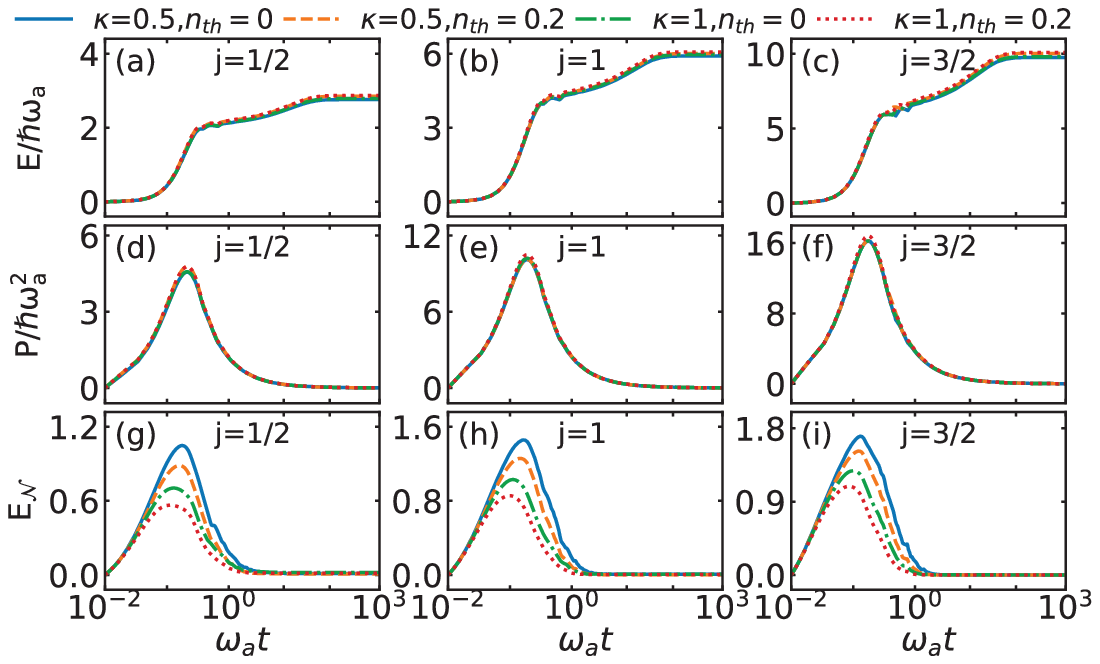}
\caption{Optimized Results: (a)-(c) The dependence of the stored energy $E(t)$ (in units of $\hbar\omega_{a}$), (d)-(f) average charging power $P(t)$ (in units of $\hbar\omega^{2}_{a}$), and (g)-(i) logarithmic negativity $E_\mathcal{N}(t)$ of open system QB as a function of $\omega_a t$ for different spin $j$. The parameter is chosen as $J=-1$.}
\label{fig10}
\end{figure}
\begin{figure}[htbp]
\centering
\includegraphics[width=0.46\textwidth]{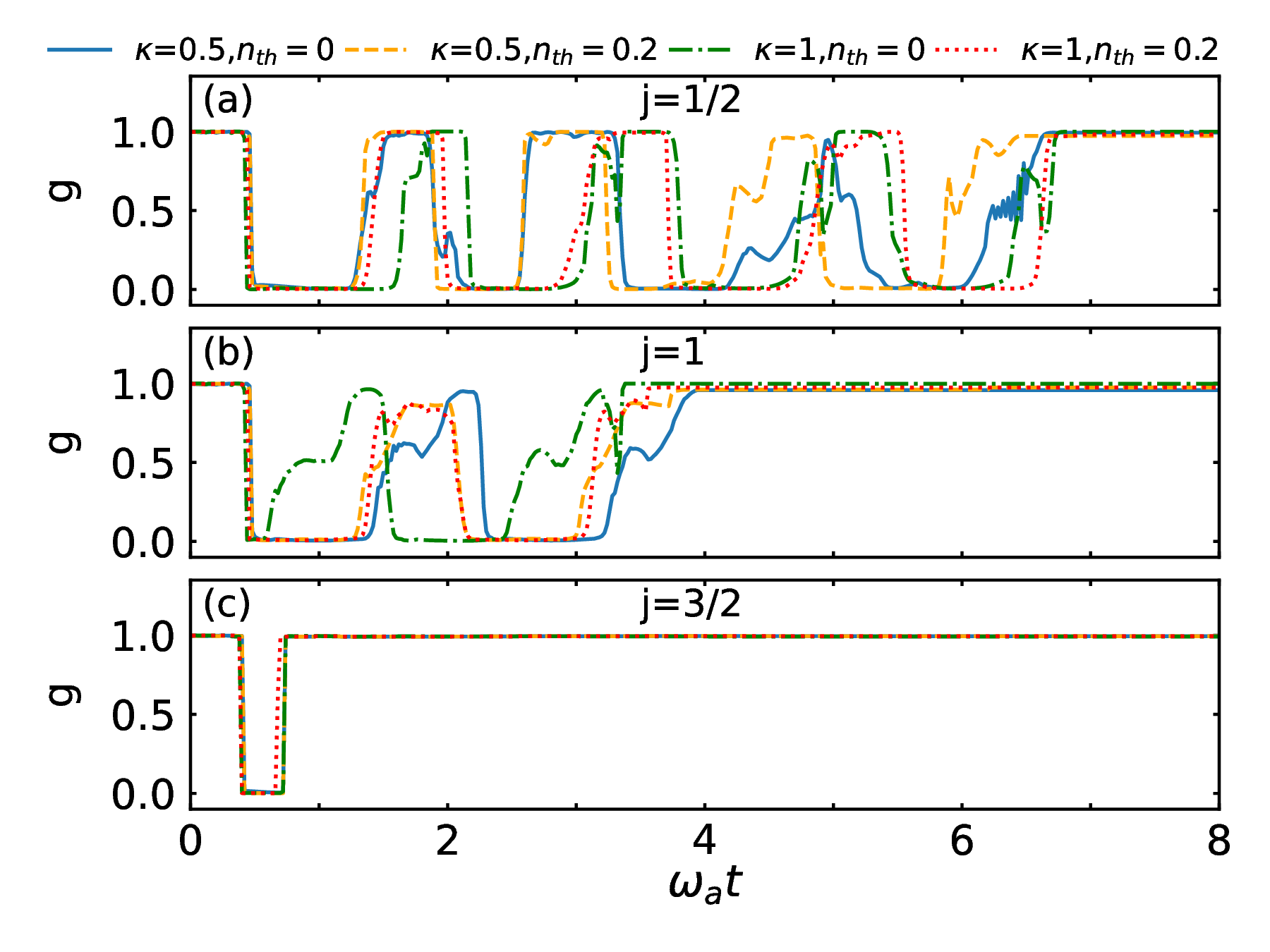}
\caption{Pathways under different ambient temperatures and cavity dissipation conditions for (a) $j=1/2$, (b) $j=1$, (c) $j=3/2$.}
\label{fig11}
\end{figure}
%In order to investigate the interaction between cavity-spin entanglement and the stable stored energy of QBs, we analyze the QBs with different spins under both ferromagnetic and antiferromagnetic interactions, where the $E(t)$ and the $E_\mathcal{N}(t)$ as the functions of the ambient temperature and the cavity field dissipation which are presented in Fig.~\ref{fig8} and Fig.~\ref{fig9}, respectively. The results suggest that the ambient temperature and the cavity field dissipation can modulate the stable energy, and larger-spin QB exhibit notably heightened stable energy. They also illustrates the higher ambient temperature increased stable storage energy while reduced cavity-spin entanglement for the QBs, and the enhanced cavity field dissipation increased the cavity-spin entanglement. Moreover, the stable storage energy and cavity-spin entanglement exhibit a opposite behavior with the evolution of ambient temperature and cavity field dissipation, which is consistent with Fig.~\ref{fig7}.

We further employ the SAC algorithm to optimize the QB performance by adjusting the cavity-spin coupling $g$ which ranges in $[0,1]$. The optimized stored energy $E(t)$, average charging power $P(t)$, and corresponding cavity-spin entanglement $E_\mathcal{N}(t)$ are presented in Fig.~\ref{fig10}. Since the QB performance can be improved regardless of the spin-spin interaction, here we show the ferromagnetic interaction and take $J=-1$. Obviously, the optimized stored energy and the average charging power have increased, and even the stored energy can reach the upper bound of the charging process without spin-spin interaction. Figure~\ref{fig11} illustrates the optimization pathways of the cavity-spin coupling $g$ for different QB configurations (We take $\omega_a t=8$, and then the coupling stays at the same value). In the open system, the effect of environment makes the system to tend to be in
a stable state. As a consequence, the adjustment of cavity-spin coupling also becomes stable. The actual optimization process begins when stored energy reaches its maximum before optimization, at which point the cavity-battery coupling is turned off, i.e., $g=0$. In this way, the environment and the cavity interact, weakening the energy reflux from the battery to the cavity while allowing the cavity to be replenished with energy from the environment. The coupling is then restored, enabling the cavity to act as a charger that continue supplying energy to the battery. During the process, the energy population of QB will be at higher energy levels. As a result, the stored energy in the battery continues to increase. This process is repeated until the stored energy reaches its maximum value. For large spins, such as spin $3/2$, the process is even simpler, and the coupling only needs to be adjusted once to achieve the purpose. Likewise, the stable stored energy corresponds to the minimum cavity-battery entanglement. It should be pointed out that, in real applications, high charging efficiency is more relevant in the early stage. Our optimization allows the system to achieve a significant improvement and reach a higher stable stored energy in the early stage. In open systems, this time is in the order of $100$ ($\omega_a t\approx 120$ for spin-{3/2}), while in closed systems this time is earlier. In all our calculations, for consistency and comparison, we take a longer time, $\omega_a t=1000$, which is the time required for the pre-optimized stored energy to reach a stable maximum of large spin in the open system.

%The stored energy $E(t)$ and average charging power $P(t)$ In comparison with the performance of the QB in Fig.~\ref{fig7}, it displays the optimized QB with spin-1/2 can achieve higher energy $E(t)$ with the similar charging power $P(t)$. In the case of the larger spins, both $E(t)$ and $P(t)$ are improved. Additionally, we can see the decreased entanglement $E_\mathcal{N}(t)$ in the optimized QB, which coincides with our previous finding that the QB charging energy is negative association with the cavity-spin entanglement. This property means that the performance optimization of the QB is realized by adjusting the cavity-spin entanglement. Specifically, the decreasing entanglement between the charger and the battery intensifies energy exchange rate between the battery and the environment, thereby facilitating greater energy replenishment from the environment. This process ultimately enables the QB to reach a higher storage energy rapidly.

\section{CONCLUSIONS} \label{section5}
We have proposed a cavity-Heisenberg spin chain QB model with spin-$j~(j=1/2,1,3/2)$ configurations and investigated the charging performance. We have shown that the stored energy and average charging power can be significantly improved with larger spin sizes. The ferromagnetic spin interaction can improve the QB performance, while the anti-ferromagnetic interaction leads to a decrease in the QB's stored energy and average charging power. Additionally, by adjusting the cavity-spin coupling and spin-spin interaction, the QB can achieve higher energy and average charging power. Further, we have considered the effects of the ambient temperature and cavity field dissipation. The open QB can achieve stable charging process and its performance is affected by ambient temperature, cavity dissipation and spin-spin interaction. For the QB with ferromagnetic interaction, the ambient temperature and cavity dissipation have positive effects on the stable stored energy, while for anti-ferromagnetic interaction QB, cavity dissipation will inhibit the stable energy. We have also employed the SAC algorithm in both closed and open systems to optimize the QB performance by adjusting the cavity-spin coupling. The optimization reduces the influence of the various parameters (including the antiferromagnetic spin interaction, the ambient temperature and cavity field dissipation) to achieve better QB performance, and its final stored energy can even exceed the upper bound without spin-spin interaction for all spin configurations. We have found that the physical mechanism of optimization process. The charger-battery entanglement can be tuned by adjusting the cavity-battery coupling parameters.
 %In the optimization process of the closed QB system, the cavity-spin entanglement is positively correlated with the stored energy.
 In the optimization process of the closed QB system, the stored energy is positively related to the cavity-spin entanglement. In contrast, in open QB, the stable stored energy reaches a maximum corresponding to low entanglement. Our result provides new insights for the construction and optimization of future QBs.

 %\textcolor{blue} {In addition, all optimization process can be carried out in the early stages, it is more advantageous to the real application.}\textcolor{blue}{It should be noted that, for comparison with the pre-optimization results, we have took a long time in our calculations. In fact, the optimization process does not require such a long time, that is, the charging energy can reach a stable and high value in the early stage of optimization. This is also crucial in real applications.}
 %\textcolor{red}{The optimization reduces the influence of the antiferromagnetic interaction to achieve better QB performance. For all spin configurations, their final stored energy can approach, even exceed the upper bound where there are no spin-spin interactions.}

%The optimization reduces the impact of the antiferromagnetic interaction, which allows the QB to achieve higher stored energy under different spin interactions. Moreover, optimization accelerates the process for the QB system to reach a higher stable energy in open system.

%Moreover, in different spin-j configurations, the higher ambient temperature increased stable storage energy while reduced cavity-spin entanglement, and the enhanced cavity field dissipation increased the cavity-spin entanglement. Unlike in closed systems, we have observed a negative correlation between cavity-spin entanglement and stored energy in open systems.

\begin{acknowledgments}
We thank Dr. P. A. Erdman for helpful discussions. The work is supported by the National Natural Science Foundation of China (Grants No. 12475026 and No. 12075193).
\end{acknowledgments}

\appendix

{\section{DETAILS ON SOFT ACTOR-CRITIC ALGORITHM FOR QB} \label{appendix1}
In this appendix we give the details on SAC algorithm for optimizing the charging process of QB, including the principles of the SAC algorithm, training method and hyperparameters setting.

\subsection{PRINCIPLES OF THE SAC ALGORITHM}
The SAC algorithm is an RL method designed for continuous action spaces \cite{PhysRevLett.133.243602,haarnoja2019softactorcriticalgorithmsapplications,haarnoja2018softactorcriticoffpolicymaximum}. In our consideration, we perform a discretization of the time interval $[0, \tau ]$, with each time-step having a duration of $\Delta t=\tau/(K - 1)$. For $k\in{0,1,\ldots,K - 1}$,  the discrete times $t_k = k\Delta t$ span the entire time interval $[0, \tau ]$. Here, $K$ represents the total number of discrete time steps within the time interval $[0, \tau ]$.} SAC's objective is to maximize both the expected reward and the policy entropy. The inclusion of policy entropy promotes randomness in the policy, which enhances exploration and prevents the algorithm from settling into suboptimal solutions.
The objective function of SAC is given by
\begin{equation}
J(\pi) = \sum_k \mathbb{E}_{(s_k, a_k) \sim \rho_\pi} \left[ r(s_k, a_k) + \alpha \mathcal{H}(\pi(\cdot|s_k)) \right],
\end{equation}
where \(J(\pi)\) is the policy objective. In QB system, \(s_k\) represents the density matrix \(\rho(t_k)\) of the system at time \(t_k\), with the initial state \(s_0\) corresponding to \(|\psi(0)\rangle = |G\rangle_{B}\otimes|N\rangle_{C}\). The action \(a_k\) is the cavity-spin coupling value selected at each time step. The state-action distribution \(\rho_\pi\) is determined by the current policy \(\pi\). The reward \(r_k\) is a key quantity, and in our study, it is calculated as \(E(t_{k + 1})-E(t_k)\) and \(P(t_{k + 1})-P(t_k)\), which can be computed using Eqs.~\eqref{E}-\eqref{P}. The entropy coefficient \(\alpha\) controls the balance between the reward and the policy entropy \(\mathcal{H}(\pi(\cdot|s_k))\). The expectation \(\mathbb{E}_{(s_k, a_k) \sim \rho_\pi}\) denotes the weighted average over state-action pairs sampled from the state-action distribution \( \rho_\pi \).

SAC estimates the value of state-action pairs using the soft value function (Q-network) \( Q(s_k, a_k) \), which is updated using the Bellman equation
\begin{equation}
Q(s_k, a_k) = r(s_k, a_k) + \gamma \mathbb{E}_{s_{k + 1} \sim p} \left[ V(s_{k + 1}) \right],
\end{equation}
where $\gamma$ is the discount factor. In our research, we use two soft value functions $Q_{\phi_i}(s_k, a_k) (i=1, 2)$, which are depicted by a set of learnable parameters $\phi_i$. They are determined by minimizing a loss function. The next state $s_{k + 1}$ is obtained by applying the action $a_k$ (cavity-spin coupling value) to the current state $s_k$ and evolving the system using the Lindblad master equation. The state transition probability $p(s_{k + 1}|s_k, a_k)$ is implicitly defined by the physical evolution of the QB system. The state value function $V(s_{k + 1})$is computed through the target Q-network.

The policy update is aimed at maximizing both the Q-value and policy entropy. The policy optimization objective is
\begin{equation}
J_\pi = \mathbb{E}_{s_k \sim D} \left[ \mathbb{E}_{a_k \sim \pi} \left[ \alpha \log(\pi(a_k|s_k)) - Q(s_k, a_k) \right] \right],
\end{equation}
where \( J_\pi \) is the policy optimization objective, and \( D \) is the experience replay buffer. We set the batch size of samples drawn from \(D\) to \(256\). This size balances the stability of the gradient estimates and the computational efficiency. The policy \(\pi(a_k|s_k)\) represents the probability of taking action \(a_k\) on state \(s_k\). The \( \log(\pi(a_k|s_k)) \) represents the log probability of action \( a_k \), which contributes to the entropy of the policy. The expectation \( \mathbb{E}_{s_k \sim D} \) is the average over samples drawn from the replay buffer \( D \), and \(\mathbb{E}_{a_k \sim \pi} \) is the average over the action distribution \( \pi \) for each state \( s_k \), which ensures the policy maximizes the expected Q-value and entropy.

\subsection{HYPERPARAMETERS SETTING AND TRAINING}
We use the adaptive moment estimation (ADAM) optimizer for training with a learning rate \(LR = 0.001\). This value is selected to ensure a balance between the speed of convergence and the stability of the optimization process. A learning rate that is excessively high may lead to overshooting the optimal solution, while a rate that is too low may result in slow convergence.

For optimizing the temperature parameter \(\alpha\), we employ a learning rate \(LR_{\alpha}=0.003\). This rate is crucial for fine-tuning the exploration-exploitation trade off. By adjusting \(\alpha\) based on the SAC's entropy-adjustment objective
\begin{equation}
J(\alpha) = \mathbb{E}_{a_k \sim \pi_k} \left[ -\alpha \log(\pi_k(a_k|s_k)) - \alpha \overline{\mathcal{H}} \right],
\end{equation}
where $\pi_k$ refers to the discretization policy, and \(\overline{\mathcal{H}}\) is the target entropy, which can control how much randomness is injected into the policy. This scheduling of the target entropy allows the agent to start with more exploration (higher entropy) and gradually transition to a more deterministic policy (lower entropy) as training progresses.

The parameters of the target Q-network are updated according to Polyak averaging and using soft updates to maintain stability, with the soft update formula given by
\begin{equation}
\phi_{\text{tar}_i} \leftarrow \beta \phi_i + (1 - \beta) \phi_{\text{tar}_i},
\end{equation}
with $i=1, 2$. Here \(\phi_i\) is the previous learnable parameters, \( \phi_{\text{tar}_i} \) is the target learnable parameters, and \( \beta \) is the soft update coefficient, typically set to a small value to ensure smooth updates. %Since we use two Q-functions, there are correspondingly two targets learnable parameters, which are $\phi_{\text{target}_1}$ and $\phi_{\text{target}_2}$.

The expectation values of Q-network is
\begin{equation}
\begin{split}
\overline{Q}(r_{k-1}, s_k)=r_{k-1}+\\
&\hspace{-2cm}\gamma\mathbb{E}_{a_{k}\sim\pi(\cdot|s_{k})}\left[\min_{i = 1,2}Q_{\phi_{tar_i}}(s_{k},a_{k})-\alpha\log(\pi(a_{k}|s_{k}))\right],
\end{split}
\end{equation}
where \(\mathbb{E}_{a_{k}\sim\pi(\cdot|s_{k})}\) represents the expectation based on actions, which means we take into account all possible actions \(a_{k}\) that can be taken in the state \(s_{k}\) according to the policy \(\pi\). The target Q-network $Q_{\phi_{tar_i}}(s_{k},a_{k}) (i=1, 2)$ take the minimum value of the outputs of these two Q-functions when evaluated the state-action pair. Thus, we obtain a more stable and accurate estimate of the action value.

The discount factor \(\gamma\) is set to \(0.993\). This value determines the relative importance of future rewards in the RL algorithm. In the context of QB, where the charging process unfolds over multiple time steps, a discount factor close to \(1\) ensures that the agent considers long term rewards, which is essential for optimizing the overall charging performance.

The replay buffer \(D\) has a size of \(180000\). A larger buffer size allows for more diverse sampling of past experiences, which helps in reducing the correlation between consecutive updates and improves the generalization ability of the algorithm. The training of the SAC algorithm for QB charging occurs over multiple episodes, with each episode consisting of a number of time steps. The total number of training steps is set to \(900000\). All hyperparameters used in our numerical calculations are provided in Table \ref{Table1}.

During the training, we employ the multi-objective optimization framework. We switch from optimizing the $E(t)$  to optimizing the $P(t)$ by
using a weighting function proportional to the Fermi distribution centered around \(c_{m}=50000\) with characteristic width \(c_{w}=20000\). When the number of training steps is close to \(c_{m}\), the weights assigned to maximizing the stored energy and the charging power start to shift. At this moment, the optimization process finds a series of paths that can make the stored energy reach the maximum value, and begins to look for relatively high power results from these paths. The \(c_{w}\) determines the rate of the weights evolution in training progress. Finally, we get the path to the relatively high power under the maximum energy. This setup allows a smoothly change between the two training goals, enabling the SAC algorithm to optimize both aspects of the QB charging performance, i.e., the RL agent learns to maximize the $E(t)$ and subsequently ensure the maximum $P(t)$.%
\begin{table}[htbp]
\centering
\caption{Hyperparameters used in all numerical calculations, and the letter $k$ stands for thousand.}\label{Table1}
\begin{tabular}{l|r}
\hline
Hyperparameter & Value \\ \hline
Training steps & 900k \\
Learning rate for $Q$-network and policy $LR$ & 0.001 \\
Learning rate for entropy coefficient $LR_{\alpha}$ & 0.003 \\
Discount factor $\gamma$ & 0.993 \\
Size of replay buffer $D$ & 180k \\
Polyak coefficient $\beta$ & 0.995 \\
Units in first hidden layer & 512 \\
Units in second hidden layer & 256 \\
Initial random steps & 5k \\
Update networks after this number of steps & 1k \\
Initial value of the policy entropy & 0.7 \\
Final value of the policy entropy & -2.8 \\
Number of steps where policy entropy transition & 200k \\
Batch size & 256 \\
Centered position $c_m$ & 50k \\
Characteristic width $c_w$ & 20k \\
Maximum number of photons $N_{\text{Fock}}$ & $4N+1$ \\ \hline
\end{tabular}
\end{table}

%\vspace{2em}
%\subsection*{}
The SAC algorithm for QB can be summarized in the following steps:
\renewcommand{\labelenumi}{\alph{enumi}.}  % DT???aD?D¡ä¡Á???D¨°o?
\begin{enumerate}
\item Policy initialization stage: Initialize the policy network $\pi$, the double Q-functions $Q_{\phi_1}(s_k, a_k)$ and $Q_{\phi_2}(s_k, a_k)$, and the target learnable parameters $\phi_{\text{tar}_1}$ and $\phi_{\text{tar}_2}$.
\item Data collection stage: Store interaction data in the replay buffer $D$.
\begin{itemize}
\item Observe the state $s_k$ and take an action $a_k$. The $s_k$ of the QB is represented by the density matrix \(\rho(t_k)\), and the $a_k$ is defined as the cavity-spin coupling value.
\item Receive a reward $r_k$ and transition to the next state $s_{k+1}$. The $r_k$ is the change in stored energy and charging power. After receiving the reward, the system evolves to the $s_{k+1}$ according to the Lindblad master equation.
\item Store $(s_k, a_k, r_k, s_{k+1})$ in the replay buffer $D$.
\end{itemize}
\item Policy evaluation stage: Randomly sample a batch of data from the replay buffer $D$. Update the value functions $Q_{\phi_i}(s_k, a_k)$.
\begin{itemize}
\item Compute the expectation values of Q-network $\overline{Q}$, which takes into account the future states and actions of the QB system.
\item Update the learnable parameters $\phi_1$ and $\phi_2$ by minimizing the loss function. This adjustment of the learnable parameters can better predict the rewards associated with different actions in the QB charging process.
\end{itemize}

\item Policy learning (improvement) stage: Update the policy network to optimize the stored energy and average charging power of QB.
\begin{itemize}
\item Update the policy network by maximizing the $\overline{Q}$ and policy entropy $\mathcal{H}(\pi(\cdot|s_k))$. The policy network for the QB determines the optimal cavity-spin coupling actions at each time step to maximize the stored energy and average charging power.
\end{itemize}
\item Adjust the entropy coefficient $\alpha$ to make the policy entropy close to the target.
\item Perform soft update of the target learnable parameters $\phi_{tar_i}$, which makes the learning process more stable.
\item Repeat steps $b$ to $h$ until convergence or the maximum number of iterations is reached.
\end{enumerate}
\section{THE ENERGY DISTRIBUTION OF QB IN CHARGING PROCESS FOR CLOSED AND OPEN SYSTEMS} \label{appendix2}
\begin{figure}[tp]
\centering
%\vspace{0.3cm}
\includegraphics[width=0.445\textwidth]{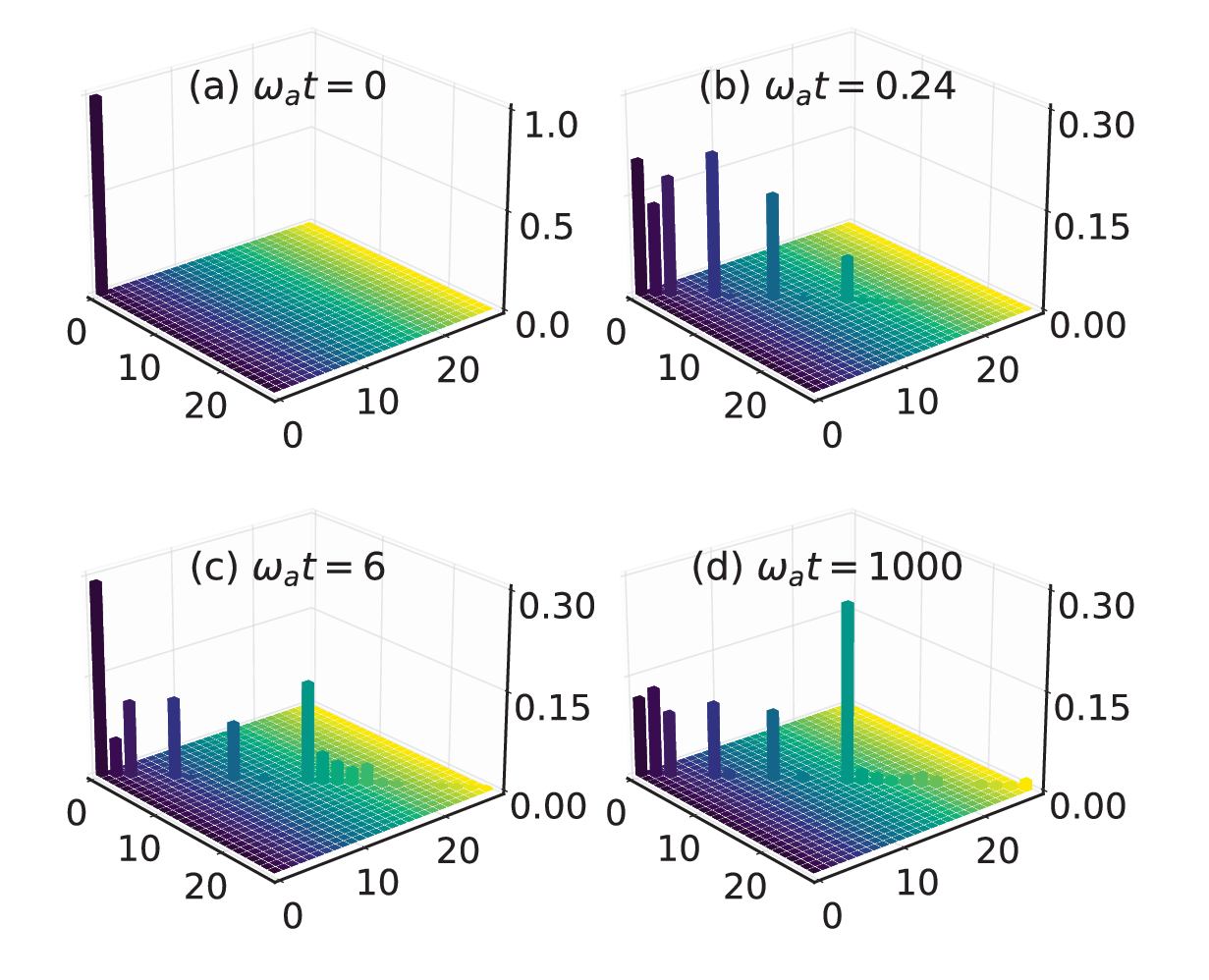}
\caption{Projection of $\rho_B(t)$ in the eigenenergy representation of $H_B$ for closed system at different times: (a) $\omega_{a}t = 0$, (b) $\omega_{a}t = 0.24$, (c) $\omega_{a}t = 6$, and (d) $\omega_{a}t = 1000$. The parameter is chosen as $J=-1$.}
\label{fig12}
\end{figure}

\begin{figure}[tp]
\centering
\includegraphics[width=0.445\textwidth]{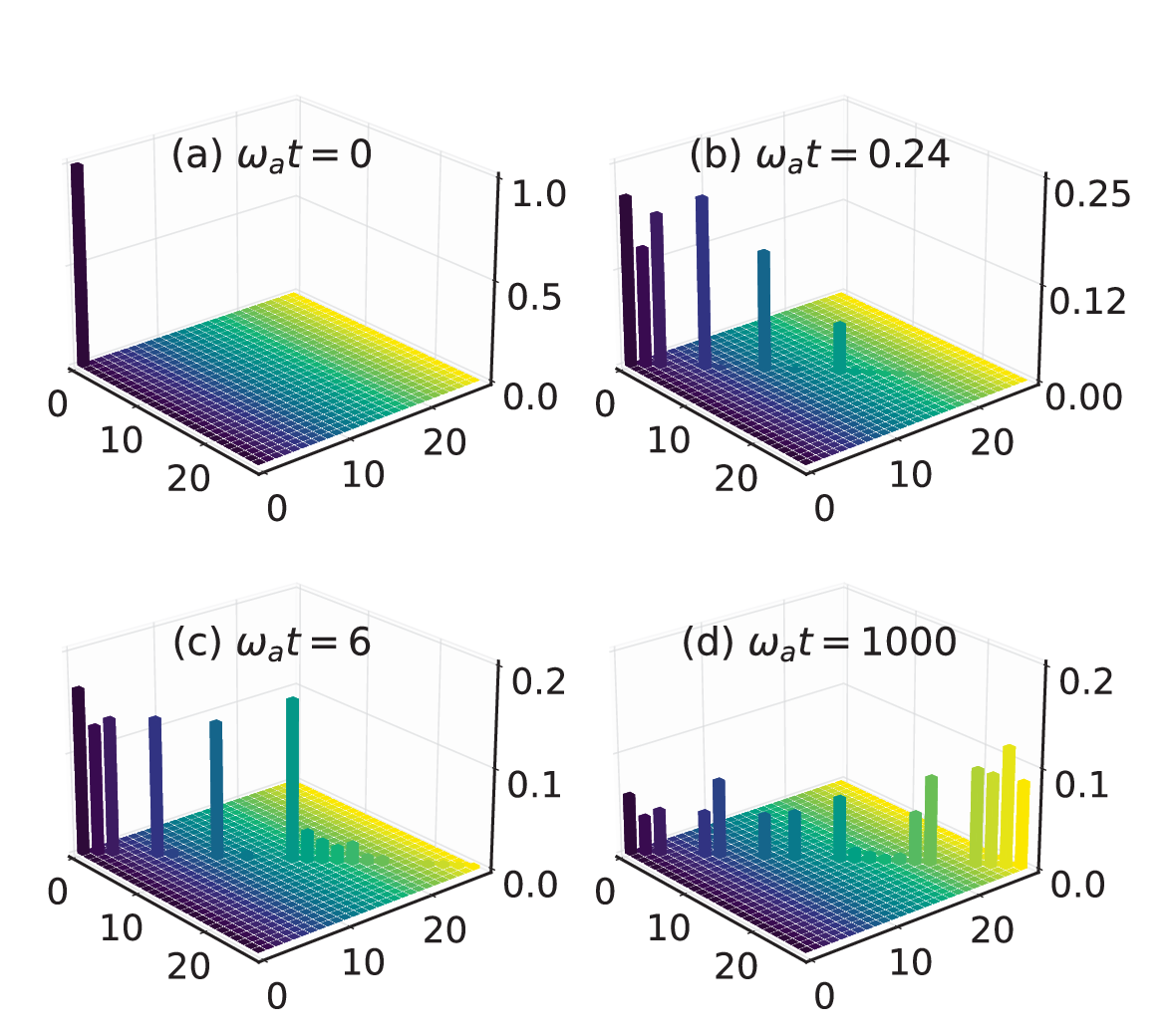}
\caption{Projection of $\rho_B(t)$ in the eigenenergy representation of $H_B$ for open system at different times: (a) $\omega_{a}t=0$, (b) $\omega_{a}t = 0.24$, (c) $\omega_{a}t = 6$, and (d) $\omega_{a}t = 1000$. The parameters are chosen as $J=-1, \kappa=0.5$, and $n_{th}=0.2$.}
\label{fig13}
\end{figure}

In this appendix, we calculate the projection of the $\rho_B(t)$ in the eigenenergy representation of the $H_B$ at different times, which represents the population in each energy eigenstate. The energy levels of $H_B$ are organized in ascending order, beginning with the lowest energy state and extending to increasingly higher energy states. These results display in Figs.~\ref{fig12}-\ref{fig13} for closed and open system, respectively. For simplicity, we only show the case with spin-$1$ before optimization. The horizontal axes represent the eigenstate orders of $H_B$, with the diagonal elements referring to the energy levels of the system, and the vertical axis indicates the occupation probability of each energy eigenstate.

At the initial time in the closed system, the system is in the lowest energy state with no population in higher energy states. As time progresses, the system transfers population from the lowest energy state to higher energy eigenstates, leading to an increase in energy. In contrast to the closed system, in the open system, at the final stage of the charging process, the energy tends to occupy much higher energy levels, which results in a further rise in the charging energy. Further calculations showed that the similar behaviour emerges in the optimized case, where the population exhibits a more pronounced distribution across higher energy levels, contributing to the increase in charging energy.

\bibliography{reference}

%merlin.mbs apsrev4-1.bst 2010-07-25 4.21a (PWD, AO, DPC) hacked
%Control: key (0)
%Control: author (8) initials jnrlst
%Control: editor formatted (1) identically to author
%Control: production of article title (-1) disabled
%Control: page (0) single
%Control: year (1) truncated
%Control: production of eprint (0) enabled
\begin{thebibliography}{104}%
\makeatletter
\providecommand \@ifxundefined [1]{%
 \@ifx{#1\undefined}
}%
\providecommand \@ifnum [1]{%
 \ifnum #1\expandafter \@firstoftwo
 \else \expandafter \@secondoftwo
 \fi
}%
\providecommand \@ifx [1]{%
 \ifx #1\expandafter \@firstoftwo
 \else \expandafter \@secondoftwo
 \fi
}%
\providecommand \natexlab [1]{#1}%
\providecommand \enquote  [1]{``#1''}%
\providecommand \bibnamefont  [1]{#1}%
\providecommand \bibfnamefont [1]{#1}%
\providecommand \citenamefont [1]{#1}%
\providecommand \href@noop [0]{\@secondoftwo}%
\providecommand \href [0]{\begingroup \@sanitize@url \@href}%
\providecommand \@href[1]{\@@startlink{#1}\@@href}%
\providecommand \@@href[1]{\endgroup#1\@@endlink}%
\providecommand \@sanitize@url [0]{\catcode `\\12\catcode `\$12\catcode
  `\&12\catcode `\#12\catcode `\^12\catcode `\_12\catcode `\%12\relax}%
\providecommand \@@startlink[1]{}%
\providecommand \@@endlink[0]{}%
\providecommand \url  [0]{\begingroup\@sanitize@url \@url }%
\providecommand \@url [1]{\endgroup\@href {#1}{\urlprefix }}%
\providecommand \urlprefix  [0]{URL }%
\providecommand \Eprint [0]{\href }%
\providecommand \doibase [0]{http://dx.doi.org/}%
\providecommand \selectlanguage [0]{\@gobble}%
\providecommand \bibinfo  [0]{\@secondoftwo}%
\providecommand \bibfield  [0]{\@secondoftwo}%
\providecommand \translation [1]{[#1]}%
\providecommand \BibitemOpen [0]{}%
\providecommand \bibitemStop [0]{}%
\providecommand \bibitemNoStop [0]{.\EOS\space}%
\providecommand \EOS [0]{\spacefactor3000\relax}%
\providecommand \BibitemShut  [1]{\csname bibitem#1\endcsname}%
\let\auto@bib@innerbib\@empty
%</preamble>
\bibitem [{\citenamefont {Nayak}\ \emph {et~al.}(2008)\citenamefont {Nayak},
  \citenamefont {Simon}, \citenamefont {Stern}, \citenamefont {Freedman},\ and\
  \citenamefont {Das~Sarma}}]{RevModPhys.80.1083}%
  \BibitemOpen
  \bibfield  {author} {\bibinfo {author} {\bibfnamefont {C.}~\bibnamefont
  {Nayak}}, \bibinfo {author} {\bibfnamefont {S.~H.}\ \bibnamefont {Simon}},
  \bibinfo {author} {\bibfnamefont {A.}~\bibnamefont {Stern}}, \bibinfo
  {author} {\bibfnamefont {M.}~\bibnamefont {Freedman}}, \ and\ \bibinfo
  {author} {\bibfnamefont {S.}~\bibnamefont {Das~Sarma}},\ }\href {\doibase
  10.1103/RevModPhys.80.1083} {\bibfield  {journal} {\bibinfo  {journal} {Rev.
  Mod. Phys.}\ }\textbf {\bibinfo {volume} {80}},\ \bibinfo {pages} {1083}
  (\bibinfo {year} {2008})}\BibitemShut {NoStop}%
\bibitem [{\citenamefont {Eisert}\ and\ \citenamefont
  {Gross}(2009)}]{PhysRevLett.102.240501}%
  \BibitemOpen
  \bibfield  {author} {\bibinfo {author} {\bibfnamefont {J.}~\bibnamefont
  {Eisert}}\ and\ \bibinfo {author} {\bibfnamefont {D.}~\bibnamefont {Gross}},\
  }\href {\doibase 10.1103/PhysRevLett.102.240501} {\bibfield  {journal}
  {\bibinfo  {journal} {Phys. Rev. Lett.}\ }\textbf {\bibinfo {volume} {102}},\
  \bibinfo {pages} {240501} (\bibinfo {year} {2009})}\BibitemShut {NoStop}%
\bibitem [{\citenamefont {Reilly}\ \emph {et~al.}(2023)\citenamefont {Reilly},
  \citenamefont {Wilson}, \citenamefont {J\"ager}, \citenamefont {Wilson},\
  and\ \citenamefont {Holland}}]{PhysRevLett.131.150802}%
  \BibitemOpen
  \bibfield  {author} {\bibinfo {author} {\bibfnamefont {J.~T.}\ \bibnamefont
  {Reilly}}, \bibinfo {author} {\bibfnamefont {J.~D.}\ \bibnamefont {Wilson}},
  \bibinfo {author} {\bibfnamefont {S.~B.}\ \bibnamefont {J\"ager}}, \bibinfo
  {author} {\bibfnamefont {C.}~\bibnamefont {Wilson}}, \ and\ \bibinfo {author}
  {\bibfnamefont {M.~J.}\ \bibnamefont {Holland}},\ }\href {\doibase
  10.1103/PhysRevLett.131.150802} {\bibfield  {journal} {\bibinfo  {journal}
  {Phys. Rev. Lett.}\ }\textbf {\bibinfo {volume} {131}},\ \bibinfo {pages}
  {150802} (\bibinfo {year} {2023})}\BibitemShut {NoStop}%
\bibitem [{\citenamefont {Degen}\ \emph {et~al.}(2017)\citenamefont {Degen},
  \citenamefont {Reinhard},\ and\ \citenamefont
  {Cappellaro}}]{RevModPhys.89.035002}%
  \BibitemOpen
  \bibfield  {author} {\bibinfo {author} {\bibfnamefont {C.~L.}\ \bibnamefont
  {Degen}}, \bibinfo {author} {\bibfnamefont {F.}~\bibnamefont {Reinhard}}, \
  and\ \bibinfo {author} {\bibfnamefont {P.}~\bibnamefont {Cappellaro}},\
  }\href {\doibase 10.1103/RevModPhys.89.035002} {\bibfield  {journal}
  {\bibinfo  {journal} {Rev. Mod. Phys.}\ }\textbf {\bibinfo {volume} {89}},\
  \bibinfo {pages} {035002} (\bibinfo {year} {2017})}\BibitemShut {NoStop}%
\bibitem [{\citenamefont {Pogorelov}\ \emph {et~al.}(2021)\citenamefont
  {Pogorelov}, \citenamefont {Feldker}, \citenamefont {Marciniak},
  \citenamefont {Postler}, \citenamefont {Jacob}, \citenamefont
  {Krieglsteiner}, \citenamefont {Podlesnic}, \citenamefont {Meth},
  \citenamefont {Negnevitsky}, \citenamefont {Stadler}, \citenamefont
  {H\"ofer}, \citenamefont {W\"achter}, \citenamefont {Lakhmanskiy},
  \citenamefont {Blatt}, \citenamefont {Schindler},\ and\ \citenamefont
  {Monz}}]{PRXQuantum.2.020343}%
  \BibitemOpen
  \bibfield  {author} {\bibinfo {author} {\bibfnamefont {I.}~\bibnamefont
  {Pogorelov}}, \bibinfo {author} {\bibfnamefont {T.}~\bibnamefont {Feldker}},
  \bibinfo {author} {\bibfnamefont {C.~D.}\ \bibnamefont {Marciniak}}, \bibinfo
  {author} {\bibfnamefont {L.}~\bibnamefont {Postler}}, \bibinfo {author}
  {\bibfnamefont {G.}~\bibnamefont {Jacob}}, \bibinfo {author} {\bibfnamefont
  {O.}~\bibnamefont {Krieglsteiner}}, \bibinfo {author} {\bibfnamefont
  {V.}~\bibnamefont {Podlesnic}}, \bibinfo {author} {\bibfnamefont
  {M.}~\bibnamefont {Meth}}, \bibinfo {author} {\bibfnamefont {V.}~\bibnamefont
  {Negnevitsky}}, \bibinfo {author} {\bibfnamefont {M.}~\bibnamefont
  {Stadler}}, \bibinfo {author} {\bibfnamefont {B.}~\bibnamefont {H\"ofer}},
  \bibinfo {author} {\bibfnamefont {C.}~\bibnamefont {W\"achter}}, \bibinfo
  {author} {\bibfnamefont {K.}~\bibnamefont {Lakhmanskiy}}, \bibinfo {author}
  {\bibfnamefont {R.}~\bibnamefont {Blatt}}, \bibinfo {author} {\bibfnamefont
  {P.}~\bibnamefont {Schindler}}, \ and\ \bibinfo {author} {\bibfnamefont
  {T.}~\bibnamefont {Monz}},\ }\href {\doibase 10.1103/PRXQuantum.2.020343}
  {\bibfield  {journal} {\bibinfo  {journal} {PRX Quantum}\ }\textbf {\bibinfo
  {volume} {2}},\ \bibinfo {pages} {020343} (\bibinfo {year}
  {2021})}\BibitemShut {NoStop}%
\bibitem [{\citenamefont {Kok}\ \emph {et~al.}(2007)\citenamefont {Kok},
  \citenamefont {Munro}, \citenamefont {Nemoto}, \citenamefont {Ralph},
  \citenamefont {Dowling},\ and\ \citenamefont {Milburn}}]{RevModPhys.79.135}%
  \BibitemOpen
  \bibfield  {author} {\bibinfo {author} {\bibfnamefont {P.}~\bibnamefont
  {Kok}}, \bibinfo {author} {\bibfnamefont {W.~J.}\ \bibnamefont {Munro}},
  \bibinfo {author} {\bibfnamefont {K.}~\bibnamefont {Nemoto}}, \bibinfo
  {author} {\bibfnamefont {T.~C.}\ \bibnamefont {Ralph}}, \bibinfo {author}
  {\bibfnamefont {J.~P.}\ \bibnamefont {Dowling}}, \ and\ \bibinfo {author}
  {\bibfnamefont {G.~J.}\ \bibnamefont {Milburn}},\ }\href {\doibase
  10.1103/RevModPhys.79.135} {\bibfield  {journal} {\bibinfo  {journal} {Rev.
  Mod. Phys.}\ }\textbf {\bibinfo {volume} {79}},\ \bibinfo {pages} {135}
  (\bibinfo {year} {2007})}\BibitemShut {NoStop}%
\bibitem [{\citenamefont {Landi}\ and\ \citenamefont
  {Paternostro}(2021)}]{RevModPhys.93.035008}%
  \BibitemOpen
  \bibfield  {author} {\bibinfo {author} {\bibfnamefont {G.~T.}\ \bibnamefont
  {Landi}}\ and\ \bibinfo {author} {\bibfnamefont {M.}~\bibnamefont
  {Paternostro}},\ }\href {\doibase 10.1103/RevModPhys.93.035008} {\bibfield
  {journal} {\bibinfo  {journal} {Rev. Mod. Phys.}\ }\textbf {\bibinfo {volume}
  {93}},\ \bibinfo {pages} {035008} (\bibinfo {year} {2021})}\BibitemShut
  {NoStop}%
\bibitem [{\citenamefont {Talkner}\ and\ \citenamefont
  {H\"anggi}(2020)}]{RevModPhys.92.041002}%
  \BibitemOpen
  \bibfield  {author} {\bibinfo {author} {\bibfnamefont {P.}~\bibnamefont
  {Talkner}}\ and\ \bibinfo {author} {\bibfnamefont {P.}~\bibnamefont
  {H\"anggi}},\ }\href {\doibase 10.1103/RevModPhys.92.041002} {\bibfield
  {journal} {\bibinfo  {journal} {Rev. Mod. Phys.}\ }\textbf {\bibinfo {volume}
  {92}},\ \bibinfo {pages} {041002} (\bibinfo {year} {2020})}\BibitemShut
  {NoStop}%
\bibitem [{\citenamefont {Skrzypczyk}\ \emph {et~al.}(2014)\citenamefont
  {Skrzypczyk}, \citenamefont {Short},\ and\ \citenamefont
  {Popescu}}]{Skrzypczyk2014}%
  \BibitemOpen
  \bibfield  {author} {\bibinfo {author} {\bibfnamefont {P.}~\bibnamefont
  {Skrzypczyk}}, \bibinfo {author} {\bibfnamefont {A.~J.}\ \bibnamefont
  {Short}}, \ and\ \bibinfo {author} {\bibfnamefont {S.}~\bibnamefont
  {Popescu}},\ }\href {\doibase 10.1038/ncomms5185} {\bibfield  {journal}
  {\bibinfo  {journal} {Nat. Commun.}\ }\textbf {\bibinfo {volume} {5}},\
  \bibinfo {pages} {4185} (\bibinfo {year} {2014})}\BibitemShut {NoStop}%
\bibitem [{\citenamefont {Campaioli}\ \emph {et~al.}(2018)\citenamefont
  {Campaioli}, \citenamefont {Pollock},\ and\ \citenamefont
  {Vinjanampathy}}]{campaioli2018quantum}%
  \BibitemOpen
  \bibfield  {author} {\bibinfo {author} {\bibfnamefont {F.}~\bibnamefont
  {Campaioli}}, \bibinfo {author} {\bibfnamefont {F.~A.}\ \bibnamefont
  {Pollock}}, \ and\ \bibinfo {author} {\bibfnamefont {S.}~\bibnamefont
  {Vinjanampathy}},\ }\enquote {\bibinfo {title} {Quantum batteries},}\ in\
  \href {\doibase 10.1007/978-3-319-99046-0_8} {\emph {\bibinfo {booktitle}
  {Thermodynamics in the Quantum Regime: Fundamental Aspects and New
  Directions}}}\ (\bibinfo {year} {2018})\ pp.\ \bibinfo {pages}
  {207--225}\BibitemShut {NoStop}%
\bibitem [{\citenamefont {Alicki}\ and\ \citenamefont
  {Fannes}(2013)}]{PhysRevE.87.042123}%
  \BibitemOpen
  \bibfield  {author} {\bibinfo {author} {\bibfnamefont {R.}~\bibnamefont
  {Alicki}}\ and\ \bibinfo {author} {\bibfnamefont {M.}~\bibnamefont
  {Fannes}},\ }\href {\doibase 10.1103/PhysRevE.87.042123} {\bibfield
  {journal} {\bibinfo  {journal} {Phys. Rev. E}\ }\textbf {\bibinfo {volume}
  {87}},\ \bibinfo {pages} {042123} (\bibinfo {year} {2013})}\BibitemShut
  {NoStop}%
\bibitem [{\citenamefont {Campaioli}\ \emph {et~al.}(2024)\citenamefont
  {Campaioli}, \citenamefont {Gherardini}, \citenamefont {Quach}, \citenamefont
  {Polini},\ and\ \citenamefont {Andolina}}]{RevModPhys.96.031001}%
  \BibitemOpen
  \bibfield  {author} {\bibinfo {author} {\bibfnamefont {F.}~\bibnamefont
  {Campaioli}}, \bibinfo {author} {\bibfnamefont {S.}~\bibnamefont
  {Gherardini}}, \bibinfo {author} {\bibfnamefont {J.~Q.}\ \bibnamefont
  {Quach}}, \bibinfo {author} {\bibfnamefont {M.}~\bibnamefont {Polini}}, \
  and\ \bibinfo {author} {\bibfnamefont {G.~M.}\ \bibnamefont {Andolina}},\
  }\href {\doibase 10.1103/RevModPhys.96.031001} {\bibfield  {journal}
  {\bibinfo  {journal} {Rev. Mod. Phys.}\ }\textbf {\bibinfo {volume} {96}},\
  \bibinfo {pages} {031001} (\bibinfo {year} {2024})}\BibitemShut {NoStop}%
\bibitem [{\citenamefont {Hu}\ \emph {et~al.}(2022)\citenamefont {Hu},
  \citenamefont {Qiu}, \citenamefont {Souza}, \citenamefont {Yuan},
  \citenamefont {Zhou}, \citenamefont {Zhang}, \citenamefont {Chu},
  \citenamefont {Pan}, \citenamefont {Hu}, \citenamefont {Li}, \citenamefont
  {Xu}, \citenamefont {Zhong}, \citenamefont {Liu}, \citenamefont {Yan},
  \citenamefont {Tan}, \citenamefont {Bachelard}, \citenamefont {Villas-Boas},
  \citenamefont {Santos},\ and\ \citenamefont {Yu}}]{Hu_2022}%
  \BibitemOpen
  \bibfield  {author} {\bibinfo {author} {\bibfnamefont {C.-K.}\ \bibnamefont
  {Hu}}, \bibinfo {author} {\bibfnamefont {J.}~\bibnamefont {Qiu}}, \bibinfo
  {author} {\bibfnamefont {P.~J.~P.}\ \bibnamefont {Souza}}, \bibinfo {author}
  {\bibfnamefont {J.}~\bibnamefont {Yuan}}, \bibinfo {author} {\bibfnamefont
  {Y.}~\bibnamefont {Zhou}}, \bibinfo {author} {\bibfnamefont {L.}~\bibnamefont
  {Zhang}}, \bibinfo {author} {\bibfnamefont {J.}~\bibnamefont {Chu}}, \bibinfo
  {author} {\bibfnamefont {X.}~\bibnamefont {Pan}}, \bibinfo {author}
  {\bibfnamefont {L.}~\bibnamefont {Hu}}, \bibinfo {author} {\bibfnamefont
  {J.}~\bibnamefont {Li}}, \bibinfo {author} {\bibfnamefont {Y.}~\bibnamefont
  {Xu}}, \bibinfo {author} {\bibfnamefont {Y.}~\bibnamefont {Zhong}}, \bibinfo
  {author} {\bibfnamefont {S.}~\bibnamefont {Liu}}, \bibinfo {author}
  {\bibfnamefont {F.}~\bibnamefont {Yan}}, \bibinfo {author} {\bibfnamefont
  {D.}~\bibnamefont {Tan}}, \bibinfo {author} {\bibfnamefont {R.}~\bibnamefont
  {Bachelard}}, \bibinfo {author} {\bibfnamefont {C.~J.}\ \bibnamefont
  {Villas-Boas}}, \bibinfo {author} {\bibfnamefont {A.~C.}\ \bibnamefont
  {Santos}}, \ and\ \bibinfo {author} {\bibfnamefont {D.}~\bibnamefont {Yu}},\
  }\href {\doibase 10.1088/2058-9565/ac8444} {\bibfield  {journal} {\bibinfo
  {journal} {Quantum Sci. Technol.}\ }\textbf {\bibinfo {volume} {7}},\
  \bibinfo {pages} {045018} (\bibinfo {year} {2022})}\BibitemShut {NoStop}%
\bibitem [{\citenamefont {Quach}\ \emph {et~al.}(2022)\citenamefont {Quach},
  \citenamefont {McGhee}, \citenamefont {Ganzer}, \citenamefont {Rouse},
  \citenamefont {Lovett}, \citenamefont {Gauger}, \citenamefont {Keeling},
  \citenamefont {Cerullo}, \citenamefont {Lidzey},\ and\ \citenamefont
  {Virgili}}]{quach2020}%
  \BibitemOpen
  \bibfield  {author} {\bibinfo {author} {\bibfnamefont {J.~Q.}\ \bibnamefont
  {Quach}}, \bibinfo {author} {\bibfnamefont {K.~E.}\ \bibnamefont {McGhee}},
  \bibinfo {author} {\bibfnamefont {L.}~\bibnamefont {Ganzer}}, \bibinfo
  {author} {\bibfnamefont {D.~M.}\ \bibnamefont {Rouse}}, \bibinfo {author}
  {\bibfnamefont {B.~W.}\ \bibnamefont {Lovett}}, \bibinfo {author}
  {\bibfnamefont {E.~M.}\ \bibnamefont {Gauger}}, \bibinfo {author}
  {\bibfnamefont {J.}~\bibnamefont {Keeling}}, \bibinfo {author} {\bibfnamefont
  {G.}~\bibnamefont {Cerullo}}, \bibinfo {author} {\bibfnamefont {D.~G.}\
  \bibnamefont {Lidzey}}, \ and\ \bibinfo {author} {\bibfnamefont
  {T.}~\bibnamefont {Virgili}},\ }\href {\doibase 10.1126/sciadv.abk3160}
  {\bibfield  {journal} {\bibinfo  {journal} {Sci. Adv.}\ }\textbf {\bibinfo
  {volume} {8}},\ \bibinfo {pages} {eabk3160} (\bibinfo {year}
  {2022})}\BibitemShut {NoStop}%
\bibitem [{\citenamefont {Zheng}\ \emph {et~al.}(2022)\citenamefont {Zheng},
  \citenamefont {Ning}, \citenamefont {Yang}, \citenamefont {Xia},\ and\
  \citenamefont {Zheng}}]{Zheng_2022}%
  \BibitemOpen
  \bibfield  {author} {\bibinfo {author} {\bibfnamefont {R.-H.}\ \bibnamefont
  {Zheng}}, \bibinfo {author} {\bibfnamefont {W.}~\bibnamefont {Ning}},
  \bibinfo {author} {\bibfnamefont {Z.-B.}\ \bibnamefont {Yang}}, \bibinfo
  {author} {\bibfnamefont {Y.}~\bibnamefont {Xia}}, \ and\ \bibinfo {author}
  {\bibfnamefont {S.-B.}\ \bibnamefont {Zheng}},\ }\href {\doibase
  10.1088/1367-2630/ac788f} {\bibfield  {journal} {\bibinfo  {journal} {New J.
  Phys.}\ }\textbf {\bibinfo {volume} {24}},\ \bibinfo {pages} {063031}
  (\bibinfo {year} {2022})}\BibitemShut {NoStop}%
\bibitem [{\citenamefont {Gemme}\ \emph {et~al.}(2022)\citenamefont {Gemme},
  \citenamefont {Grossi}, \citenamefont {Ferraro}, \citenamefont {Vallecorsa},\
  and\ \citenamefont {Sassetti}}]{batteries8050043}%
  \BibitemOpen
  \bibfield  {author} {\bibinfo {author} {\bibfnamefont {G.}~\bibnamefont
  {Gemme}}, \bibinfo {author} {\bibfnamefont {M.}~\bibnamefont {Grossi}},
  \bibinfo {author} {\bibfnamefont {D.}~\bibnamefont {Ferraro}}, \bibinfo
  {author} {\bibfnamefont {S.}~\bibnamefont {Vallecorsa}}, \ and\ \bibinfo
  {author} {\bibfnamefont {M.}~\bibnamefont {Sassetti}},\ }\href
  {https://www.mdpi.com/2313-0105/8/5/43} {\bibfield  {journal} {\bibinfo
  {journal} {Batteries}\ }\textbf {\bibinfo {volume} {8}},\ \bibinfo {pages}
  {43} (\bibinfo {year} {2022})}\BibitemShut {NoStop}%
\bibitem [{\citenamefont {Maillette~de Buy~Wenniger}\ \emph
  {et~al.}(2023)\citenamefont {Maillette~de Buy~Wenniger}, \citenamefont
  {Thomas}, \citenamefont {Maffei}, \citenamefont {Wein}, \citenamefont {Pont},
  \citenamefont {Belabas}, \citenamefont {Prasad}, \citenamefont {Harouri},
  \citenamefont {Lema\^{\i}tre}, \citenamefont {Sagnes}, \citenamefont
  {Somaschi}, \citenamefont {Auff\`eves},\ and\ \citenamefont
  {Senellart}}]{PhysRevLett.131.260401}%
  \BibitemOpen
  \bibfield  {author} {\bibinfo {author} {\bibfnamefont {I.}~\bibnamefont
  {Maillette~de Buy~Wenniger}}, \bibinfo {author} {\bibfnamefont {S.~E.}\
  \bibnamefont {Thomas}}, \bibinfo {author} {\bibfnamefont {M.}~\bibnamefont
  {Maffei}}, \bibinfo {author} {\bibfnamefont {S.~C.}\ \bibnamefont {Wein}},
  \bibinfo {author} {\bibfnamefont {M.}~\bibnamefont {Pont}}, \bibinfo {author}
  {\bibfnamefont {N.}~\bibnamefont {Belabas}}, \bibinfo {author} {\bibfnamefont
  {S.}~\bibnamefont {Prasad}}, \bibinfo {author} {\bibfnamefont
  {A.}~\bibnamefont {Harouri}}, \bibinfo {author} {\bibfnamefont
  {A.}~\bibnamefont {Lema\^{\i}tre}}, \bibinfo {author} {\bibfnamefont
  {I.}~\bibnamefont {Sagnes}}, \bibinfo {author} {\bibfnamefont
  {N.}~\bibnamefont {Somaschi}}, \bibinfo {author} {\bibfnamefont
  {A.}~\bibnamefont {Auff\`eves}}, \ and\ \bibinfo {author} {\bibfnamefont
  {P.}~\bibnamefont {Senellart}},\ }\href {\doibase
  10.1103/PhysRevLett.131.260401} {\bibfield  {journal} {\bibinfo  {journal}
  {Phys. Rev. Lett.}\ }\textbf {\bibinfo {volume} {131}},\ \bibinfo {pages}
  {260401} (\bibinfo {year} {2023})}\BibitemShut {NoStop}%
\bibitem [{\citenamefont {Joshi}\ and\ \citenamefont
  {Mahesh}(2022)}]{PhysRevA.106.042601}%
  \BibitemOpen
  \bibfield  {author} {\bibinfo {author} {\bibfnamefont {J.}~\bibnamefont
  {Joshi}}\ and\ \bibinfo {author} {\bibfnamefont {T.~S.}\ \bibnamefont
  {Mahesh}},\ }\href {\doibase 10.1103/PhysRevA.106.042601} {\bibfield
  {journal} {\bibinfo  {journal} {Phys. Rev. A}\ }\textbf {\bibinfo {volume}
  {106}},\ \bibinfo {pages} {042601} (\bibinfo {year} {2022})}\BibitemShut
  {NoStop}%
\bibitem [{\citenamefont {Guo}\ \emph {et~al.}(2024)\citenamefont {Guo},
  \citenamefont {Yang},\ and\ \citenamefont {Dou}}]{PhysRevA.109.032201}%
  \BibitemOpen
  \bibfield  {author} {\bibinfo {author} {\bibfnamefont {W.-X.}\ \bibnamefont
  {Guo}}, \bibinfo {author} {\bibfnamefont {F.-M.}\ \bibnamefont {Yang}}, \
  and\ \bibinfo {author} {\bibfnamefont {F.-Q.}\ \bibnamefont {Dou}},\ }\href
  {\doibase 10.1103/PhysRevA.109.032201} {\bibfield  {journal} {\bibinfo
  {journal} {Phys. Rev. A}\ }\textbf {\bibinfo {volume} {109}},\ \bibinfo
  {pages} {032201} (\bibinfo {year} {2024})}\BibitemShut {NoStop}%
\bibitem [{\citenamefont {Caravelli}\ \emph {et~al.}(2020)\citenamefont
  {Caravelli}, \citenamefont {Coulter-De~Wit}, \citenamefont
  {Garc\'{\i}a-Pintos},\ and\ \citenamefont
  {Hamma}}]{PhysRevResearch.2.023095}%
  \BibitemOpen
  \bibfield  {author} {\bibinfo {author} {\bibfnamefont {F.}~\bibnamefont
  {Caravelli}}, \bibinfo {author} {\bibfnamefont {G.}~\bibnamefont
  {Coulter-De~Wit}}, \bibinfo {author} {\bibfnamefont {L.~P.}\ \bibnamefont
  {Garc\'{\i}a-Pintos}}, \ and\ \bibinfo {author} {\bibfnamefont
  {A.}~\bibnamefont {Hamma}},\ }\href {\doibase
  10.1103/PhysRevResearch.2.023095} {\bibfield  {journal} {\bibinfo  {journal}
  {Phys. Rev. Res.}\ }\textbf {\bibinfo {volume} {2}},\ \bibinfo {pages}
  {023095} (\bibinfo {year} {2020})}\BibitemShut {NoStop}%
\bibitem [{\citenamefont {Rojo-Franc\`as}\ \emph {et~al.}(2024)\citenamefont
  {Rojo-Franc\`as}, \citenamefont {Isaule}, \citenamefont {Santos},
  \citenamefont {Juli\'a-D\'{\i}az},\ and\ \citenamefont
  {Zinner}}]{PhysRevA.110.032205}%
  \BibitemOpen
  \bibfield  {author} {\bibinfo {author} {\bibfnamefont {A.}~\bibnamefont
  {Rojo-Franc\`as}}, \bibinfo {author} {\bibfnamefont {F.}~\bibnamefont
  {Isaule}}, \bibinfo {author} {\bibfnamefont {A.~C.}\ \bibnamefont {Santos}},
  \bibinfo {author} {\bibfnamefont {B.}~\bibnamefont {Juli\'a-D\'{\i}az}}, \
  and\ \bibinfo {author} {\bibfnamefont {N.~T.}\ \bibnamefont {Zinner}},\
  }\href {\doibase 10.1103/PhysRevA.110.032205} {\bibfield  {journal} {\bibinfo
   {journal} {Phys. Rev. A}\ }\textbf {\bibinfo {volume} {110}},\ \bibinfo
  {pages} {032205} (\bibinfo {year} {2024})}\BibitemShut {NoStop}%
\bibitem [{\citenamefont {Beleño}\ \emph {et~al.}(2024)\citenamefont
  {Beleño}, \citenamefont {Santos},\ and\ \citenamefont {Barra}}]{Beleo_2024}%
  \BibitemOpen
  \bibfield  {author} {\bibinfo {author} {\bibfnamefont {Z.}~\bibnamefont
  {Beleño}}, \bibinfo {author} {\bibfnamefont {M.~F.}\ \bibnamefont {Santos}},
  \ and\ \bibinfo {author} {\bibfnamefont {F.}~\bibnamefont {Barra}},\ }\href
  {\doibase 10.1088/1367-2630/ad6348} {\bibfield  {journal} {\bibinfo
  {journal} {New J. Phys.}\ }\textbf {\bibinfo {volume} {26}},\ \bibinfo
  {pages} {073049} (\bibinfo {year} {2024})}\BibitemShut {NoStop}%
\bibitem [{\citenamefont {Andolina}\ \emph {et~al.}(2018)\citenamefont
  {Andolina}, \citenamefont {Farina}, \citenamefont {Mari}, \citenamefont
  {Pellegrini}, \citenamefont {Giovannetti},\ and\ \citenamefont
  {Polini}}]{PhysRevB.98.205423}%
  \BibitemOpen
  \bibfield  {author} {\bibinfo {author} {\bibfnamefont {G.~M.}\ \bibnamefont
  {Andolina}}, \bibinfo {author} {\bibfnamefont {D.}~\bibnamefont {Farina}},
  \bibinfo {author} {\bibfnamefont {A.}~\bibnamefont {Mari}}, \bibinfo {author}
  {\bibfnamefont {V.}~\bibnamefont {Pellegrini}}, \bibinfo {author}
  {\bibfnamefont {V.}~\bibnamefont {Giovannetti}}, \ and\ \bibinfo {author}
  {\bibfnamefont {M.}~\bibnamefont {Polini}},\ }\href {\doibase
  10.1103/PhysRevB.98.205423} {\bibfield  {journal} {\bibinfo  {journal} {Phys.
  Rev. B}\ }\textbf {\bibinfo {volume} {98}},\ \bibinfo {pages} {205423}
  (\bibinfo {year} {2018})}\BibitemShut {NoStop}%
\bibitem [{\citenamefont {Yang}\ \emph
  {et~al.}(2024{\natexlab{a}})\citenamefont {Yang}, \citenamefont {Yang},\ and\
  \citenamefont {Dou}}]{PhysRevB.109.235432}%
  \BibitemOpen
  \bibfield  {author} {\bibinfo {author} {\bibfnamefont {D.-L.}\ \bibnamefont
  {Yang}}, \bibinfo {author} {\bibfnamefont {F.-M.}\ \bibnamefont {Yang}}, \
  and\ \bibinfo {author} {\bibfnamefont {F.-Q.}\ \bibnamefont {Dou}},\ }\href
  {\doibase 10.1103/PhysRevB.109.235432} {\bibfield  {journal} {\bibinfo
  {journal} {Phys. Rev. B}\ }\textbf {\bibinfo {volume} {109}},\ \bibinfo
  {pages} {235432} (\bibinfo {year} {2024}{\natexlab{a}})}\BibitemShut
  {NoStop}%
\bibitem [{\citenamefont {Rossini}\ \emph {et~al.}(2020)\citenamefont
  {Rossini}, \citenamefont {Andolina}, \citenamefont {Rosa}, \citenamefont
  {Carrega},\ and\ \citenamefont {Polini}}]{PhysRevLett.125.236402}%
  \BibitemOpen
  \bibfield  {author} {\bibinfo {author} {\bibfnamefont {D.}~\bibnamefont
  {Rossini}}, \bibinfo {author} {\bibfnamefont {G.~M.}\ \bibnamefont
  {Andolina}}, \bibinfo {author} {\bibfnamefont {D.}~\bibnamefont {Rosa}},
  \bibinfo {author} {\bibfnamefont {M.}~\bibnamefont {Carrega}}, \ and\
  \bibinfo {author} {\bibfnamefont {M.}~\bibnamefont {Polini}},\ }\href
  {\doibase 10.1103/PhysRevLett.125.236402} {\bibfield  {journal} {\bibinfo
  {journal} {Phys. Rev. Lett.}\ }\textbf {\bibinfo {volume} {125}},\ \bibinfo
  {pages} {236402} (\bibinfo {year} {2020})}\BibitemShut {NoStop}%
\bibitem [{\citenamefont {Santos}\ \emph {et~al.}(2019)\citenamefont {Santos},
  \citenamefont {Cakmak}, \citenamefont {Campbell},\ and\ \citenamefont
  {Zinner}}]{PhysRevE.100.032107}%
  \BibitemOpen
  \bibfield  {author} {\bibinfo {author} {\bibfnamefont {A.~C.}\ \bibnamefont
  {Santos}}, \bibinfo {author} {\bibfnamefont {B.}~\bibnamefont {Cakmak}},
  \bibinfo {author} {\bibfnamefont {S.}~\bibnamefont {Campbell}}, \ and\
  \bibinfo {author} {\bibfnamefont {N.~T.}\ \bibnamefont {Zinner}},\ }\href
  {\doibase 10.1103/PhysRevE.100.032107} {\bibfield  {journal} {\bibinfo
  {journal} {Phys. Rev. E}\ }\textbf {\bibinfo {volume} {100}},\ \bibinfo
  {pages} {032107} (\bibinfo {year} {2019})}\BibitemShut {NoStop}%
\bibitem [{\citenamefont {Dou}\ \emph {et~al.}(2020)\citenamefont {Dou},
  \citenamefont {Wang},\ and\ \citenamefont {Sun}}]{Dou2020}%
  \BibitemOpen
  \bibfield  {author} {\bibinfo {author} {\bibfnamefont {F.~Q.}\ \bibnamefont
  {Dou}}, \bibinfo {author} {\bibfnamefont {Y.~J.}\ \bibnamefont {Wang}}, \
  and\ \bibinfo {author} {\bibfnamefont {J.~A.}\ \bibnamefont {Sun}},\ }\href
  {\doibase 10.1209/0295-5075/131/43001} {\bibfield  {journal} {\bibinfo
  {journal} {Europhys. Lett.}\ }\textbf {\bibinfo {volume} {131}},\ \bibinfo
  {pages} {43001} (\bibinfo {year} {2020})}\BibitemShut {NoStop}%
\bibitem [{\citenamefont {Ahmadi}\ \emph {et~al.}(2024)\citenamefont {Ahmadi},
  \citenamefont {Mazurek}, \citenamefont {Horodecki},\ and\ \citenamefont
  {Barzanjeh}}]{PhysRevLett.132.210402}%
  \BibitemOpen
  \bibfield  {author} {\bibinfo {author} {\bibfnamefont {B.}~\bibnamefont
  {Ahmadi}}, \bibinfo {author} {\bibfnamefont {P.}~\bibnamefont {Mazurek}},
  \bibinfo {author} {\bibfnamefont {P.}~\bibnamefont {Horodecki}}, \ and\
  \bibinfo {author} {\bibfnamefont {S.}~\bibnamefont {Barzanjeh}},\ }\href
  {\doibase 10.1103/PhysRevLett.132.210402} {\bibfield  {journal} {\bibinfo
  {journal} {Phys. Rev. Lett.}\ }\textbf {\bibinfo {volume} {132}},\ \bibinfo
  {pages} {210402} (\bibinfo {year} {2024})}\BibitemShut {NoStop}%
\bibitem [{\citenamefont {Konar}\ \emph {et~al.}(2022)\citenamefont {Konar},
  \citenamefont {Lakkaraju}, \citenamefont {Ghosh},\ and\ \citenamefont
  {Sen(De)}}]{PhysRevA.106.022618}%
  \BibitemOpen
  \bibfield  {author} {\bibinfo {author} {\bibfnamefont {T.~K.}\ \bibnamefont
  {Konar}}, \bibinfo {author} {\bibfnamefont {L.~G.~C.}\ \bibnamefont
  {Lakkaraju}}, \bibinfo {author} {\bibfnamefont {S.}~\bibnamefont {Ghosh}}, \
  and\ \bibinfo {author} {\bibfnamefont {A.}~\bibnamefont {Sen(De)}},\ }\href
  {\doibase 10.1103/PhysRevA.106.022618} {\bibfield  {journal} {\bibinfo
  {journal} {Phys. Rev. A}\ }\textbf {\bibinfo {volume} {106}},\ \bibinfo
  {pages} {022618} (\bibinfo {year} {2022})}\BibitemShut {NoStop}%
\bibitem [{\citenamefont {Lu}\ \emph {et~al.}(2024)\citenamefont {Lu},
  \citenamefont {Tian}, \citenamefont {Lü},\ and\ \citenamefont
  {Shang}}]{lu2024topologicalquantumbatteries}%
  \BibitemOpen
  \bibfield  {author} {\bibinfo {author} {\bibfnamefont {Z.-G.}\ \bibnamefont
  {Lu}}, \bibinfo {author} {\bibfnamefont {G.}~\bibnamefont {Tian}}, \bibinfo
  {author} {\bibfnamefont {X.-Y.}\ \bibnamefont {Lü}}, \ and\ \bibinfo
  {author} {\bibfnamefont {C.}~\bibnamefont {Shang}},\ }\href@noop {} {}
  (\bibinfo {year} {2024}),\ \Eprint {http://arxiv.org/abs/2405.03675}
  {2405.03675} \BibitemShut {NoStop}%
\bibitem [{\citenamefont {Pirmoradian}\ and\ \citenamefont
  {M\o{}lmer}(2019)}]{PhysRevA.100.043833}%
  \BibitemOpen
  \bibfield  {author} {\bibinfo {author} {\bibfnamefont {F.}~\bibnamefont
  {Pirmoradian}}\ and\ \bibinfo {author} {\bibfnamefont {K.}~\bibnamefont
  {M\o{}lmer}},\ }\href {\doibase 10.1103/PhysRevA.100.043833} {\bibfield
  {journal} {\bibinfo  {journal} {Phys. Rev. A}\ }\textbf {\bibinfo {volume}
  {100}},\ \bibinfo {pages} {043833} (\bibinfo {year} {2019})}\BibitemShut
  {NoStop}%
\bibitem [{\citenamefont {Yao}\ and\ \citenamefont
  {Shao}(2021)}]{PhysRevE.104.044116}%
  \BibitemOpen
  \bibfield  {author} {\bibinfo {author} {\bibfnamefont {Y.}~\bibnamefont
  {Yao}}\ and\ \bibinfo {author} {\bibfnamefont {X.~Q.}\ \bibnamefont {Shao}},\
  }\href {\doibase 10.1103/PhysRevE.104.044116} {\bibfield  {journal} {\bibinfo
   {journal} {Phys. Rev. E}\ }\textbf {\bibinfo {volume} {104}},\ \bibinfo
  {pages} {044116} (\bibinfo {year} {2021})}\BibitemShut {NoStop}%
\bibitem [{\citenamefont {Fusco}\ \emph {et~al.}(2016)\citenamefont {Fusco},
  \citenamefont {Paternostro},\ and\ \citenamefont
  {De~Chiara}}]{PhysRevE.94.052122}%
  \BibitemOpen
  \bibfield  {author} {\bibinfo {author} {\bibfnamefont {L.}~\bibnamefont
  {Fusco}}, \bibinfo {author} {\bibfnamefont {M.}~\bibnamefont {Paternostro}},
  \ and\ \bibinfo {author} {\bibfnamefont {G.}~\bibnamefont {De~Chiara}},\
  }\href {\doibase 10.1103/PhysRevE.94.052122} {\bibfield  {journal} {\bibinfo
  {journal} {Phys. Rev. E}\ }\textbf {\bibinfo {volume} {94}},\ \bibinfo
  {pages} {052122} (\bibinfo {year} {2016})}\BibitemShut {NoStop}%
\bibitem [{\citenamefont {Andolina}\ \emph {et~al.}(2019)\citenamefont
  {Andolina}, \citenamefont {Keck}, \citenamefont {Mari}, \citenamefont
  {Campisi}, \citenamefont {Giovannetti},\ and\ \citenamefont
  {Polini}}]{PhysRevLett.122.047702}%
  \BibitemOpen
  \bibfield  {author} {\bibinfo {author} {\bibfnamefont {G.~M.}\ \bibnamefont
  {Andolina}}, \bibinfo {author} {\bibfnamefont {M.}~\bibnamefont {Keck}},
  \bibinfo {author} {\bibfnamefont {A.}~\bibnamefont {Mari}}, \bibinfo {author}
  {\bibfnamefont {M.}~\bibnamefont {Campisi}}, \bibinfo {author} {\bibfnamefont
  {V.}~\bibnamefont {Giovannetti}}, \ and\ \bibinfo {author} {\bibfnamefont
  {M.}~\bibnamefont {Polini}},\ }\href {\doibase
  10.1103/PhysRevLett.122.047702} {\bibfield  {journal} {\bibinfo  {journal}
  {Phys. Rev. Lett.}\ }\textbf {\bibinfo {volume} {122}},\ \bibinfo {pages}
  {047702} (\bibinfo {year} {2019})}\BibitemShut {NoStop}%
\bibitem [{\citenamefont {Ferraro}\ \emph {et~al.}(2018)\citenamefont
  {Ferraro}, \citenamefont {Campisi}, \citenamefont {Andolina}, \citenamefont
  {Pellegrini},\ and\ \citenamefont {Polini}}]{PhysRevLett.120.117702}%
  \BibitemOpen
  \bibfield  {author} {\bibinfo {author} {\bibfnamefont {D.}~\bibnamefont
  {Ferraro}}, \bibinfo {author} {\bibfnamefont {M.}~\bibnamefont {Campisi}},
  \bibinfo {author} {\bibfnamefont {G.~M.}\ \bibnamefont {Andolina}}, \bibinfo
  {author} {\bibfnamefont {V.}~\bibnamefont {Pellegrini}}, \ and\ \bibinfo
  {author} {\bibfnamefont {M.}~\bibnamefont {Polini}},\ }\href {\doibase
  10.1103/PhysRevLett.120.117702} {\bibfield  {journal} {\bibinfo  {journal}
  {Phys. Rev. Lett.}\ }\textbf {\bibinfo {volume} {120}},\ \bibinfo {pages}
  {117702} (\bibinfo {year} {2018})}\BibitemShut {NoStop}%
\bibitem [{\citenamefont {Zhang}\ and\ \citenamefont
  {Blaauboer}(2023)}]{zhang2018enhanced}%
  \BibitemOpen
  \bibfield  {author} {\bibinfo {author} {\bibfnamefont {X.}~\bibnamefont
  {Zhang}}\ and\ \bibinfo {author} {\bibfnamefont {M.}~\bibnamefont
  {Blaauboer}},\ }\href
  {https://www.frontiersin.org/journals/physics/articles/10.3389/fphy.2022.1097564}
  {\bibfield  {journal} {\bibinfo  {journal} {Front. Phys.}\ }\textbf {\bibinfo
  {volume} {10}},\ \bibinfo {pages} {1097564} (\bibinfo {year}
  {2023})}\BibitemShut {NoStop}%
\bibitem [{\citenamefont {Crescente}\ \emph {et~al.}(2020)\citenamefont
  {Crescente}, \citenamefont {Carrega}, \citenamefont {Sassetti},\ and\
  \citenamefont {Ferraro}}]{PhysRevB.102.245407}%
  \BibitemOpen
  \bibfield  {author} {\bibinfo {author} {\bibfnamefont {A.}~\bibnamefont
  {Crescente}}, \bibinfo {author} {\bibfnamefont {M.}~\bibnamefont {Carrega}},
  \bibinfo {author} {\bibfnamefont {M.}~\bibnamefont {Sassetti}}, \ and\
  \bibinfo {author} {\bibfnamefont {D.}~\bibnamefont {Ferraro}},\ }\href
  {\doibase 10.1103/PhysRevB.102.245407} {\bibfield  {journal} {\bibinfo
  {journal} {Phys. Rev. B}\ }\textbf {\bibinfo {volume} {102}},\ \bibinfo
  {pages} {245407} (\bibinfo {year} {2020})}\BibitemShut {NoStop}%
\bibitem [{\citenamefont {Dou}\ \emph {et~al.}(2022{\natexlab{a}})\citenamefont
  {Dou}, \citenamefont {Lu}, \citenamefont {Wang},\ and\ \citenamefont
  {Sun}}]{PhysRevB.105.115405}%
  \BibitemOpen
  \bibfield  {author} {\bibinfo {author} {\bibfnamefont {F.-Q.}\ \bibnamefont
  {Dou}}, \bibinfo {author} {\bibfnamefont {Y.-Q.}\ \bibnamefont {Lu}},
  \bibinfo {author} {\bibfnamefont {Y.-J.}\ \bibnamefont {Wang}}, \ and\
  \bibinfo {author} {\bibfnamefont {J.-A.}\ \bibnamefont {Sun}},\ }\href
  {\doibase 10.1103/PhysRevB.105.115405} {\bibfield  {journal} {\bibinfo
  {journal} {Phys. Rev. B}\ }\textbf {\bibinfo {volume} {105}},\ \bibinfo
  {pages} {115405} (\bibinfo {year} {2022}{\natexlab{a}})}\BibitemShut
  {NoStop}%
\bibitem [{\citenamefont {Hadipour}\ \emph {et~al.}(2024)\citenamefont
  {Hadipour}, \citenamefont {Haseli}, \citenamefont {Wang},\ and\ \citenamefont
  {Haddadi}}]{https://doi.org/10.1002/qute.202400115}%
  \BibitemOpen
  \bibfield  {author} {\bibinfo {author} {\bibfnamefont {M.}~\bibnamefont
  {Hadipour}}, \bibinfo {author} {\bibfnamefont {S.}~\bibnamefont {Haseli}},
  \bibinfo {author} {\bibfnamefont {D.}~\bibnamefont {Wang}}, \ and\ \bibinfo
  {author} {\bibfnamefont {S.}~\bibnamefont {Haddadi}},\ }\href {\doibase
  https://doi.org/10.1002/qute.202400115} {\bibfield  {journal} {\bibinfo
  {journal} {Adv. Quantum Technol.}\ }\textbf {\bibinfo {volume} {7}},\
  \bibinfo {pages} {2400115} (\bibinfo {year} {2024})}\BibitemShut {NoStop}%
\bibitem [{\citenamefont {Wang}\ \emph {et~al.}(2023)\citenamefont {Wang},
  \citenamefont {Liu}, \citenamefont {Wu}, \citenamefont {Fan},\ and\
  \citenamefont {Liu}}]{PhysRevA.108.062402}%
  \BibitemOpen
  \bibfield  {author} {\bibinfo {author} {\bibfnamefont {L.}~\bibnamefont
  {Wang}}, \bibinfo {author} {\bibfnamefont {S.-Q.}\ \bibnamefont {Liu}},
  \bibinfo {author} {\bibfnamefont {F.-L.}\ \bibnamefont {Wu}}, \bibinfo
  {author} {\bibfnamefont {H.}~\bibnamefont {Fan}}, \ and\ \bibinfo {author}
  {\bibfnamefont {S.-Y.}\ \bibnamefont {Liu}},\ }\href {\doibase
  10.1103/PhysRevA.108.062402} {\bibfield  {journal} {\bibinfo  {journal}
  {Phys. Rev. A}\ }\textbf {\bibinfo {volume} {108}},\ \bibinfo {pages}
  {062402} (\bibinfo {year} {2023})}\BibitemShut {NoStop}%
\bibitem [{\citenamefont {Grazi}\ \emph {et~al.}(2024)\citenamefont {Grazi},
  \citenamefont {Sacco~Shaikh}, \citenamefont {Sassetti}, \citenamefont
  {Traverso~Ziani},\ and\ \citenamefont {Ferraro}}]{PhysRevLett.133.197001}%
  \BibitemOpen
  \bibfield  {author} {\bibinfo {author} {\bibfnamefont {R.}~\bibnamefont
  {Grazi}}, \bibinfo {author} {\bibfnamefont {D.}~\bibnamefont {Sacco~Shaikh}},
  \bibinfo {author} {\bibfnamefont {M.}~\bibnamefont {Sassetti}}, \bibinfo
  {author} {\bibfnamefont {N.}~\bibnamefont {Traverso~Ziani}}, \ and\ \bibinfo
  {author} {\bibfnamefont {D.}~\bibnamefont {Ferraro}},\ }\href {\doibase
  10.1103/PhysRevLett.133.197001} {\bibfield  {journal} {\bibinfo  {journal}
  {Phys. Rev. Lett.}\ }\textbf {\bibinfo {volume} {133}},\ \bibinfo {pages}
  {197001} (\bibinfo {year} {2024})}\BibitemShut {NoStop}%
\bibitem [{\citenamefont {Yao}\ and\ \citenamefont
  {Shao}(2022)}]{PhysRevE.106.014138}%
  \BibitemOpen
  \bibfield  {author} {\bibinfo {author} {\bibfnamefont {Y.}~\bibnamefont
  {Yao}}\ and\ \bibinfo {author} {\bibfnamefont {X.~Q.}\ \bibnamefont {Shao}},\
  }\href {\doibase 10.1103/PhysRevE.106.014138} {\bibfield  {journal} {\bibinfo
   {journal} {Phys. Rev. E}\ }\textbf {\bibinfo {volume} {106}},\ \bibinfo
  {pages} {014138} (\bibinfo {year} {2022})}\BibitemShut {NoStop}%
\bibitem [{\citenamefont {Zhao}\ \emph {et~al.}(2021)\citenamefont {Zhao},
  \citenamefont {Dou},\ and\ \citenamefont {Zhao}}]{PhysRevA.103.033715}%
  \BibitemOpen
  \bibfield  {author} {\bibinfo {author} {\bibfnamefont {F.}~\bibnamefont
  {Zhao}}, \bibinfo {author} {\bibfnamefont {F.-Q.}\ \bibnamefont {Dou}}, \
  and\ \bibinfo {author} {\bibfnamefont {Q.}~\bibnamefont {Zhao}},\ }\href
  {\doibase 10.1103/PhysRevA.103.033715} {\bibfield  {journal} {\bibinfo
  {journal} {Phys. Rev. A}\ }\textbf {\bibinfo {volume} {103}},\ \bibinfo
  {pages} {033715} (\bibinfo {year} {2021})}\BibitemShut {NoStop}%
\bibitem [{\citenamefont {Evangelakos}\ \emph {et~al.}(2024)\citenamefont
  {Evangelakos}, \citenamefont {Paspalakis},\ and\ \citenamefont
  {Stefanatos}}]{PhysRevA.110.052601}%
  \BibitemOpen
  \bibfield  {author} {\bibinfo {author} {\bibfnamefont {V.}~\bibnamefont
  {Evangelakos}}, \bibinfo {author} {\bibfnamefont {E.}~\bibnamefont
  {Paspalakis}}, \ and\ \bibinfo {author} {\bibfnamefont {D.}~\bibnamefont
  {Stefanatos}},\ }\href {\doibase 10.1103/PhysRevA.110.052601} {\bibfield
  {journal} {\bibinfo  {journal} {Phys. Rev. A}\ }\textbf {\bibinfo {volume}
  {110}},\ \bibinfo {pages} {052601} (\bibinfo {year} {2024})}\BibitemShut
  {NoStop}%
\bibitem [{\citenamefont {Salvia}\ \emph {et~al.}(2023)\citenamefont {Salvia},
  \citenamefont {Perarnau-Llobet}, \citenamefont {Haack}, \citenamefont
  {Brunner},\ and\ \citenamefont {Nimmrichter}}]{PhysRevResearch.5.013155}%
  \BibitemOpen
  \bibfield  {author} {\bibinfo {author} {\bibfnamefont {R.}~\bibnamefont
  {Salvia}}, \bibinfo {author} {\bibfnamefont {M.}~\bibnamefont
  {Perarnau-Llobet}}, \bibinfo {author} {\bibfnamefont {G.}~\bibnamefont
  {Haack}}, \bibinfo {author} {\bibfnamefont {N.}~\bibnamefont {Brunner}}, \
  and\ \bibinfo {author} {\bibfnamefont {S.}~\bibnamefont {Nimmrichter}},\
  }\href {\doibase 10.1103/PhysRevResearch.5.013155} {\bibfield  {journal}
  {\bibinfo  {journal} {Phys. Rev. Res.}\ }\textbf {\bibinfo {volume} {5}},\
  \bibinfo {pages} {013155} (\bibinfo {year} {2023})}\BibitemShut {NoStop}%
\bibitem [{\citenamefont {Peng}\ \emph {et~al.}(2021)\citenamefont {Peng},
  \citenamefont {He}, \citenamefont {Chesi}, \citenamefont {Lin},\ and\
  \citenamefont {Guan}}]{PhysRevA.103.052220}%
  \BibitemOpen
  \bibfield  {author} {\bibinfo {author} {\bibfnamefont {L.}~\bibnamefont
  {Peng}}, \bibinfo {author} {\bibfnamefont {W.~B.}\ \bibnamefont {He}},
  \bibinfo {author} {\bibfnamefont {S.}~\bibnamefont {Chesi}}, \bibinfo
  {author} {\bibfnamefont {H.~Q.}\ \bibnamefont {Lin}}, \ and\ \bibinfo
  {author} {\bibfnamefont {X.~W.}\ \bibnamefont {Guan}},\ }\href {\doibase
  10.1103/PhysRevA.103.052220} {\bibfield  {journal} {\bibinfo  {journal}
  {Phys. Rev. A}\ }\textbf {\bibinfo {volume} {103}},\ \bibinfo {pages}
  {052220} (\bibinfo {year} {2021})}\BibitemShut {NoStop}%
\bibitem [{\citenamefont {Shi}\ \emph {et~al.}(2022)\citenamefont {Shi},
  \citenamefont {Ding}, \citenamefont {Wan}, \citenamefont {Wang},\ and\
  \citenamefont {Yang}}]{PhysRevLett.129.130602}%
  \BibitemOpen
  \bibfield  {author} {\bibinfo {author} {\bibfnamefont {H.-L.}\ \bibnamefont
  {Shi}}, \bibinfo {author} {\bibfnamefont {S.}~\bibnamefont {Ding}}, \bibinfo
  {author} {\bibfnamefont {Q.-K.}\ \bibnamefont {Wan}}, \bibinfo {author}
  {\bibfnamefont {X.-H.}\ \bibnamefont {Wang}}, \ and\ \bibinfo {author}
  {\bibfnamefont {W.-L.}\ \bibnamefont {Yang}},\ }\href {\doibase
  10.1103/PhysRevLett.129.130602} {\bibfield  {journal} {\bibinfo  {journal}
  {Phys. Rev. Lett.}\ }\textbf {\bibinfo {volume} {129}},\ \bibinfo {pages}
  {130602} (\bibinfo {year} {2022})}\BibitemShut {NoStop}%
\bibitem [{\citenamefont {Le}\ \emph {et~al.}(2018)\citenamefont {Le},
  \citenamefont {Levinsen}, \citenamefont {Modi}, \citenamefont {Parish},\ and\
  \citenamefont {Pollock}}]{PhysRevA.97.022106}%
  \BibitemOpen
  \bibfield  {author} {\bibinfo {author} {\bibfnamefont {T.~P.}\ \bibnamefont
  {Le}}, \bibinfo {author} {\bibfnamefont {J.}~\bibnamefont {Levinsen}},
  \bibinfo {author} {\bibfnamefont {K.}~\bibnamefont {Modi}}, \bibinfo {author}
  {\bibfnamefont {M.~M.}\ \bibnamefont {Parish}}, \ and\ \bibinfo {author}
  {\bibfnamefont {F.~A.}\ \bibnamefont {Pollock}},\ }\href {\doibase
  10.1103/PhysRevA.97.022106} {\bibfield  {journal} {\bibinfo  {journal} {Phys.
  Rev. A}\ }\textbf {\bibinfo {volume} {97}},\ \bibinfo {pages} {022106}
  (\bibinfo {year} {2018})}\BibitemShut {NoStop}%
\bibitem [{\citenamefont {Dou}\ \emph {et~al.}(2022{\natexlab{b}})\citenamefont
  {Dou}, \citenamefont {Wang},\ and\ \citenamefont {Sun}}]{DouLMG2022}%
  \BibitemOpen
  \bibfield  {author} {\bibinfo {author} {\bibfnamefont {F.-Q.}\ \bibnamefont
  {Dou}}, \bibinfo {author} {\bibfnamefont {Y.-J.}\ \bibnamefont {Wang}}, \
  and\ \bibinfo {author} {\bibfnamefont {J.-A.}\ \bibnamefont {Sun}},\
  }\href@noop {} {} (\bibinfo {year} {2022}{\natexlab{b}}),\ \Eprint
  {http://arxiv.org/abs/arXiv:2208.04831} {arXiv:2208.04831} \BibitemShut
  {NoStop}%
\bibitem [{\citenamefont {Ghosh}\ and\ \citenamefont
  {Sen(De)}(2022)}]{PhysRevA.105.022628}%
  \BibitemOpen
  \bibfield  {author} {\bibinfo {author} {\bibfnamefont {S.}~\bibnamefont
  {Ghosh}}\ and\ \bibinfo {author} {\bibfnamefont {A.}~\bibnamefont
  {Sen(De)}},\ }\href {\doibase 10.1103/PhysRevA.105.022628} {\bibfield
  {journal} {\bibinfo  {journal} {Phys. Rev. A}\ }\textbf {\bibinfo {volume}
  {105}},\ \bibinfo {pages} {022628} (\bibinfo {year} {2022})}\BibitemShut
  {NoStop}%
\bibitem [{\citenamefont {Ali}\ \emph {et~al.}(2024)\citenamefont {Ali},
  \citenamefont {Al-Kuwari}, \citenamefont {Hussain}, \citenamefont {Byrnes},
  \citenamefont {Rahim}, \citenamefont {Quach}, \citenamefont {Ghominejad},\
  and\ \citenamefont {Haddadi}}]{PhysRevA.110.052404}%
  \BibitemOpen
  \bibfield  {author} {\bibinfo {author} {\bibfnamefont {A.}~\bibnamefont
  {Ali}}, \bibinfo {author} {\bibfnamefont {S.}~\bibnamefont {Al-Kuwari}},
  \bibinfo {author} {\bibfnamefont {M.~I.}\ \bibnamefont {Hussain}}, \bibinfo
  {author} {\bibfnamefont {T.}~\bibnamefont {Byrnes}}, \bibinfo {author}
  {\bibfnamefont {M.~T.}\ \bibnamefont {Rahim}}, \bibinfo {author}
  {\bibfnamefont {J.~Q.}\ \bibnamefont {Quach}}, \bibinfo {author}
  {\bibfnamefont {M.}~\bibnamefont {Ghominejad}}, \ and\ \bibinfo {author}
  {\bibfnamefont {S.}~\bibnamefont {Haddadi}},\ }\href {\doibase
  10.1103/PhysRevA.110.052404} {\bibfield  {journal} {\bibinfo  {journal}
  {Phys. Rev. A}\ }\textbf {\bibinfo {volume} {110}},\ \bibinfo {pages}
  {052404} (\bibinfo {year} {2024})}\BibitemShut {NoStop}%
\bibitem [{\citenamefont {Ghosh}\ \emph {et~al.}(2020)\citenamefont {Ghosh},
  \citenamefont {Chanda},\ and\ \citenamefont {Sen(De)}}]{PhysRevA.101.032115}%
  \BibitemOpen
  \bibfield  {author} {\bibinfo {author} {\bibfnamefont {S.}~\bibnamefont
  {Ghosh}}, \bibinfo {author} {\bibfnamefont {T.}~\bibnamefont {Chanda}}, \
  and\ \bibinfo {author} {\bibfnamefont {A.}~\bibnamefont {Sen(De)}},\ }\href
  {\doibase 10.1103/PhysRevA.101.032115} {\bibfield  {journal} {\bibinfo
  {journal} {Phys. Rev. A}\ }\textbf {\bibinfo {volume} {101}},\ \bibinfo
  {pages} {032115} (\bibinfo {year} {2020})}\BibitemShut {NoStop}%
\bibitem [{\citenamefont {Huangfu}\ and\ \citenamefont
  {Jing}(2021)}]{PhysRevE.104.024129}%
  \BibitemOpen
  \bibfield  {author} {\bibinfo {author} {\bibfnamefont {Y.}~\bibnamefont
  {Huangfu}}\ and\ \bibinfo {author} {\bibfnamefont {J.}~\bibnamefont {Jing}},\
  }\href {\doibase 10.1103/PhysRevE.104.024129} {\bibfield  {journal} {\bibinfo
   {journal} {Phys. Rev. E}\ }\textbf {\bibinfo {volume} {104}},\ \bibinfo
  {pages} {024129} (\bibinfo {year} {2021})}\BibitemShut {NoStop}%
\bibitem [{\citenamefont {Rossini}\ \emph {et~al.}(2019)\citenamefont
  {Rossini}, \citenamefont {Andolina},\ and\ \citenamefont
  {Polini}}]{PhysRevB.100.115142}%
  \BibitemOpen
  \bibfield  {author} {\bibinfo {author} {\bibfnamefont {D.}~\bibnamefont
  {Rossini}}, \bibinfo {author} {\bibfnamefont {G.~M.}\ \bibnamefont
  {Andolina}}, \ and\ \bibinfo {author} {\bibfnamefont {M.}~\bibnamefont
  {Polini}},\ }\href {\doibase 10.1103/PhysRevB.100.115142} {\bibfield
  {journal} {\bibinfo  {journal} {Phys. Rev. B}\ }\textbf {\bibinfo {volume}
  {100}},\ \bibinfo {pages} {115142} (\bibinfo {year} {2019})}\BibitemShut
  {NoStop}%
\bibitem [{\citenamefont {de~Moraes}\ \emph {et~al.}(2024)\citenamefont
  {de~Moraes}, \citenamefont {Duriez}, \citenamefont {Saguia}, \citenamefont
  {Santos},\ and\ \citenamefont {Sarandy}}]{deMoraes_2024}%
  \BibitemOpen
  \bibfield  {author} {\bibinfo {author} {\bibfnamefont {L.~F.~C.}\
  \bibnamefont {de~Moraes}}, \bibinfo {author} {\bibfnamefont {A.~C.}\
  \bibnamefont {Duriez}}, \bibinfo {author} {\bibfnamefont {A.}~\bibnamefont
  {Saguia}}, \bibinfo {author} {\bibfnamefont {A.~C.}\ \bibnamefont {Santos}},
  \ and\ \bibinfo {author} {\bibfnamefont {M.~S.}\ \bibnamefont {Sarandy}},\
  }\href {\doibase 10.1088/2058-9565/ad71ed} {\bibfield  {journal} {\bibinfo
  {journal} {Quantum Sci. Technol.}\ }\textbf {\bibinfo {volume} {9}},\
  \bibinfo {pages} {045033} (\bibinfo {year} {2024})}\BibitemShut {NoStop}%
\bibitem [{\citenamefont {Konar}\ \emph {et~al.}(2024)\citenamefont {Konar},
  \citenamefont {Lakkaraju},\ and\ \citenamefont
  {Sen~(De)}}]{PhysRevA.109.042207}%
  \BibitemOpen
  \bibfield  {author} {\bibinfo {author} {\bibfnamefont {T.~K.}\ \bibnamefont
  {Konar}}, \bibinfo {author} {\bibfnamefont {L.~G.~C.}\ \bibnamefont
  {Lakkaraju}}, \ and\ \bibinfo {author} {\bibfnamefont {A.}~\bibnamefont
  {Sen~(De)}},\ }\href {\doibase 10.1103/PhysRevA.109.042207} {\bibfield
  {journal} {\bibinfo  {journal} {Phys. Rev. A}\ }\textbf {\bibinfo {volume}
  {109}},\ \bibinfo {pages} {042207} (\bibinfo {year} {2024})}\BibitemShut
  {NoStop}%
\bibitem [{\citenamefont {Kamin}\ \emph {et~al.}(2020)\citenamefont {Kamin},
  \citenamefont {Tabesh}, \citenamefont {Salimi},\ and\ \citenamefont
  {Santos}}]{PhysRevE.102.052109}%
  \BibitemOpen
  \bibfield  {author} {\bibinfo {author} {\bibfnamefont {F.~H.}\ \bibnamefont
  {Kamin}}, \bibinfo {author} {\bibfnamefont {F.~T.}\ \bibnamefont {Tabesh}},
  \bibinfo {author} {\bibfnamefont {S.}~\bibnamefont {Salimi}}, \ and\ \bibinfo
  {author} {\bibfnamefont {A.~C.}\ \bibnamefont {Santos}},\ }\href {\doibase
  10.1103/PhysRevE.102.052109} {\bibfield  {journal} {\bibinfo  {journal}
  {Phys. Rev. E}\ }\textbf {\bibinfo {volume} {102}},\ \bibinfo {pages}
  {052109} (\bibinfo {year} {2020})}\BibitemShut {NoStop}%
\bibitem [{\citenamefont {Mojaveri}\ \emph {et~al.}(2024)\citenamefont
  {Mojaveri}, \citenamefont {Jafarzadeh~Bahrbeig},\ and\ \citenamefont
  {Fasihi}}]{PhysRevA.109.042619}%
  \BibitemOpen
  \bibfield  {author} {\bibinfo {author} {\bibfnamefont {B.}~\bibnamefont
  {Mojaveri}}, \bibinfo {author} {\bibfnamefont {R.}~\bibnamefont
  {Jafarzadeh~Bahrbeig}}, \ and\ \bibinfo {author} {\bibfnamefont {M.~A.}\
  \bibnamefont {Fasihi}},\ }\href {\doibase 10.1103/PhysRevA.109.042619}
  {\bibfield  {journal} {\bibinfo  {journal} {Phys. Rev. A}\ }\textbf {\bibinfo
  {volume} {109}},\ \bibinfo {pages} {042619} (\bibinfo {year}
  {2024})}\BibitemShut {NoStop}%
\bibitem [{\citenamefont {Liu}\ \emph {et~al.}(2024)\citenamefont {Liu},
  \citenamefont {Wang}, \citenamefont {Fan}, \citenamefont {Wu},\ and\
  \citenamefont {Liu}}]{PhysRevA.109.042411}%
  \BibitemOpen
  \bibfield  {author} {\bibinfo {author} {\bibfnamefont {S.-Q.}\ \bibnamefont
  {Liu}}, \bibinfo {author} {\bibfnamefont {L.}~\bibnamefont {Wang}}, \bibinfo
  {author} {\bibfnamefont {H.}~\bibnamefont {Fan}}, \bibinfo {author}
  {\bibfnamefont {F.-L.}\ \bibnamefont {Wu}}, \ and\ \bibinfo {author}
  {\bibfnamefont {S.-Y.}\ \bibnamefont {Liu}},\ }\href {\doibase
  10.1103/PhysRevA.109.042411} {\bibfield  {journal} {\bibinfo  {journal}
  {Phys. Rev. A}\ }\textbf {\bibinfo {volume} {109}},\ \bibinfo {pages}
  {042411} (\bibinfo {year} {2024})}\BibitemShut {NoStop}%
\bibitem [{\citenamefont {Chen}\ \emph {et~al.}(2022)\citenamefont {Chen},
  \citenamefont {Yin}, \citenamefont {Jiang},\ and\ \citenamefont
  {Jin}}]{PhysRevE.106.054119}%
  \BibitemOpen
  \bibfield  {author} {\bibinfo {author} {\bibfnamefont {P.}~\bibnamefont
  {Chen}}, \bibinfo {author} {\bibfnamefont {T.~S.}\ \bibnamefont {Yin}},
  \bibinfo {author} {\bibfnamefont {Z.~Q.}\ \bibnamefont {Jiang}}, \ and\
  \bibinfo {author} {\bibfnamefont {G.~R.}\ \bibnamefont {Jin}},\ }\href
  {\doibase 10.1103/PhysRevE.106.054119} {\bibfield  {journal} {\bibinfo
  {journal} {Phys. Rev. E}\ }\textbf {\bibinfo {volume} {106}},\ \bibinfo
  {pages} {054119} (\bibinfo {year} {2022})}\BibitemShut {NoStop}%
\bibitem [{\citenamefont {Gao}\ \emph {et~al.}(2022)\citenamefont {Gao},
  \citenamefont {Cheng}, \citenamefont {He}, \citenamefont {Mondaini},
  \citenamefont {Guan},\ and\ \citenamefont {Lin}}]{PhysRevResearch.4.043150}%
  \BibitemOpen
  \bibfield  {author} {\bibinfo {author} {\bibfnamefont {L.}~\bibnamefont
  {Gao}}, \bibinfo {author} {\bibfnamefont {C.}~\bibnamefont {Cheng}}, \bibinfo
  {author} {\bibfnamefont {W.-B.}\ \bibnamefont {He}}, \bibinfo {author}
  {\bibfnamefont {R.}~\bibnamefont {Mondaini}}, \bibinfo {author}
  {\bibfnamefont {X.-W.}\ \bibnamefont {Guan}}, \ and\ \bibinfo {author}
  {\bibfnamefont {H.-Q.}\ \bibnamefont {Lin}},\ }\href {\doibase
  10.1103/PhysRevResearch.4.043150} {\bibfield  {journal} {\bibinfo  {journal}
  {Phys. Rev. Res.}\ }\textbf {\bibinfo {volume} {4}},\ \bibinfo {pages}
  {043150} (\bibinfo {year} {2022})}\BibitemShut {NoStop}%
\bibitem [{\citenamefont {Dou}\ \emph {et~al.}(2022{\natexlab{c}})\citenamefont
  {Dou}, \citenamefont {Zhou},\ and\ \citenamefont
  {Sun}}]{PhysRevA.106.032212}%
  \BibitemOpen
  \bibfield  {author} {\bibinfo {author} {\bibfnamefont {F.-Q.}\ \bibnamefont
  {Dou}}, \bibinfo {author} {\bibfnamefont {H.}~\bibnamefont {Zhou}}, \ and\
  \bibinfo {author} {\bibfnamefont {J.-A.}\ \bibnamefont {Sun}},\ }\href
  {\doibase 10.1103/PhysRevA.106.032212} {\bibfield  {journal} {\bibinfo
  {journal} {Phys. Rev. A}\ }\textbf {\bibinfo {volume} {106}},\ \bibinfo
  {pages} {032212} (\bibinfo {year} {2022}{\natexlab{c}})}\BibitemShut
  {NoStop}%
\bibitem [{\citenamefont {Xu}\ \emph {et~al.}(2021)\citenamefont {Xu},
  \citenamefont {Zhu}, \citenamefont {Zhang},\ and\ \citenamefont
  {Liu}}]{PhysRevE.104.064143}%
  \BibitemOpen
  \bibfield  {author} {\bibinfo {author} {\bibfnamefont {K.}~\bibnamefont
  {Xu}}, \bibinfo {author} {\bibfnamefont {H.-J.}\ \bibnamefont {Zhu}},
  \bibinfo {author} {\bibfnamefont {G.-F.}\ \bibnamefont {Zhang}}, \ and\
  \bibinfo {author} {\bibfnamefont {W.-M.}\ \bibnamefont {Liu}},\ }\href
  {\doibase 10.1103/PhysRevE.104.064143} {\bibfield  {journal} {\bibinfo
  {journal} {Phys. Rev. E}\ }\textbf {\bibinfo {volume} {104}},\ \bibinfo
  {pages} {064143} (\bibinfo {year} {2021})}\BibitemShut {NoStop}%
\bibitem [{\citenamefont {Farina}\ \emph {et~al.}(2019)\citenamefont {Farina},
  \citenamefont {Andolina}, \citenamefont {Mari}, \citenamefont {Polini},\ and\
  \citenamefont {Giovannetti}}]{PhysRevB.99.035421}%
  \BibitemOpen
  \bibfield  {author} {\bibinfo {author} {\bibfnamefont {D.}~\bibnamefont
  {Farina}}, \bibinfo {author} {\bibfnamefont {G.~M.}\ \bibnamefont
  {Andolina}}, \bibinfo {author} {\bibfnamefont {A.}~\bibnamefont {Mari}},
  \bibinfo {author} {\bibfnamefont {M.}~\bibnamefont {Polini}}, \ and\ \bibinfo
  {author} {\bibfnamefont {V.}~\bibnamefont {Giovannetti}},\ }\href {\doibase
  10.1103/PhysRevB.99.035421} {\bibfield  {journal} {\bibinfo  {journal} {Phys.
  Rev. B}\ }\textbf {\bibinfo {volume} {99}},\ \bibinfo {pages} {035421}
  (\bibinfo {year} {2019})}\BibitemShut {NoStop}%
\bibitem [{\citenamefont {Yang}\ and\ \citenamefont
  {Dou}(2024)}]{PhysRevA.109.062432}%
  \BibitemOpen
  \bibfield  {author} {\bibinfo {author} {\bibfnamefont {F.-M.}\ \bibnamefont
  {Yang}}\ and\ \bibinfo {author} {\bibfnamefont {F.-Q.}\ \bibnamefont {Dou}},\
  }\href {\doibase 10.1103/PhysRevA.109.062432} {\bibfield  {journal} {\bibinfo
   {journal} {Phys. Rev. A}\ }\textbf {\bibinfo {volume} {109}},\ \bibinfo
  {pages} {062432} (\bibinfo {year} {2024})}\BibitemShut {NoStop}%
\bibitem [{\citenamefont {Dou}\ and\ \citenamefont
  {Yang}(2023)}]{PhysRevA.107.023725}%
  \BibitemOpen
  \bibfield  {author} {\bibinfo {author} {\bibfnamefont {F.-Q.}\ \bibnamefont
  {Dou}}\ and\ \bibinfo {author} {\bibfnamefont {F.-M.}\ \bibnamefont {Yang}},\
  }\href {\doibase 10.1103/PhysRevA.107.023725} {\bibfield  {journal} {\bibinfo
   {journal} {Phys. Rev. A}\ }\textbf {\bibinfo {volume} {107}},\ \bibinfo
  {pages} {023725} (\bibinfo {year} {2023})}\BibitemShut {NoStop}%
\bibitem [{\citenamefont {Tabesh}\ \emph {et~al.}(2020)\citenamefont {Tabesh},
  \citenamefont {Kamin},\ and\ \citenamefont {Salimi}}]{PhysRevA.102.052223}%
  \BibitemOpen
  \bibfield  {author} {\bibinfo {author} {\bibfnamefont {F.~T.}\ \bibnamefont
  {Tabesh}}, \bibinfo {author} {\bibfnamefont {F.~H.}\ \bibnamefont {Kamin}}, \
  and\ \bibinfo {author} {\bibfnamefont {S.}~\bibnamefont {Salimi}},\ }\href
  {\doibase 10.1103/PhysRevA.102.052223} {\bibfield  {journal} {\bibinfo
  {journal} {Phys. Rev. A}\ }\textbf {\bibinfo {volume} {102}},\ \bibinfo
  {pages} {052223} (\bibinfo {year} {2020})}\BibitemShut {NoStop}%
\bibitem [{\citenamefont {Caravelli}\ \emph {et~al.}(2021)\citenamefont
  {Caravelli}, \citenamefont {Yan}, \citenamefont {Garc{\'{i}}a-Pintos},\ and\
  \citenamefont {Hamma}}]{Caravelli2021energystorage}%
  \BibitemOpen
  \bibfield  {author} {\bibinfo {author} {\bibfnamefont {F.}~\bibnamefont
  {Caravelli}}, \bibinfo {author} {\bibfnamefont {B.}~\bibnamefont {Yan}},
  \bibinfo {author} {\bibfnamefont {L.~P.}\ \bibnamefont
  {Garc{\'{i}}a-Pintos}}, \ and\ \bibinfo {author} {\bibfnamefont
  {A.}~\bibnamefont {Hamma}},\ }\href {\doibase 10.22331/q-2021-07-15-505}
  {\bibfield  {journal} {\bibinfo  {journal} {{Quantum}}\ }\textbf {\bibinfo
  {volume} {5}},\ \bibinfo {pages} {505} (\bibinfo {year} {2021})}\BibitemShut
  {NoStop}%
\bibitem [{\citenamefont {Zakavati}\ \emph {et~al.}(2021)\citenamefont
  {Zakavati}, \citenamefont {Tabesh},\ and\ \citenamefont
  {Salimi}}]{PhysRevE.104.054117}%
  \BibitemOpen
  \bibfield  {author} {\bibinfo {author} {\bibfnamefont {S.}~\bibnamefont
  {Zakavati}}, \bibinfo {author} {\bibfnamefont {F.~T.}\ \bibnamefont
  {Tabesh}}, \ and\ \bibinfo {author} {\bibfnamefont {S.}~\bibnamefont
  {Salimi}},\ }\href {\doibase 10.1103/PhysRevE.104.054117} {\bibfield
  {journal} {\bibinfo  {journal} {Phys. Rev. E}\ }\textbf {\bibinfo {volume}
  {104}},\ \bibinfo {pages} {054117} (\bibinfo {year} {2021})}\BibitemShut
  {NoStop}%
\bibitem [{\citenamefont {Zhang}\ \emph {et~al.}(2024)\citenamefont {Zhang},
  \citenamefont {Ma}, \citenamefont {Jiang}, \citenamefont {Yu}, \citenamefont
  {Jin},\ and\ \citenamefont {Chen}}]{PhysRevA.110.032211}%
  \BibitemOpen
  \bibfield  {author} {\bibinfo {author} {\bibfnamefont {D.-Y.}\ \bibnamefont
  {Zhang}}, \bibinfo {author} {\bibfnamefont {S.-Q.}\ \bibnamefont {Ma}},
  \bibinfo {author} {\bibfnamefont {Y.-X.}\ \bibnamefont {Jiang}}, \bibinfo
  {author} {\bibfnamefont {Y.-B.}\ \bibnamefont {Yu}}, \bibinfo {author}
  {\bibfnamefont {G.-R.}\ \bibnamefont {Jin}}, \ and\ \bibinfo {author}
  {\bibfnamefont {A.-X.}\ \bibnamefont {Chen}},\ }\href {\doibase
  10.1103/PhysRevA.110.032211} {\bibfield  {journal} {\bibinfo  {journal}
  {Phys. Rev. A}\ }\textbf {\bibinfo {volume} {110}},\ \bibinfo {pages}
  {032211} (\bibinfo {year} {2024})}\BibitemShut {NoStop}%
\bibitem [{\citenamefont {Hovhannisyan}\ \emph {et~al.}(2013)\citenamefont
  {Hovhannisyan}, \citenamefont {Perarnau-Llobet}, \citenamefont {Huber},\ and\
  \citenamefont {Ac\'{\i}n}}]{PhysRevLett.111.240401}%
  \BibitemOpen
  \bibfield  {author} {\bibinfo {author} {\bibfnamefont {K.~V.}\ \bibnamefont
  {Hovhannisyan}}, \bibinfo {author} {\bibfnamefont {M.}~\bibnamefont
  {Perarnau-Llobet}}, \bibinfo {author} {\bibfnamefont {M.}~\bibnamefont
  {Huber}}, \ and\ \bibinfo {author} {\bibfnamefont {A.}~\bibnamefont
  {Ac\'{\i}n}},\ }\href {\doibase 10.1103/PhysRevLett.111.240401} {\bibfield
  {journal} {\bibinfo  {journal} {Phys. Rev. Lett.}\ }\textbf {\bibinfo
  {volume} {111}},\ \bibinfo {pages} {240401} (\bibinfo {year}
  {2013})}\BibitemShut {NoStop}%
\bibitem [{\citenamefont {Gumberidze}\ \emph {et~al.}(2019)\citenamefont
  {Gumberidze}, \citenamefont {Kolář},\ and\ \citenamefont
  {Filip}}]{Gumberidze2019}%
  \BibitemOpen
  \bibfield  {author} {\bibinfo {author} {\bibfnamefont {M.}~\bibnamefont
  {Gumberidze}}, \bibinfo {author} {\bibfnamefont {M.}~\bibnamefont {Kolář}},
  \ and\ \bibinfo {author} {\bibfnamefont {R.}~\bibnamefont {Filip}},\ }\href
  {\doibase 10.1038/s41598-019-56158-8} {\bibfield  {journal} {\bibinfo
  {journal} {Sci. Rep.}\ }\textbf {\bibinfo {volume} {9}},\ \bibinfo {pages}
  {19628} (\bibinfo {year} {2019})}\BibitemShut {NoStop}%
\bibitem [{\citenamefont {Gyhm}\ and\ \citenamefont
  {Fischer}(2024)}]{10.1116/5.0184903}%
  \BibitemOpen
  \bibfield  {author} {\bibinfo {author} {\bibfnamefont {J.-Y.}\ \bibnamefont
  {Gyhm}}\ and\ \bibinfo {author} {\bibfnamefont {U.~R.}\ \bibnamefont
  {Fischer}},\ }\href {\doibase 10.1116/5.0184903} {\bibfield  {journal}
  {\bibinfo  {journal} {AVS Quantum Sci.}\ }\textbf {\bibinfo {volume} {6}},\
  \bibinfo {pages} {012001} (\bibinfo {year} {2024})}\BibitemShut {NoStop}%
\bibitem [{\citenamefont {Zhu}\ \emph {et~al.}(2023)\citenamefont {Zhu},
  \citenamefont {Chen}, \citenamefont {Hasegawa},\ and\ \citenamefont
  {Xue}}]{PhysRevLett.131.240401}%
  \BibitemOpen
  \bibfield  {author} {\bibinfo {author} {\bibfnamefont {G.}~\bibnamefont
  {Zhu}}, \bibinfo {author} {\bibfnamefont {Y.}~\bibnamefont {Chen}}, \bibinfo
  {author} {\bibfnamefont {Y.}~\bibnamefont {Hasegawa}}, \ and\ \bibinfo
  {author} {\bibfnamefont {P.}~\bibnamefont {Xue}},\ }\href {\doibase
  10.1103/PhysRevLett.131.240401} {\bibfield  {journal} {\bibinfo  {journal}
  {Phys. Rev. Lett.}\ }\textbf {\bibinfo {volume} {131}},\ \bibinfo {pages}
  {240401} (\bibinfo {year} {2023})}\BibitemShut {NoStop}%
\bibitem [{\citenamefont {Dou}\ \emph {et~al.}(2021)\citenamefont {Dou},
  \citenamefont {Wang},\ and\ \citenamefont {Sun}}]{Dou2021}%
  \BibitemOpen
  \bibfield  {author} {\bibinfo {author} {\bibfnamefont {F.-Q.}\ \bibnamefont
  {Dou}}, \bibinfo {author} {\bibfnamefont {Y.-J.}\ \bibnamefont {Wang}}, \
  and\ \bibinfo {author} {\bibfnamefont {J.-A.}\ \bibnamefont {Sun}},\ }\href
  {\doibase 10.1007/s11467-021-1130-5} {\bibfield  {journal} {\bibinfo
  {journal} {Front. Phys.}\ }\textbf {\bibinfo {volume} {17}},\ \bibinfo
  {pages} {31503} (\bibinfo {year} {2021})}\BibitemShut {NoStop}%
\bibitem [{\citenamefont {Gyhm}\ \emph {et~al.}(2022)\citenamefont {Gyhm},
  \citenamefont {\ifmmode~\check{S}\else \v{S}\fi{}afr\'anek},\ and\
  \citenamefont {Rosa}}]{PhysRevLett.128.140501}%
  \BibitemOpen
  \bibfield  {author} {\bibinfo {author} {\bibfnamefont {J.-Y.}\ \bibnamefont
  {Gyhm}}, \bibinfo {author} {\bibfnamefont {D.}~\bibnamefont
  {\ifmmode~\check{S}\else \v{S}\fi{}afr\'anek}}, \ and\ \bibinfo {author}
  {\bibfnamefont {D.}~\bibnamefont {Rosa}},\ }\href {\doibase
  10.1103/PhysRevLett.128.140501} {\bibfield  {journal} {\bibinfo  {journal}
  {Phys. Rev. Lett.}\ }\textbf {\bibinfo {volume} {128}},\ \bibinfo {pages}
  {140501} (\bibinfo {year} {2022})}\BibitemShut {NoStop}%
\bibitem [{\citenamefont {Zhang}\ \emph {et~al.}(2019)\citenamefont {Zhang},
  \citenamefont {Yang}, \citenamefont {Fu},\ and\ \citenamefont
  {Wang}}]{PhysRevE.99.052106}%
  \BibitemOpen
  \bibfield  {author} {\bibinfo {author} {\bibfnamefont {Y.-Y.}\ \bibnamefont
  {Zhang}}, \bibinfo {author} {\bibfnamefont {T.-R.}\ \bibnamefont {Yang}},
  \bibinfo {author} {\bibfnamefont {L.}~\bibnamefont {Fu}}, \ and\ \bibinfo
  {author} {\bibfnamefont {X.}~\bibnamefont {Wang}},\ }\href {\doibase
  10.1103/PhysRevE.99.052106} {\bibfield  {journal} {\bibinfo  {journal} {Phys.
  Rev. E}\ }\textbf {\bibinfo {volume} {99}},\ \bibinfo {pages} {052106}
  (\bibinfo {year} {2019})}\BibitemShut {NoStop}%
\bibitem [{\citenamefont {Downing}\ and\ \citenamefont
  {Ukhtary}(2024)}]{PhysRevA.109.052206}%
  \BibitemOpen
  \bibfield  {author} {\bibinfo {author} {\bibfnamefont {C.~A.}\ \bibnamefont
  {Downing}}\ and\ \bibinfo {author} {\bibfnamefont {M.~S.}\ \bibnamefont
  {Ukhtary}},\ }\href {\doibase 10.1103/PhysRevA.109.052206} {\bibfield
  {journal} {\bibinfo  {journal} {Phys. Rev. A}\ }\textbf {\bibinfo {volume}
  {109}},\ \bibinfo {pages} {052206} (\bibinfo {year} {2024})}\BibitemShut
  {NoStop}%
\bibitem [{\citenamefont {Mazzoncini}\ \emph {et~al.}(2023)\citenamefont
  {Mazzoncini}, \citenamefont {Cavina}, \citenamefont {Andolina}, \citenamefont
  {Erdman},\ and\ \citenamefont {Giovannetti}}]{PhysRevA.107.032218}%
  \BibitemOpen
  \bibfield  {author} {\bibinfo {author} {\bibfnamefont {F.}~\bibnamefont
  {Mazzoncini}}, \bibinfo {author} {\bibfnamefont {V.}~\bibnamefont {Cavina}},
  \bibinfo {author} {\bibfnamefont {G.~M.}\ \bibnamefont {Andolina}}, \bibinfo
  {author} {\bibfnamefont {P.~A.}\ \bibnamefont {Erdman}}, \ and\ \bibinfo
  {author} {\bibfnamefont {V.}~\bibnamefont {Giovannetti}},\ }\href {\doibase
  10.1103/PhysRevA.107.032218} {\bibfield  {journal} {\bibinfo  {journal}
  {Phys. Rev. A}\ }\textbf {\bibinfo {volume} {107}},\ \bibinfo {pages}
  {032218} (\bibinfo {year} {2023})}\BibitemShut {NoStop}%
\bibitem [{\citenamefont {Rodríguez}\ \emph {et~al.}(2024)\citenamefont
  {Rodríguez}, \citenamefont {Ahmadi}, \citenamefont {Suárez}, \citenamefont
  {Mazurek}, \citenamefont {Barzanjeh},\ and\ \citenamefont
  {Horodecki}}]{Rodrguez_2024}%
  \BibitemOpen
  \bibfield  {author} {\bibinfo {author} {\bibfnamefont {R.~R.}\ \bibnamefont
  {Rodríguez}}, \bibinfo {author} {\bibfnamefont {B.}~\bibnamefont {Ahmadi}},
  \bibinfo {author} {\bibfnamefont {G.}~\bibnamefont {Suárez}}, \bibinfo
  {author} {\bibfnamefont {P.}~\bibnamefont {Mazurek}}, \bibinfo {author}
  {\bibfnamefont {S.}~\bibnamefont {Barzanjeh}}, \ and\ \bibinfo {author}
  {\bibfnamefont {P.}~\bibnamefont {Horodecki}},\ }\href {\doibase
  10.1088/1367-2630/ad3843} {\bibfield  {journal} {\bibinfo  {journal} {New J.
  Phys.}\ }\textbf {\bibinfo {volume} {26}},\ \bibinfo {pages} {043004}
  (\bibinfo {year} {2024})}\BibitemShut {NoStop}%
\bibitem [{\citenamefont {Gemme}\ \emph {et~al.}(2024)\citenamefont {Gemme},
  \citenamefont {Grossi}, \citenamefont {Vallecorsa}, \citenamefont
  {Sassetti},\ and\ \citenamefont {Ferraro}}]{PhysRevResearch.6.023091}%
  \BibitemOpen
  \bibfield  {author} {\bibinfo {author} {\bibfnamefont {G.}~\bibnamefont
  {Gemme}}, \bibinfo {author} {\bibfnamefont {M.}~\bibnamefont {Grossi}},
  \bibinfo {author} {\bibfnamefont {S.}~\bibnamefont {Vallecorsa}}, \bibinfo
  {author} {\bibfnamefont {M.}~\bibnamefont {Sassetti}}, \ and\ \bibinfo
  {author} {\bibfnamefont {D.}~\bibnamefont {Ferraro}},\ }\href {\doibase
  10.1103/PhysRevResearch.6.023091} {\bibfield  {journal} {\bibinfo  {journal}
  {Phys. Rev. Res.}\ }\textbf {\bibinfo {volume} {6}},\ \bibinfo {pages}
  {023091} (\bibinfo {year} {2024})}\BibitemShut {NoStop}%
\bibitem [{\citenamefont {Mitra}\ and\ \citenamefont
  {Srivastava}(2024)}]{PhysRevA.110.012227}%
  \BibitemOpen
  \bibfield  {author} {\bibinfo {author} {\bibfnamefont {A.}~\bibnamefont
  {Mitra}}\ and\ \bibinfo {author} {\bibfnamefont {S.~C.~L.}\ \bibnamefont
  {Srivastava}},\ }\href {\doibase 10.1103/PhysRevA.110.012227} {\bibfield
  {journal} {\bibinfo  {journal} {Phys. Rev. A}\ }\textbf {\bibinfo {volume}
  {110}},\ \bibinfo {pages} {012227} (\bibinfo {year} {2024})}\BibitemShut
  {NoStop}%
\bibitem [{\citenamefont {Yang}\ \emph
  {et~al.}(2024{\natexlab{b}})\citenamefont {Yang}, \citenamefont {Shi},
  \citenamefont {Wan}, \citenamefont {Zhang}, \citenamefont {Wang},\ and\
  \citenamefont {Yang}}]{PhysRevA.109.012204}%
  \BibitemOpen
  \bibfield  {author} {\bibinfo {author} {\bibfnamefont {H.-Y.}\ \bibnamefont
  {Yang}}, \bibinfo {author} {\bibfnamefont {H.-L.}\ \bibnamefont {Shi}},
  \bibinfo {author} {\bibfnamefont {Q.-K.}\ \bibnamefont {Wan}}, \bibinfo
  {author} {\bibfnamefont {K.}~\bibnamefont {Zhang}}, \bibinfo {author}
  {\bibfnamefont {X.-H.}\ \bibnamefont {Wang}}, \ and\ \bibinfo {author}
  {\bibfnamefont {W.-L.}\ \bibnamefont {Yang}},\ }\href {\doibase
  10.1103/PhysRevA.109.012204} {\bibfield  {journal} {\bibinfo  {journal}
  {Phys. Rev. A}\ }\textbf {\bibinfo {volume} {109}},\ \bibinfo {pages}
  {012204} (\bibinfo {year} {2024}{\natexlab{b}})}\BibitemShut {NoStop}%
\bibitem [{\citenamefont {Song}\ \emph {et~al.}(2024)\citenamefont {Song},
  \citenamefont {Liu}, \citenamefont {Zhou}, \citenamefont {Yang},\ and\
  \citenamefont {An}}]{PhysRevLett.132.090401}%
  \BibitemOpen
  \bibfield  {author} {\bibinfo {author} {\bibfnamefont {W.-L.}\ \bibnamefont
  {Song}}, \bibinfo {author} {\bibfnamefont {H.-B.}\ \bibnamefont {Liu}},
  \bibinfo {author} {\bibfnamefont {B.}~\bibnamefont {Zhou}}, \bibinfo {author}
  {\bibfnamefont {W.-L.}\ \bibnamefont {Yang}}, \ and\ \bibinfo {author}
  {\bibfnamefont {J.-H.}\ \bibnamefont {An}},\ }\href {\doibase
  10.1103/PhysRevLett.132.090401} {\bibfield  {journal} {\bibinfo  {journal}
  {Phys. Rev. Lett.}\ }\textbf {\bibinfo {volume} {132}},\ \bibinfo {pages}
  {090401} (\bibinfo {year} {2024})}\BibitemShut {NoStop}%
\bibitem [{\citenamefont {Xu}\ \emph {et~al.}(2024)\citenamefont {Xu},
  \citenamefont {Li}, \citenamefont {Zhu},\ and\ \citenamefont
  {Liu}}]{PhysRevE.109.054132}%
  \BibitemOpen
  \bibfield  {author} {\bibinfo {author} {\bibfnamefont {K.}~\bibnamefont
  {Xu}}, \bibinfo {author} {\bibfnamefont {H.-G.}\ \bibnamefont {Li}}, \bibinfo
  {author} {\bibfnamefont {H.-J.}\ \bibnamefont {Zhu}}, \ and\ \bibinfo
  {author} {\bibfnamefont {W.-M.}\ \bibnamefont {Liu}},\ }\href {\doibase
  10.1103/PhysRevE.109.054132} {\bibfield  {journal} {\bibinfo  {journal}
  {Phys. Rev. E}\ }\textbf {\bibinfo {volume} {109}},\ \bibinfo {pages}
  {054132} (\bibinfo {year} {2024})}\BibitemShut {NoStop}%
\bibitem [{\citenamefont {Bhanja}\ \emph {et~al.}(2024)\citenamefont {Bhanja},
  \citenamefont {Tiwari},\ and\ \citenamefont
  {Banerjee}}]{PhysRevA.109.012224}%
  \BibitemOpen
  \bibfield  {author} {\bibinfo {author} {\bibfnamefont {G.}~\bibnamefont
  {Bhanja}}, \bibinfo {author} {\bibfnamefont {D.}~\bibnamefont {Tiwari}}, \
  and\ \bibinfo {author} {\bibfnamefont {S.}~\bibnamefont {Banerjee}},\ }\href
  {\doibase 10.1103/PhysRevA.109.012224} {\bibfield  {journal} {\bibinfo
  {journal} {Phys. Rev. A}\ }\textbf {\bibinfo {volume} {109}},\ \bibinfo
  {pages} {012224} (\bibinfo {year} {2024})}\BibitemShut {NoStop}%
\bibitem [{\citenamefont {Carleo}\ \emph {et~al.}(2019)\citenamefont {Carleo},
  \citenamefont {Cirac}, \citenamefont {Cranmer}, \citenamefont {Daudet},
  \citenamefont {Schuld}, \citenamefont {Tishby}, \citenamefont
  {Vogt-Maranto},\ and\ \citenamefont {Zdeborov\'a}}]{RevModPhys.91.045002}%
  \BibitemOpen
  \bibfield  {author} {\bibinfo {author} {\bibfnamefont {G.}~\bibnamefont
  {Carleo}}, \bibinfo {author} {\bibfnamefont {I.}~\bibnamefont {Cirac}},
  \bibinfo {author} {\bibfnamefont {K.}~\bibnamefont {Cranmer}}, \bibinfo
  {author} {\bibfnamefont {L.}~\bibnamefont {Daudet}}, \bibinfo {author}
  {\bibfnamefont {M.}~\bibnamefont {Schuld}}, \bibinfo {author} {\bibfnamefont
  {N.}~\bibnamefont {Tishby}}, \bibinfo {author} {\bibfnamefont
  {L.}~\bibnamefont {Vogt-Maranto}}, \ and\ \bibinfo {author} {\bibfnamefont
  {L.}~\bibnamefont {Zdeborov\'a}},\ }\href {\doibase
  10.1103/RevModPhys.91.045002} {\bibfield  {journal} {\bibinfo  {journal}
  {Rev. Mod. Phys.}\ }\textbf {\bibinfo {volume} {91}},\ \bibinfo {pages}
  {045002} (\bibinfo {year} {2019})}\BibitemShut {NoStop}%
\bibitem [{\citenamefont {Zhang}\ \emph {et~al.}(2018)\citenamefont {Zhang},
  \citenamefont {Shen},\ and\ \citenamefont {Zhai}}]{PhysRevLett.120.066401}%
  \BibitemOpen
  \bibfield  {author} {\bibinfo {author} {\bibfnamefont {P.}~\bibnamefont
  {Zhang}}, \bibinfo {author} {\bibfnamefont {H.}~\bibnamefont {Shen}}, \ and\
  \bibinfo {author} {\bibfnamefont {H.}~\bibnamefont {Zhai}},\ }\href {\doibase
  10.1103/PhysRevLett.120.066401} {\bibfield  {journal} {\bibinfo  {journal}
  {Phys. Rev. Lett.}\ }\textbf {\bibinfo {volume} {120}},\ \bibinfo {pages}
  {066401} (\bibinfo {year} {2018})}\BibitemShut {NoStop}%
\bibitem [{\citenamefont {Kottmann}\ \emph {et~al.}(2021)\citenamefont
  {Kottmann}, \citenamefont {Metz}, \citenamefont {Fraxanet},\ and\
  \citenamefont {Baldelli}}]{PhysRevResearch.3.043184}%
  \BibitemOpen
  \bibfield  {author} {\bibinfo {author} {\bibfnamefont {K.}~\bibnamefont
  {Kottmann}}, \bibinfo {author} {\bibfnamefont {F.}~\bibnamefont {Metz}},
  \bibinfo {author} {\bibfnamefont {J.}~\bibnamefont {Fraxanet}}, \ and\
  \bibinfo {author} {\bibfnamefont {N.}~\bibnamefont {Baldelli}},\ }\href
  {\doibase 10.1103/PhysRevResearch.3.043184} {\bibfield  {journal} {\bibinfo
  {journal} {Phys. Rev. Res.}\ }\textbf {\bibinfo {volume} {3}},\ \bibinfo
  {pages} {043184} (\bibinfo {year} {2021})}\BibitemShut {NoStop}%
\bibitem [{\citenamefont {Jasinski}\ \emph {et~al.}(2020)\citenamefont
  {Jasinski}, \citenamefont {Montaner}, \citenamefont {Forrey}, \citenamefont
  {Yang}, \citenamefont {Stancil}, \citenamefont {Balakrishnan}, \citenamefont
  {Dai}, \citenamefont {Vargas-Hern\'andez},\ and\ \citenamefont
  {Krems}}]{PhysRevResearch.2.032051}%
  \BibitemOpen
  \bibfield  {author} {\bibinfo {author} {\bibfnamefont {A.}~\bibnamefont
  {Jasinski}}, \bibinfo {author} {\bibfnamefont {J.}~\bibnamefont {Montaner}},
  \bibinfo {author} {\bibfnamefont {R.~C.}\ \bibnamefont {Forrey}}, \bibinfo
  {author} {\bibfnamefont {B.~H.}\ \bibnamefont {Yang}}, \bibinfo {author}
  {\bibfnamefont {P.~C.}\ \bibnamefont {Stancil}}, \bibinfo {author}
  {\bibfnamefont {N.}~\bibnamefont {Balakrishnan}}, \bibinfo {author}
  {\bibfnamefont {J.}~\bibnamefont {Dai}}, \bibinfo {author} {\bibfnamefont
  {R.~A.}\ \bibnamefont {Vargas-Hern\'andez}}, \ and\ \bibinfo {author}
  {\bibfnamefont {R.~V.}\ \bibnamefont {Krems}},\ }\href {\doibase
  10.1103/PhysRevResearch.2.032051} {\bibfield  {journal} {\bibinfo  {journal}
  {Phys. Rev. Res.}\ }\textbf {\bibinfo {volume} {2}},\ \bibinfo {pages}
  {032051} (\bibinfo {year} {2020})}\BibitemShut {NoStop}%
\bibitem [{\citenamefont {Jerbi}\ \emph {et~al.}(2021)\citenamefont {Jerbi},
  \citenamefont {Trenkwalder}, \citenamefont {Poulsen~Nautrup}, \citenamefont
  {Briegel},\ and\ \citenamefont {Dunjko}}]{PRXQuantum.2.010328}%
  \BibitemOpen
  \bibfield  {author} {\bibinfo {author} {\bibfnamefont {S.}~\bibnamefont
  {Jerbi}}, \bibinfo {author} {\bibfnamefont {L.~M.}\ \bibnamefont
  {Trenkwalder}}, \bibinfo {author} {\bibfnamefont {H.}~\bibnamefont
  {Poulsen~Nautrup}}, \bibinfo {author} {\bibfnamefont {H.~J.}\ \bibnamefont
  {Briegel}}, \ and\ \bibinfo {author} {\bibfnamefont {V.}~\bibnamefont
  {Dunjko}},\ }\href {\doibase 10.1103/PRXQuantum.2.010328} {\bibfield
  {journal} {\bibinfo  {journal} {PRX Quantum}\ }\textbf {\bibinfo {volume}
  {2}},\ \bibinfo {pages} {010328} (\bibinfo {year} {2021})}\BibitemShut
  {NoStop}%
\bibitem [{\citenamefont {F\"osel}\ \emph {et~al.}(2018)\citenamefont
  {F\"osel}, \citenamefont {Tighineanu}, \citenamefont {Weiss},\ and\
  \citenamefont {Marquardt}}]{PhysRevX.8.031084}%
  \BibitemOpen
  \bibfield  {author} {\bibinfo {author} {\bibfnamefont {T.}~\bibnamefont
  {F\"osel}}, \bibinfo {author} {\bibfnamefont {P.}~\bibnamefont {Tighineanu}},
  \bibinfo {author} {\bibfnamefont {T.}~\bibnamefont {Weiss}}, \ and\ \bibinfo
  {author} {\bibfnamefont {F.}~\bibnamefont {Marquardt}},\ }\href {\doibase
  10.1103/PhysRevX.8.031084} {\bibfield  {journal} {\bibinfo  {journal} {Phys.
  Rev. X}\ }\textbf {\bibinfo {volume} {8}},\ \bibinfo {pages} {031084}
  (\bibinfo {year} {2018})}\BibitemShut {NoStop}%
\bibitem [{\citenamefont {Borah}\ \emph {et~al.}(2021)\citenamefont {Borah},
  \citenamefont {Sarma}, \citenamefont {Kewming}, \citenamefont {Milburn},\
  and\ \citenamefont {Twamley}}]{PhysRevLett.127.190403}%
  \BibitemOpen
  \bibfield  {author} {\bibinfo {author} {\bibfnamefont {S.}~\bibnamefont
  {Borah}}, \bibinfo {author} {\bibfnamefont {B.}~\bibnamefont {Sarma}},
  \bibinfo {author} {\bibfnamefont {M.}~\bibnamefont {Kewming}}, \bibinfo
  {author} {\bibfnamefont {G.~J.}\ \bibnamefont {Milburn}}, \ and\ \bibinfo
  {author} {\bibfnamefont {J.}~\bibnamefont {Twamley}},\ }\href {\doibase
  10.1103/PhysRevLett.127.190403} {\bibfield  {journal} {\bibinfo  {journal}
  {Phys. Rev. Lett.}\ }\textbf {\bibinfo {volume} {127}},\ \bibinfo {pages}
  {190403} (\bibinfo {year} {2021})}\BibitemShut {NoStop}%
\bibitem [{\citenamefont {Zhang}\ \emph {et~al.}(2020)\citenamefont {Zhang},
  \citenamefont {Zheng}, \citenamefont {Zhang},\ and\ \citenamefont
  {Deng}}]{PhysRevLett.125.170501}%
  \BibitemOpen
  \bibfield  {author} {\bibinfo {author} {\bibfnamefont {Y.-H.}\ \bibnamefont
  {Zhang}}, \bibinfo {author} {\bibfnamefont {P.-L.}\ \bibnamefont {Zheng}},
  \bibinfo {author} {\bibfnamefont {Y.}~\bibnamefont {Zhang}}, \ and\ \bibinfo
  {author} {\bibfnamefont {D.-L.}\ \bibnamefont {Deng}},\ }\href {\doibase
  10.1103/PhysRevLett.125.170501} {\bibfield  {journal} {\bibinfo  {journal}
  {Phys. Rev. Lett.}\ }\textbf {\bibinfo {volume} {125}},\ \bibinfo {pages}
  {170501} (\bibinfo {year} {2020})}\BibitemShut {NoStop}%
\bibitem [{\citenamefont {Bolens}\ and\ \citenamefont
  {Heyl}(2021)}]{PhysRevLett.127.110502}%
  \BibitemOpen
  \bibfield  {author} {\bibinfo {author} {\bibfnamefont {A.}~\bibnamefont
  {Bolens}}\ and\ \bibinfo {author} {\bibfnamefont {M.}~\bibnamefont {Heyl}},\
  }\href {\doibase 10.1103/PhysRevLett.127.110502} {\bibfield  {journal}
  {\bibinfo  {journal} {Phys. Rev. Lett.}\ }\textbf {\bibinfo {volume} {127}},\
  \bibinfo {pages} {110502} (\bibinfo {year} {2021})}\BibitemShut {NoStop}%
\bibitem [{\citenamefont {Erdman}\ \emph {et~al.}(2024)\citenamefont {Erdman},
  \citenamefont {Andolina}, \citenamefont {Giovannetti},\ and\ \citenamefont
  {No\'e}}]{PhysRevLett.133.243602}%
  \BibitemOpen
  \bibfield  {author} {\bibinfo {author} {\bibfnamefont {P.~A.}\ \bibnamefont
  {Erdman}}, \bibinfo {author} {\bibfnamefont {G.~M.}\ \bibnamefont
  {Andolina}}, \bibinfo {author} {\bibfnamefont {V.}~\bibnamefont
  {Giovannetti}}, \ and\ \bibinfo {author} {\bibfnamefont {F.}~\bibnamefont
  {No\'e}},\ }\href {\doibase 10.1103/PhysRevLett.133.243602} {\bibfield
  {journal} {\bibinfo  {journal} {Phys. Rev. Lett.}\ }\textbf {\bibinfo
  {volume} {133}},\ \bibinfo {pages} {243602} (\bibinfo {year}
  {2024})}\BibitemShut {NoStop}%
\bibitem [{\citenamefont {Rodr\'{\i}guez}\ \emph {et~al.}(2023)\citenamefont
  {Rodr\'{\i}guez}, \citenamefont {Rosa},\ and\ \citenamefont
  {Olle}}]{PhysRevA.108.042618}%
  \BibitemOpen
  \bibfield  {author} {\bibinfo {author} {\bibfnamefont {C.}~\bibnamefont
  {Rodr\'{\i}guez}}, \bibinfo {author} {\bibfnamefont {D.}~\bibnamefont
  {Rosa}}, \ and\ \bibinfo {author} {\bibfnamefont {J.}~\bibnamefont {Olle}},\
  }\href {\doibase 10.1103/PhysRevA.108.042618} {\bibfield  {journal} {\bibinfo
   {journal} {Phys. Rev. A}\ }\textbf {\bibinfo {volume} {108}},\ \bibinfo
  {pages} {042618} (\bibinfo {year} {2023})}\BibitemShut {NoStop}%
\bibitem [{\citenamefont {Haarnoja}\ \emph
  {et~al.}(2018{\natexlab{a}})\citenamefont {Haarnoja}, \citenamefont {Zhou},
  \citenamefont {Hartikainen}, \citenamefont {Tucker}, \citenamefont {Ha},
  \citenamefont {Tan}, \citenamefont {Kumar}, \citenamefont {Zhu},
  \citenamefont {Gupta}, \citenamefont {Abbeel},\ and\ \citenamefont
  {Levine}}]{haarnoja2019softactorcriticalgorithmsapplications}%
  \BibitemOpen
  \bibfield  {author} {\bibinfo {author} {\bibfnamefont {T.}~\bibnamefont
  {Haarnoja}}, \bibinfo {author} {\bibfnamefont {A.}~\bibnamefont {Zhou}},
  \bibinfo {author} {\bibfnamefont {K.}~\bibnamefont {Hartikainen}}, \bibinfo
  {author} {\bibfnamefont {G.}~\bibnamefont {Tucker}}, \bibinfo {author}
  {\bibfnamefont {S.}~\bibnamefont {Ha}}, \bibinfo {author} {\bibfnamefont
  {J.}~\bibnamefont {Tan}}, \bibinfo {author} {\bibfnamefont {V.}~\bibnamefont
  {Kumar}}, \bibinfo {author} {\bibfnamefont {H.}~\bibnamefont {Zhu}}, \bibinfo
  {author} {\bibfnamefont {A.}~\bibnamefont {Gupta}}, \bibinfo {author}
  {\bibfnamefont {P.}~\bibnamefont {Abbeel}}, \ and\ \bibinfo {author}
  {\bibfnamefont {S.}~\bibnamefont {Levine}},\ }\href
  {https://arxiv.org/abs/1812.05905} {} (\bibinfo {year}
  {2018}{\natexlab{a}}),\ \Eprint {http://arxiv.org/abs/1812.05905}
  {arXiv:1812.05905} \BibitemShut {NoStop}%
\bibitem [{\citenamefont {Haarnoja}\ \emph
  {et~al.}(2018{\natexlab{b}})\citenamefont {Haarnoja}, \citenamefont {Zhou},
  \citenamefont {Abbeel},\ and\ \citenamefont
  {Levine}}]{haarnoja2018softactorcriticoffpolicymaximum}%
  \BibitemOpen
  \bibfield  {author} {\bibinfo {author} {\bibfnamefont {T.}~\bibnamefont
  {Haarnoja}}, \bibinfo {author} {\bibfnamefont {A.}~\bibnamefont {Zhou}},
  \bibinfo {author} {\bibfnamefont {P.}~\bibnamefont {Abbeel}}, \ and\ \bibinfo
  {author} {\bibfnamefont {S.}~\bibnamefont {Levine}},\ }\href
  {https://arxiv.org/abs/1801.01290} {} (\bibinfo {year}
  {2018}{\natexlab{b}}),\ \Eprint {http://arxiv.org/abs/1801.01290}
  {arXiv:1801.01290} \BibitemShut {NoStop}%
\bibitem [{\citenamefont {Paszke}\ \emph {et~al.}(2019)\citenamefont {Paszke},
  \citenamefont {Gross}, \citenamefont {Massa}, \citenamefont {Lerer},
  \citenamefont {Bradbury}, \citenamefont {Chanan}, \citenamefont {Killeen},
  \citenamefont {Lin}, \citenamefont {Gimelshein}, \citenamefont {Antiga},
  \citenamefont {Desmaison}, \citenamefont {K{\"o}pf}, \citenamefont {Yang},
  \citenamefont {DeVito}, \citenamefont {Raison}, \citenamefont {Tejani},
  \citenamefont {Chilamkurthy}, \citenamefont {Steiner}, \citenamefont {Fang},
  \citenamefont {Bai},\ and\ \citenamefont {Chintala}}]{Paszke2019}%
  \BibitemOpen
  \bibfield  {author} {\bibinfo {author} {\bibfnamefont {A.}~\bibnamefont
  {Paszke}}, \bibinfo {author} {\bibfnamefont {S.}~\bibnamefont {Gross}},
  \bibinfo {author} {\bibfnamefont {F.}~\bibnamefont {Massa}}, \bibinfo
  {author} {\bibfnamefont {A.}~\bibnamefont {Lerer}}, \bibinfo {author}
  {\bibfnamefont {J.}~\bibnamefont {Bradbury}}, \bibinfo {author}
  {\bibfnamefont {G.}~\bibnamefont {Chanan}}, \bibinfo {author} {\bibfnamefont
  {T.}~\bibnamefont {Killeen}}, \bibinfo {author} {\bibfnamefont
  {Z.}~\bibnamefont {Lin}}, \bibinfo {author} {\bibfnamefont {N.}~\bibnamefont
  {Gimelshein}}, \bibinfo {author} {\bibfnamefont {L.}~\bibnamefont {Antiga}},
  \bibinfo {author} {\bibfnamefont {A.}~\bibnamefont {Desmaison}}, \bibinfo
  {author} {\bibfnamefont {A.}~\bibnamefont {K{\"o}pf}}, \bibinfo {author}
  {\bibfnamefont {E.}~\bibnamefont {Yang}}, \bibinfo {author} {\bibfnamefont
  {Z.}~\bibnamefont {DeVito}}, \bibinfo {author} {\bibfnamefont
  {M.}~\bibnamefont {Raison}}, \bibinfo {author} {\bibfnamefont
  {A.}~\bibnamefont {Tejani}}, \bibinfo {author} {\bibfnamefont
  {S.}~\bibnamefont {Chilamkurthy}}, \bibinfo {author} {\bibfnamefont
  {B.}~\bibnamefont {Steiner}}, \bibinfo {author} {\bibfnamefont
  {L.}~\bibnamefont {Fang}}, \bibinfo {author} {\bibfnamefont {J.}~\bibnamefont
  {Bai}}, \ and\ \bibinfo {author} {\bibfnamefont {S.}~\bibnamefont
  {Chintala}},\ }\href
  {https://proceedings.neurips.cc/paper_files/paper/2019/file/bdbca288fee7f92f2bfa9f7012727740-Paper.pdf}
  {\bibfield  {journal} {\bibinfo  {journal} {Adv. Neural. Inf. Process.
  Syst.}\ }\textbf {\bibinfo {volume} {32}},\ \bibinfo {pages} {8026} (\bibinfo
  {year} {2019})}\BibitemShut {NoStop}%
\bibitem [{\citenamefont {Johansson}\ \emph {et~al.}(2013)\citenamefont
  {Johansson}, \citenamefont {Nation},\ and\ \citenamefont
  {Nori}}]{JOHANSSON20131234}%
  \BibitemOpen
  \bibfield  {author} {\bibinfo {author} {\bibfnamefont {J.}~\bibnamefont
  {Johansson}}, \bibinfo {author} {\bibfnamefont {P.}~\bibnamefont {Nation}}, \
  and\ \bibinfo {author} {\bibfnamefont {F.}~\bibnamefont {Nori}},\ }\href
  {\doibase https://doi.org/10.1016/j.cpc.2012.11.019} {\bibfield  {journal}
  {\bibinfo  {journal} {Comput. Phys. Commun.}\ }\textbf {\bibinfo {volume}
  {184}},\ \bibinfo {pages} {1234} (\bibinfo {year} {2013})}\BibitemShut
  {NoStop}%
\bibitem [{\citenamefont {Plenio}(2005)}]{PhysRevLett.95.090503}%
  \BibitemOpen
  \bibfield  {author} {\bibinfo {author} {\bibfnamefont {M.~B.}\ \bibnamefont
  {Plenio}},\ }\href {\doibase 10.1103/PhysRevLett.95.090503} {\bibfield
  {journal} {\bibinfo  {journal} {Phys. Rev. Lett.}\ }\textbf {\bibinfo
  {volume} {95}},\ \bibinfo {pages} {090503} (\bibinfo {year}
  {2005})}\BibitemShut {NoStop}%
\bibitem [{\citenamefont {Erdman}\ and\ \citenamefont
  {No{\'e}}(2022)}]{Erdman2022}%
  \BibitemOpen
  \bibfield  {author} {\bibinfo {author} {\bibfnamefont {P.~A.}\ \bibnamefont
  {Erdman}}\ and\ \bibinfo {author} {\bibfnamefont {F.}~\bibnamefont
  {No{\'e}}},\ }\href {\doibase 10.1038/s41534-021-00512-0} {\bibfield
  {journal} {\bibinfo  {journal} {npj Quantum Inf.}\ }\textbf {\bibinfo
  {volume} {8}},\ \bibinfo {pages} {1} (\bibinfo {year} {2022})}\BibitemShut
  {NoStop}%
\bibitem [{\citenamefont {Quach}\ and\ \citenamefont
  {Munro}(2020)}]{PhysRevApplied.14.024092}%
  \BibitemOpen
  \bibfield  {author} {\bibinfo {author} {\bibfnamefont {J.~Q.}\ \bibnamefont
  {Quach}}\ and\ \bibinfo {author} {\bibfnamefont {W.~J.}\ \bibnamefont
  {Munro}},\ }\href {\doibase 10.1103/PhysRevApplied.14.024092} {\bibfield
  {journal} {\bibinfo  {journal} {Phys. Rev. Appl.}\ }\textbf {\bibinfo
  {volume} {14}},\ \bibinfo {pages} {024092} (\bibinfo {year}
  {2020})}\BibitemShut {NoStop}%
\end{thebibliography}%
\end{document}